\documentclass{ws-rv9x6}
\usepackage{ws-rv-thm}
\usepackage[square]{ws-rv-van}
\usepackage{latexsym}
\usepackage{amsmath,amsfonts,feynmp,epsf}
\usepackage{amssymb}

\usepackage[]{hyperref}
\crop[off]{}
\makeatletter
\def\@makechapterhead#1{%
{\vbox to 110pt{
\def\thefootnote{\@fnsymbol\c@footnote}%
\vspace*{37pt}      
\parindent\z@\raggedright\reset@font
{\centering{
\vskip 0.25in
\vbox{\CTfont #1\par}\par}\par}\nobreak\vfill}}}
\makeatother

\makeindex

{\catcode`\|=\active
  \gdef\Braket#1{\left<\mathcode`\|"8000\let|\BraVert {#1}\right>}}
\def\BraVert{\egroup\,\mid@vertical\,\bgroup}
{\catcode`\|=\active
  \gdef\set#1{\mathinner{\lbrace\,{\mathcode`\|"8000\let|\midvert #1}\,\rbrace}}
  \gdef\Set#1{\left\{\:{\mathcode`\|"8000\let|\SetVert #1}\:\right\}}}
\def\midvert{\egroup\mid\bgroup}
\def\SetVert{\egroup\;\mid@vertical\;\bgroup}

\def\be{\begin{equation}}
\def\ee{\end{equation}}
\def\bea{\begin{eqnarray}}
\def\eea{\end{eqnarray}}
\def\ie{\begin{equation}\begin{aligned}}
\def\fe{\end{aligned}\end{equation}}

\newcommand{\A}{{\alpha}}
\newcommand{\B}{{\beta}}
\newcommand{\C}{{\gamma}}

\newcommand{\da}{{\dot\alpha}}
\newcommand{\db}{{\dot\beta}}

\newcommand{\dd}{{\dot\delta}}

\begin{document}

\chapter{TASI Lectures on the Higher Spin - CFT duality}\label{ch:higherSpin}

\author[Simone Giombi]{Simone Giombi}

\address{Department of Physics, Princeton University \\ Princeton, NJ 08544 \\
sgiombi@princeton.edu}

\begin{abstract}
In these lectures we give an overview of the duality between gravitational theories of massless higher spin fields 
in AdS and large $N$ vector models. We first review the original higher spin/vector 
model duality conjectured by Klebanov and Polyakov, and then discuss its generalizations involving 
vector models coupled to Chern-Simons gauge fields. We proceed to review some aspects of the theory of massless 
higher spins, starting with the Fronsdal equations for free fields and moving on 
to the fully non-linear Vasiliev equations in four dimensions. We end by reviewing some recent tests of 
the higher spin/vector model duality at the level of correlation functions and one-loop partition functions.
\end{abstract}
\body

\thispagestyle{empty}
\makeatletter
\renewcommand*\l@chapter[2]{}
\renewcommand*\l@schapter[2]{}
\renewcommand*\l@author[2]{}
\makeatother
\newpage
\setcounter{page}{1}
\setcounter{tocdepth}{2}
\smalltoc
\tableofcontents

\section{Introduction}

The aim of these notes is to provide a pedagogical introduction to some aspects of gravitational theories 
of massless higher spin (HS) fields, and in particular review recent progress in 
understanding their role in the context of the AdS/CFT correspondence. 

The problem of constructing field theories describing the consistent propagation 
and interactions of HS fields ($s>2$) has a long history and is a highly non-trivial one. 
In the case of massless fields in particular, the possible interactions are greatly restricted 
by the higher spin gauge symmetry that must be present to decouple unphysical polarizations. In flat spacetime,
powerful no-go theorems \cite{Weinberg:1964ew, Coleman:1967ad}, see e.g. \cite{Bekaert:2010hw} for a review,
severely constrain the S-matrix in the presence of massless HS particles. Essentially, having massless HS fields 
results in higher conservation laws that are too restrictive to allow, under general assumptions, 
for a non-trivial S-matrix. 

The flat spacetime no-go theorems can be circumvented if one assumes a non-zero cosmological constant, i.e. (A)dS
backgrounds, where S-matrix arguments do not apply. Consistent cubic vertices of massless HS fields in (A)dS 
were explicitly constructed by Fradkin and Vasiliev \cite{Fradkin:1987ks} and, remarkably, a fully non-linear theory of interacting 
higher spins in (A)dS was found by Vasiliev \cite{Vasiliev:1990en,Vasiliev:1992av,Vasiliev:1995dn,Vasiliev:1999ba}. 
Let us summarize some important features of the theory:
\begin{itemize}
\item Vasiliev constructed the exact non-linear equations of motion of the theory. They admit a vacuum solution which corresponds to 
the maximally symmetric space (A)dS. 
\item In the simplest bosonic version of the model in AdS$_4$, the spectrum of fluctuations around the vacuum includes a scalar 
field with $m^2=-2/\ell^2_{\rm AdS}$ (this value corresponds to a conformally coupled scalar. It is above the Breitenlohner-Freedman 
bound \cite{Breitenlohner:1982jf}), and a tower of HS massless fields of {\it all spins}, $s=1,2,3,\ldots, \infty$. A minimal truncation 
to a spectrum involving only even spins is possible. It is essential for the consistency of the theory that the spectrum includes an {\it infinite} 
tower of HS fields. 
\item While the original equations were written in 4d, generalizations to (A)dS$_{d+1}$ for all $d$ have been constructed \cite{Vasiliev:2003ev}. 
\item In all versions of the HS theory, the spectrum always includes a $s=2$ field, the graviton. Hence, HS gauge theories are in 
particular theories of gravity, which generalize the familiar Einstein theory by the inclusion of an infinite tower of massless higher spins.
\item The interactions among the higher spin fields involve higher derivatives; they carry inverse powers of the cosmological constant, 
and are singular in the flat space limit. The theory intrinsically lives in (A)dS. 
\item Quantization of the theory is not fully understood at the moment, mainly because a conventional action for the theory has not been constructed 
yet.\footnote{Interesting work on the action principle in HS theories have appeared in recent years 
\cite{Boulanger:2011dd,Doroud:2011xs,Boulanger:2012bj,Boulanger:2015kfa,Bonezzi:2016ttk} (see \cite{Vasiliev:1988sa} for earlier work), though it appears 
that an explicit action which at quadratic level reduces to the familiar free field actions is not available at present. Nevertheless, 
we believe that there should be an underlying, ordinary action in terms of the physical HS fields. For recent progress in constructing such 
actions perturbatively, see \cite{Sleight:2016dba,Bekaert:2015tva}.} However, one may speculate that due to the infinite dimensional 
HS symmetry, the theory may provide a UV finite model of quantum gravity. Some preliminary evidence towards this was recently given 
at one-loop level \cite{Giombi:2013fka, Giombi:2014iua}. 
\end{itemize} 

While the Vasiliev HS theory was constructed several years before the discovery of the AdS/CFT correspondence 
\cite{Maldacena:1997re,Gubser:1998bc,Witten:1998qj}, one can in fact see that, from the AdS/CFT viewpoint, it is natural that 
consistent theories of massless HS fields do exist: they have precisely the right structure to be dual to simple 
vector model CFTs at the boundary of AdS. Starting in Section \ref{HS-free-scalar} with a summary of the basics of HS symmetries in CFT, 
we move on to show how the various features of the HS theories listed above can be naturally understood from the CFT point of view. 
This will lead us to review the original higher spin/vector model conjecture of Klebanov and Polyakov \cite{Klebanov:2002ja} in Section \ref{CFT-AdS} and \ref{KP}, and 
its fermionic generalization \cite{Sezgin:2003pt, Leigh:2003gk} in Section \ref{ferm-HS-sec}. 
In Section \ref{CS-duality}, we review the recent extensions \cite{Giombi:2011kc, Aharony:2011jz} of these 
conjectures to Chern-Simons gauge theories coupled to vector models, and the  
``3d bosonization" duality \cite{Maldacena:2012sf, Aharony:2012nh} relating scalar and 
fermionic theories coupled to Chern-Simons. In Section \ref{Fronsdal}, we review the Fronsdal equations for free massless HS fields, and then move 
on to describe the frame-like formulation of HS fields which is at the basis of Vasiliev non-linear equations. These equations, in the 4d case,
are reviewed in Section \ref{Vas-Equations}. Finally, in Section \ref{3pt-tests} and \ref{OneLoop}, we review recent tests of the higher spin/vector 
model dualities at the level of 3-point correlation functions and vacuum partition functions. 

While these notes are meant to be pedagogical, they are by no means comprehensive, and there are a number of topics that will not be covered. 
In particular, our focus will be mainly on the AdS$_4$/CFT$_3$ dualities (and briefly on their higher dimensional generalizations). 
There is a large body of literature on the AdS$_3$/CFT$_2$ version of the higher spin/CFT duality \cite{Gaberdiel:2010pz}, 
see \cite{Gaberdiel:2012uj} for a review, which unfortunately we will not have time to cover.    

\section{Higher Spins from free CFT}
\label{HS-free-scalar}


To introduce the concept of HS currents and the corresponding symmetries, let us start with a very simple theory: a free 
massless scalar field
\begin{equation}
S = \int d^d x \frac{1}{2}\left(\partial_{\mu}\phi\right)^2\,.
\end{equation}
We keep for now the space-time dimension $d>2$ arbitrary, though the focus of these lectures will mainly be on the case $d=3$. We assume Euclidean 
signature throughout. Of course, this model is a conformal field theory. As it is well known, it admits a conserved, traceless stress-energy tensor 
(see, for instance, \cite{Osborn:1993cr})
\begin{equation}
T_{\mu\nu}= \partial_{\mu}\phi \partial_{\nu}\phi -\frac{1}{4(d-1)}\left((d-2)\partial_{\mu}\partial_{\nu}+g_{\mu\nu}\partial^2\right)\phi^2\,.
\end{equation}
This is an operator of spin 2 and conformal dimension $\Delta = d$, as one can see directly from the fact that a free scalar field has dimension 
$\Delta_{\phi}=d/2-1$. It is easy to explicitly verify (and left as an exercise) that this operator is conserved and traceless
\begin{equation}
\partial^{\mu}T_{\mu\nu} = 0\,,\qquad T^{\mu}_{\, \mu}=0
\end{equation}
up to terms that vanish by the equation of motion $\partial^2\phi=0$. Given a conformal Killing vector $\zeta^{\mu}$, which satisfies 
$\partial_{\mu}\zeta_{\nu}+\partial_{\nu}\zeta_{\mu} = \frac{2}{d}g_{\mu\nu} \partial^{\rho}\zeta_{\rho}$, one can construct from $T_{\mu\nu}$ a conserved 
current $J_{\mu}^{\zeta} = T_{\mu\nu}\zeta^{\nu}$, and from it a conserved charge in the standard way. The resulting conserved charges yield to 
the $(d+1)(d+2)/2$ generators of the conformal group 
\begin{equation}
P_{\mu}\,,\quad M_{\mu\nu}\,,\quad K_{\mu}\,,\quad D
\end{equation}
which are in one-to-one correspondence with the conformal Killing vectors (there are $d$+$d(d-1)/2$ Killing vectors respectively for translations $P_{\mu}$ and rotations 
$M_{\mu\nu}$, and $d+1$ conformal Killing vectors for special conformal generators $K_{\mu}$ and dilatations $D$). 

Now it turns out that this free CFT admits a much larger symmetry, which is an infinite dimensional extension of the conformal algebra: this is the 
HS algebra. The simplest and most explicit way to see this is to realize that the theory has a tower of HS operators, one for 
each even spin $s$, which are 
conserved. These are bilinears in the scalar field carrying $s$-derivatives, with the structure
\begin{equation}
J_{\mu_1\mu_2\cdots \mu_s} = \sum_{k=0}^s c_{sk} \partial_{\{\mu_1\cdots \mu_k}\phi \partial_{\mu_{k+1}\cdots \mu_s\}}\phi\,, \qquad s=2,4,6,\ldots
\label{Js-ind}
\end{equation}
where the curly brackets denote traceless symmetrization, so that the corresponding operator is totally symmetric and traceless, corresponding 
to an irreducible representation of $SO(d)$ of spin $s$ (note that odd spins are not present since we have a single real scalar field). The coefficients 
$c_{sk}$ can be fixed by imposing conservation 
\begin{equation}
\partial^{\mu}J_{\mu\mu_2\cdots \mu_s}=0\,.
\label{divJ}
\end{equation} 
For explicit calculations, it is often useful to employ an ``index-free" notation by introducing an auxiliary polarization vector $\epsilon^{\mu}$, which 
can be taken to be null, $\epsilon^{\mu}\epsilon_{\mu}=0$. In Euclidean signature, such null vector is complex, but this is no cause of concern 
for our purposes. Then, we may construct the index-free objects
\begin{equation}
\label{Js-no-ind}
J_s(x,\epsilon) = J_{\mu_1\mu_2\cdots \mu_s}\epsilon^{\mu_1}\cdots \epsilon^{\mu_s}\,.
\end{equation} 
Note that, since the polarization vector $\epsilon^{\mu}$ is null, trace terms are automatically projected out. One may ``free" indices with the aid of 
the differential operator in the auxiliary polarization vector space \cite{Dobrev:1975ru, Craigie:1983fb, Costa:2011mg}
\begin{equation}
D_{\mu} = \left(\frac{d}{2}-1+\epsilon^{\nu}\frac{\partial}{\partial\epsilon^{\nu}}\right) \frac{\partial}{\partial\epsilon^{\mu}} 
-\frac{1}{2}\epsilon_{\mu} \frac{\partial}{\partial\epsilon^{\nu}} \frac{\partial}{\partial\epsilon_{\nu}}\,.
\end{equation}
Acting with this operator removes the polarization tensor while keeping track of the constraint $\epsilon^{\mu}\epsilon_{\mu}=0$, i.e. one has 
\begin{equation}
J_{\mu_1\cdots\mu_s} \propto D_{\mu_1}\cdots D_{\mu_s} J_s(x,\epsilon)
\end{equation}
and the conservation equation (\ref{divJ}) may be compactly expressed as
\begin{equation}
\partial^{\mu}D_{\mu}J_s(x,\epsilon) =0\,.
\label{cons-eps}
\end{equation}
The spin-$s$ operators (\ref{Js-ind}) may be expressed in this language as
\begin{equation}
J_s(x,\epsilon) = \sum_{k=0}^s c_{sk} (\epsilon\cdot \partial)^k \phi (\epsilon\cdot \partial)^{s-k}\phi = 
\phi f_s(\epsilon\cdot \overleftarrow{\partial},\epsilon\cdot \overrightarrow{\partial})\phi \,,
\label{Js-eps}
\end{equation}
where we have encoded the coefficients $c_{sk}$ into the function of two variables $f_s(u,v)$. Using the equation of motion $\partial^2\phi=0$, 
one can then show that the conservation equation (\ref{cons-eps}) yields the differential equation
\begin{equation}
\big((d/2-1)(\partial_u+\partial_v) + u \partial_{u}^2  + v \partial_{v}^2\big) f_{s}(u,v)=0.
\label{cons-fs}
\end{equation}
Equivalently, one may obtain the same equation by requiring that the operators (\ref{Js-eps}) are primaries, i.e. they commute with the special conformal 
generators at $x=0$. This is equivalent to conservation as a consequence of the conformal algebra $[K_{\mu},P_{\nu}]=2i g_{\mu\nu}D-2i M_{\mu\nu}$ and 
the fact that $J_s$ has spin $s$ and conformal dimension $\Delta_s = d-2+s$ (since $\phi$ is a free field): this is the unitarity bound for a 
spin $s$ primary operator, and its saturation implies that $J_s$ is a conserved current \cite{Ferrara:1973yt}. 

The solution to (\ref{cons-fs}) can be expressed in terms of Gegenbauer polynomials, see e.g. \cite{Craigie:1983fb} for details. In terms of 
the representation (\ref{Js-eps}), we may write (up to overall irrelevant normalization constant)
\begin{equation}
f_s(u,v) = \sum_{k=0}^s \frac{(-1)^k}{k! (k+\frac{d-4}{2})! (s-k+\frac{d-4}{2})! (s-k)!} u^k v^{s-k}
\end{equation}
which defines the coefficients $c_{sk}$ introduced above. Let us also mention that it is often convenient to package all HS operators into 
a single generating function (this is also natural from the point of view of the bulk HS theory, where the HS fields are packaged into a 
single ``master field", as we will see later). For instance, in $d=3$ a nice form of the generating function is given by \cite{Giombi:2009wh}
\begin{equation}
{\cal J}(x,\epsilon) = \phi\, e^{\epsilon\cdot \overleftarrow{\partial}-\epsilon\cdot \overrightarrow{\partial}} 
\cos\left(2\sqrt{\epsilon\cdot \overleftarrow{\partial} \epsilon\cdot \overrightarrow{\partial}}\right)\,\phi\,.
\label{all-s-sc}
\end{equation}
The reader may verify that, expanding in powers of $\epsilon$, this correctly reproduces the HS operators obtained above. One may also derive 
analogous generating functions in general dimension $d$, but we will not give their explicit form here. 

Given the conserved spin-$s$ currents $J_s$, we can construct conserved charges $Q_s$ which generate the corresponding symmetries. 
These can be obtained 
in a canonical way that generalizes the discussion reviewed above for the stress tensor. Given a spin-($s-1$) conformal Killing tensor 
$\zeta^{\mu_1\cdots \mu_{s-1}}$, \footnote{A conformal Killing tensor is a symmetric tensor satisfying 
$\partial_{(\mu_1}\zeta_{\mu_2\cdots \mu_{s})}=\frac{s-1}{d+2s-4} g_{(\mu_1\mu_2} \partial^{\nu} \zeta_{\mu_3\cdots \mu_s)\nu}$.} one can 
write down an ordinary conserved current 
\begin{equation}
J_{\mu}^{\zeta_{s-1}}=J_{\mu\mu_2\cdots \mu_s}\zeta^{\mu_2\cdots \mu_{s}}\,,\qquad \partial^{\mu}J_{\mu}^{\zeta_{s-1}}=0
\end{equation}
and from this the corresponding conserved charge $Q_s$ in the usual way. 
For $s=2$, this leads to the generators of the conformal group. For higher spins, 
we get further generators that are essentially higher derivative symmetries. They schematically act on the scalar field as
\begin{equation}
[ Q_s,\phi] \sim \zeta^{\mu_1\cdots \mu_{s-1}} \partial_{\mu_1}\cdots \partial_{\mu_s}\phi  \,.
\end{equation}
Note that there is one charge for each conformal Killing tensor $\zeta_{s-1}$. These can be constructed from tensor products of the 
conformal Killing vectors, and they organize in representations of the conformal group $SO(d+1,1)$ given by a Young tableaux with two rows 
with $s-1$ boxes (e.g. for spin 2, this is just the adjoint representation), see for instance \cite{Giombi:2013yva} and \cite{Eastwood:2002su}. 
The number of HS generators at each spin $s$ is then the dimension of this representation. In $d=3$, for example, there are $s(4s^2-1)/3$ 
generators at spin $s$.  The algebra generated by commutators of these charges is the HS algebra. One of its distinguishing features 
is that it is intrinsically infinite dimensional; the only finite subalgebra is given by the spin 2 generators, corresponding to the conformal algebra. 
In general, commutators of charges produce charges of greater spins, and one needs the infinite tower to get a closed algebra. 
For example, the schematic structure of the commutator of spin 4 charges is
\begin{equation}
[Q_4, Q_4] \sim Q_2+Q_4+Q_6\,,
\end{equation}
which shows that once a spin 4 charge is present, we must also have a spin 6 charge, and so on. This is of course clear from the explicit construction 
in terms of the free scalar CFT, where we see that we have an infinite tower of conserved currents, but one may proceed more abstractly and 
ask whether there can be other CFTs with (exact) higher spin symmetry. This question was answered by Maldacena and Zhiboedov \cite{Maldacena:2011jn}, 
who showed that, assuming a CFT with an exactly conserved spin 4 current $J_4$, then an infinite 
tower of conserved HS operators must be present in the theory, 
and all correlation functions of local operators coincide with those of a free CFT 
(the analysis of \cite{Maldacena:2011jn} was in $d=3$; the generalization to higher $d$ was studied in \cite{Boulanger:2013zza, Stanev:2013qra,  Alba:2013yda, Alba:2015upa}). 

\subsection{The free $O(N)$ vector model}

To move towards the AdS/CFT side of the story, we can consider a simple generalization of the above free scalar CFT. We take $N$ massless free scalars
\begin{equation}
S = \int d^d x \frac{1}{2}\left(\partial_{\mu}\phi^i\right)^2
\end{equation}
with equation of motion
\begin{equation}
\partial^2\phi^i=0\,,\qquad i=1,\ldots, N\,.
\end{equation}
The model has now a global $O(N)$ symmetry under which $\phi^i$ transforms in the fundamental, or vector, representation. We will therefore refer to this 
as the (free) $O(N)$ vector model. There are now conserved higher spin currents carrying the $O(N)$ indices 
\begin{equation}
J_s^{ij}(x,\epsilon) = \sum_{k=0}^s c_{sk} (\epsilon\cdot \partial)^k \phi^i (\epsilon\cdot \partial)^{s-k}\phi^j
\end{equation}
where the coefficients $c_{sk}$ are the same as in the previous section. These operators may be decomposed into irreducible representations of $O(N)$
\begin{equation}
J_s^{ij} \rightarrow J_s+J_s^{(ij)}+J_s^{[ij]}
\end{equation}
where $J_s, s=2,4,6,\ldots$ are O(N) singlets; $J_s^{(ij)}, s=2,4,6,\ldots$ are in the symmetric traceless; and $J_s^{[ij]}, s=1,3,5,\ldots$ in the 
antisymmetric representation. The conformal stress tensor is the $O(N)$ singlet $J_2$. The spin 1 operator $J_1^{[ij]}$ is just the current in the adjoint which 
corresponds to the global $O(N)$ symmetry. 

\section{From CFT to AdS: higher spin/vector model duality}
\label{CFT-AdS}

We now consider the following truncation of the $O(N)$ vector model: we declare that we are interested only 
in correlation functions of $O(N)$ invariant operators, i.e. we truncate the model to its $O(N)$ {\it singlet sector} (as we will discuss more later, this 
can be done in practice by weakly gauging the $O(N)$ symmetry). Then, there is a meaningful separation between ``single trace" operators and ``multi-trace" 
operators, where we borrow the terminology usually employed in matrix-type theories. In this vector model context, ``single trace" just means that 
there is a single sum over the $O(N)$ indices. Hence, the single trace operators are just the bilinears in $\phi^i$. The full list of single trace 
primaries is thus exhausted by the singlet higher spin currents plus the scalar operator $J_0 = \phi^i \phi^i$ of dimension $\Delta_0 = d-2$
\begin{equation}
\begin{aligned}
{\rm single~~trace:}~~~&J_0 +\sum_{s=2,4,6,\ldots} J_s \\
(\Delta,S) = & (d-2,0)+\sum_{s=2,4,6,\ldots}(d-2+s,s)\,.
\end{aligned}
\label{single-trace}
\end{equation}
This result is essentially equivalent to the Flato-Fronsdal theorem \cite{Flato:1978qz, Flato:1980zk}. 
The $O(N)$ invariant operators containing more than two scalar fields are the equivalent of multi-trace operators. For instance, an operator of the schematic 
form 
\begin{equation}
(\phi^i \phi^i)(\phi^j \partial^s \phi^j)
\label{double-tr}
\end{equation}
should be viewed as a double-trace operator. Its dimension is equal to the sum of the dimensions of its single trace constituents (and if interactions 
are turned on, it is equal to the sum of the single trace dimension plus corrections that are small in the large $N$ limit). 

If we assume that the AdS/CFT correspondence holds in its most general form, we expect that this singlet sector CFT should be dual to some kind 
of gravitational theory in AdS, which becomes weakly coupled in the large $N$ limit. What should such AdS dual look like? According to the 
general rules of AdS/CFT 
\begin{equation}
\mbox{Single trace operators in CFT} \qquad \Leftrightarrow \qquad \mbox{Single particle states in AdS}
\nonumber
\end{equation}
Thus, the CFT single trace spectrum (\ref{single-trace}) should match the single particle spectrum of the bulk dual (multi-trace operators such as 
(\ref{double-tr}) should be dual to multi-particle states in AdS). Conserved currents in 
a CFT are dual to corresponding gauge fields in AdS. The spin 1 and spin 2 examples of this are familiar in usual applications of AdS/CFT: a conserved 
spin 1 current is dual to a spin 1 gauge field in the bulk, and the conserved stress tensor is dual to the graviton. The same 
idea generalizes to all spins. Thus, for each conserved spin $s$ current we expect a corresponding {\it massless HS gauge field} in AdS
\begin{equation}
J_{s}, \qquad \partial\cdot J_s = 0   \qquad \Leftrightarrow \qquad 
\mbox{Massless HS gauge field $\varphi_s$}\,\qquad \delta \phi_s \sim \nabla \epsilon_{s-1} 
\label{HS-spec}
\end{equation}
The global HS symmetry on the CFT side corresponds to a HS gauge symmetry in the bulk generated by a spin $s-1$ gauge parameter $\epsilon_{s-1}$. 
In addition, the scalar operator $J_0 = \phi^i\phi^i$ should be dual to a bulk scalar field 
\begin{equation}
J_0 \qquad \Leftrightarrow \qquad 
\mbox{Scalar field $\varphi$ with} \quad m^2=\Delta_0(\Delta_0-d)/\ell^2_{AdS}
\label{sc-spec}
\end{equation}
where the mass of the bulk scalar is fixed by the familiar AdS/CFT relation in terms of the conformal dimension of the dual operator. In the free scalar 
CFT, we have $\Delta_0 = d-2$ and hence $m^2 = -2(d-2)/\ell^2_{AdS}$. As it turns out, (\ref{HS-spec}) and (\ref{sc-spec}) 
is precisely the spectrum of the so-called minimal bosonic Vasiliev higher spin theory in AdS$_{d+1}$ \cite{Vasiliev:2003ev}. In particular, the mass of the scalar field is 
not a free parameter and is dictated by the structure of the HS invariant equations.

\begin{figure}
\begin{center}
\includegraphics[width=0.7\textwidth]{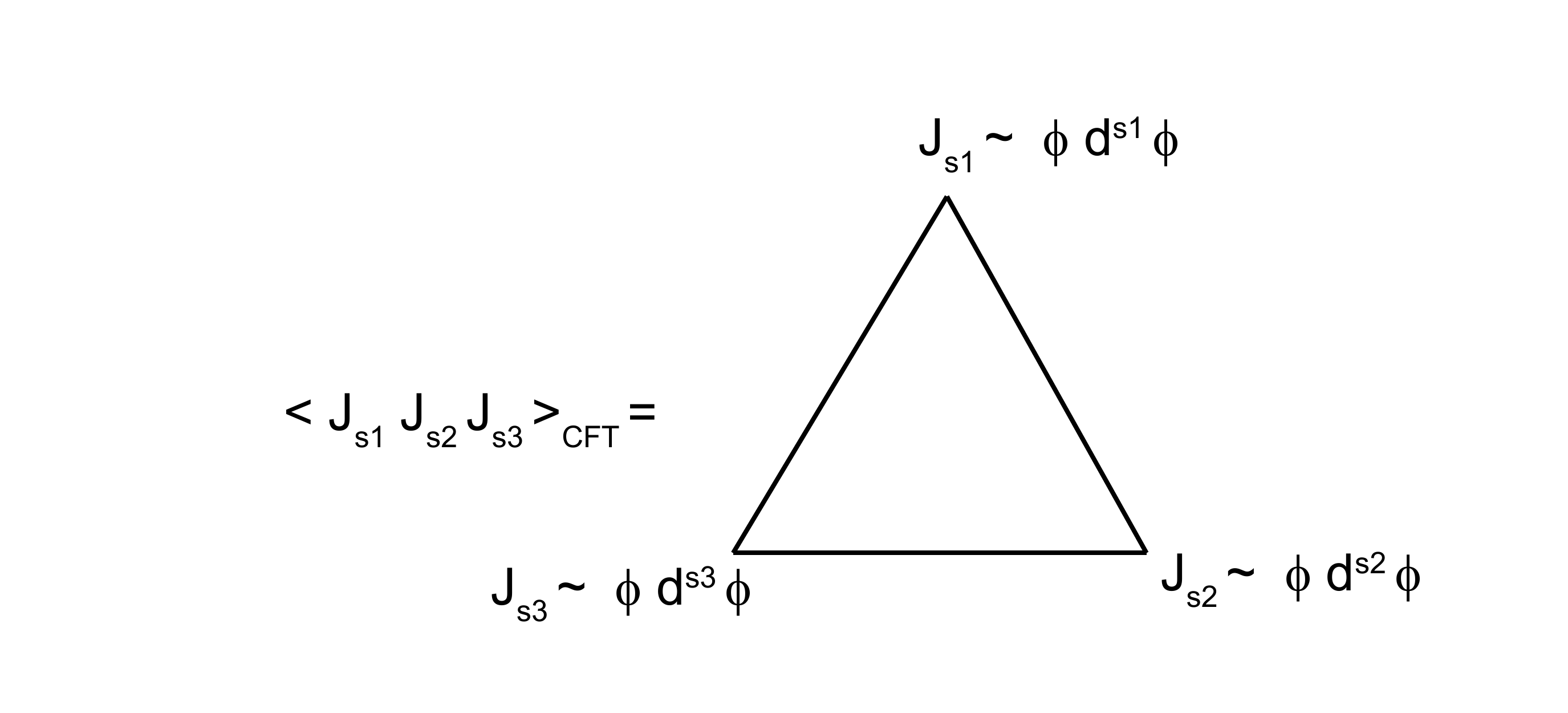}
\end{center}
\caption{Three-point function of HS currents in the free scalar CFT. Solid lines represent free scalar propagators.}
\label{3ptCFT}
\end{figure}

\begin{figure}
\begin{center}
\includegraphics[width=0.7\textwidth]{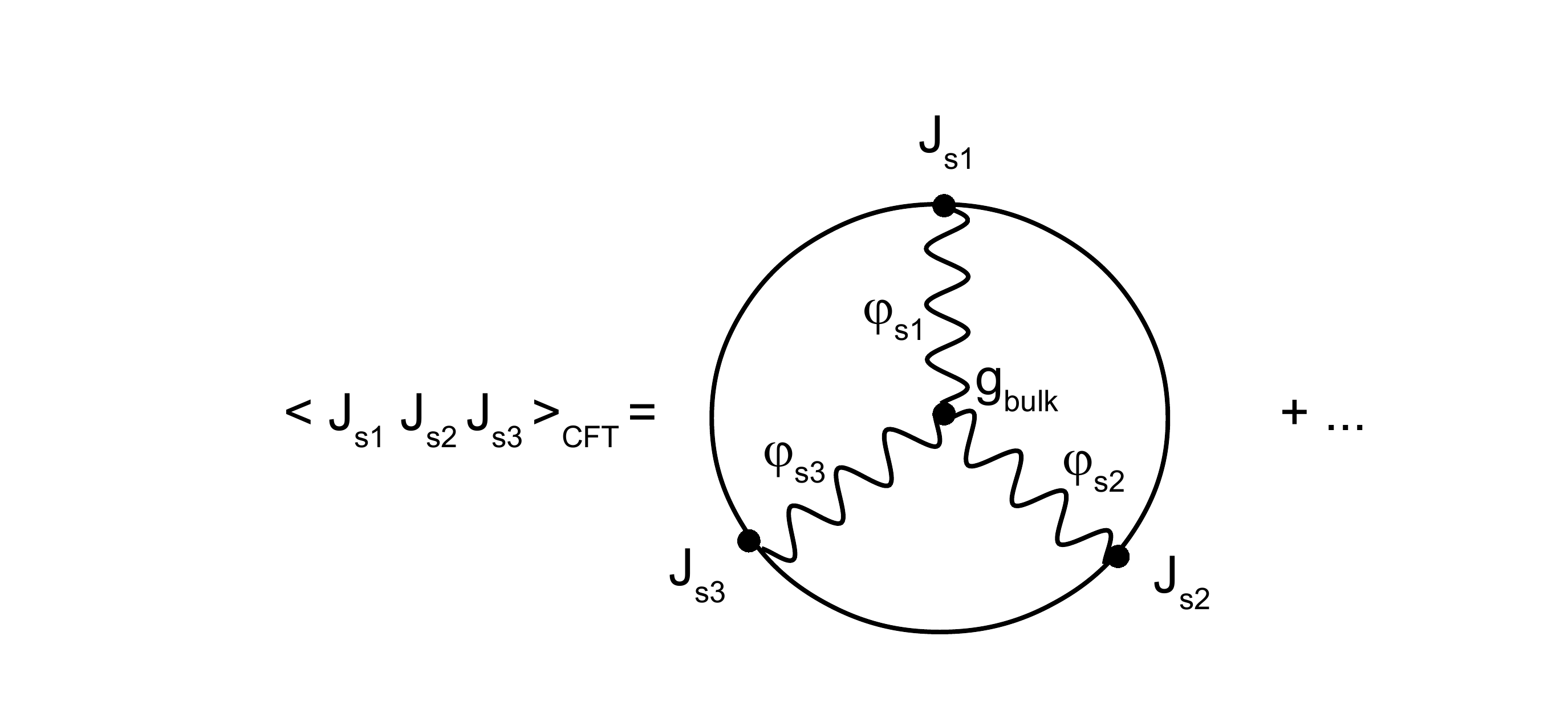}
\end{center}
\caption{Holographic 3-point function}
\label{3ptW}
\end{figure}

The CFT and AdS single trace/single particle spectra clearly match, but what about interactions? In the CFT, the basic 
object of interest are correlation functions of primary operators. In the free CFT, these are easy to compute, using the explicit 
expressions for the currents $J_s$ given above and the free field Wick contractions. For instance, the 3-point function of 
the HS operators can be computed by the triangle diagram in Figure \ref{3ptCFT}, where each line is a scalar propagator and the vertices 
include the appropriate derivatives. Explicit expressions for these correlation functions can be found for instance 
in \cite{Giombi:2009wh,Giombi:2010vg, Giombi:2011rz} (see also \cite{Didenko:2012tv} for higher point functions).
According to the AdS/CFT dictionary, these correlation functions should be matched to a dual Witten diagram calculation, as shown 
for instance in Figure \ref{3ptW} for the 3-point function. 
Clearly, since the CFT correlation functions are non-zero, there {\it must be} non-trivial interactions of the HS gauge fields in 
the AdS dual theory. Therefore, from AdS/CFT point of view, it is not surprising that consistent interactions of massless HS fields 
in AdS should exist. And indeed, Vasiliev theory provides an explicit construction of such consistent interactions (even though, 
as mentioned earlier, Vasiliev equations were in fact written down before AdS/CFT). 

The perturbative expansion in the bulk, controlled by the coupling constant denoted $g_{\rm bulk}$ in Figure \ref{3ptW}, 
is related as usual to the large $N$ expansion on the CFT side. We can see this as follows. If we normalize the currents $J_s$ so that 
they have 2-point functions of order one, i.e. $J_s \sim \frac{1}{\sqrt{N}}\phi^i \partial^s \phi^i$, then the 3-point functions scale as 
\begin{equation}
\langle J_{s_1} J_{s_2} J_{s_3}\rangle \sim \frac{1}{\sqrt{N}} \qquad {\rm with}~~\langle J_s J_s\rangle \sim O(1)\,.
\end{equation}
Therefore, we see that 
\begin{equation}
g_{\rm bulk} \sim \frac{1}{\sqrt{N}}\,.
\end{equation}
Introducing Newton's constant in the usual way as $S_{\rm bulk} \sim \frac{1}{G_N} \int d^{d+1}x {\cal L}_{\rm bulk}=
\frac{1}{g_{\rm bulk}^2} \int d^{d+1}x {\cal L}_{\rm bulk}$, we then see that the required scaling of the Newton's constant with 
$N$ in the higher spin/vector model dualities is (we set the AdS scale to one here)
\begin{equation}
G_N \sim N^{-1}\,.
\end{equation}
Note that this is different from versions of the AdS/CFT duality involving adjoint fields, where $G_N \sim N^{-2}$. In any case, 
it is clear that, as usual, the $1/N$ expansion on the CFT side is mapped to the perturbative expansion on the AdS gravity side, 
which runs in powers of the Newton's constant.  

A particularly interesting CFT correlation function is the 3-point function $\langle TTT\rangle$ of the stress-tensor. This is given, 
to leading order in the bulk expansion, by a tree-level Witten diagram as in Figure \ref{3ptW}, and it involves the cubic 
coupling of the graviton. One may ask whether this cubic coupling is perhaps the same as the one obtained from the ordinary two-derivative 
Einstein gravity
\begin{equation}
S_{\rm Einstein} = \frac{1}{G_N} \int d^{d+1}x \left(R+\Lambda\right)\,.
\end{equation} 
Let us specialize to the case of $d=3$. It is known \cite{Osborn:1993cr, Maldacena:2011nz, Giombi:2011rz} that in a general 
3d CFT the stress-tensor 3-point function is fixed by conformal invariance and conservation to be a linear combination 
of three structures
\begin{equation}
\langle TTT\rangle = a_B \langle TTT\rangle_{\rm free~sc}+a_F \langle TTT\rangle_{\rm free~fer}+a_{\rm odd} 
\langle TTT\rangle_{\rm odd}
\label{TTT}
\end{equation}
where the first two structure correspond to the stress-tensor 3-point functions in free scalar and free fermion theories (which 
yield independent tensor structures), and the third one is a parity odd tensor structure which does not arise in free theories, 
but can arise in interacting theories that break parity. Using AdS/CFT, the stress tensor 3-point function that follows from 
the ordinary Einstein gravity was computed in \cite{AF99}. It turns out to be a linear combination of scalar and fermion 
structures.\footnote{In fact, it corresponds to the 3-point function of the stress tensor in the supersymmetric CFT.} Therefore, the HS dual to 
the free $O(N)$ vector model must have a cubic graviton coupling which is different from the ordinary Einstein gravity. The coefficients 
$a_B$, $a_F$ in (\ref{TTT}) can be changed by adding higher derivative terms to the Einstein action. So we learn that the HS dual 
should be a higher derivative theory of gravity, at the level of interactions. One can verify that this is indeed a feature 
of Vasiliev theory, as was briefly mentioned above: interactions involve higher derivatives. 

From all the above arguments we see that a consistent theory of interacting HS gauge fields in AdS should exist, because it 
should provide the AdS dual of free CFTs. The known features of Vasiliev theory all agree with what we expect from CFT perspective. So 
it is very natural to conjecture \cite{Klebanov:2002ja} the higher spin/vector model duality
\begin{equation}
\begin{aligned}
&\mbox{Free $O(N)$ vector model} \quad \Leftrightarrow \quad \mbox{Higher Spin Gravity in AdS} \\
&~~~~~~\mbox{(singlet sector)}
\end{aligned}
\nonumber 
\end{equation} 
Note that, while the focus of these lectures is mainly on $d=3$, this duality between free scalar vector model and HS gravity 
makes sense in any $d$, and the corresponding non-linear HS theories are explicitly known \cite{Vasiliev:2003ev}. 

\subsection{Complex scalars and $U(N)$ vector model}

In the above discussion we focused on the theory of $N$ real scalars restricted to the $O(N)$ singlet sector. Of course, we may 
also consider $N$ complex scalars, for which we can write the $U(N)$ invariant Lagrangian 
\begin{equation}
S = \int d^d x ~\partial_{\mu}\phi_i^* \partial^{\mu}\phi^i \,,
\label{UN-sc}
\end{equation}
and restrict to the $U(N)$ singlet sector. The single trace spectrum in this case then involves also odd spin currents 
\begin{equation}
\begin{aligned}
\mbox{single trace ($U(N)$ singlets):}~~~&J_0 +\sum_{s=1}^{\infty} J_s \\
(\Delta,S) = & (d-2,0)+\sum_{s=1}^{\infty}(d-2+s,s)\,.
\end{aligned}
\label{single-trace-UN}
\end{equation}
For instance, for $s=1$ we have the familiar current $J_1 = \phi_i^* \stackrel{\leftrightarrow}{\partial}\phi^i$. The dual HS theory 
should then involve HS gauge fields of all spins, one for each integer spin. In fact, this is precisely the spectrum of the bosonic 
Vasiliev theory in AdS$_{d+1}$ \cite{Vasiliev:2003ev}, and the minimal theory involving even spins only (dual to the $O(N)$ model) 
can be obtained from it by imposing a certain consistent truncation on the master fields. Let us mention that for
the complex scalar theory (\ref{UN-sc}) with $N$ even, one can also impose a $USp(N)$ singlet constraint \cite{Giombi:2013fka}. In this 
case, the single trace spectrum includes one current for each even spin, and three currents of each odd spin. This spectrum 
is precisely dual to the Vasiliev theory based on the algebra $husp(2,0|4)$ \cite{Vasiliev:1999ba}, whose odd spin gauge fields transform 
in the adjoint of $USp(2)=SU(2)$. 

\section{Interacting $O(N)$ model and its AdS dual}
\label{KP}
The conjecture that the singlet sector of the free $O(N)$/$U(N)$ scalar CFT is dual to Vasiliev HS gravity was first made precise 
in \cite{Klebanov:2002ja}. Earlier closely related work 
\cite{HaggiMani:2000ru, Sundborg:2000wp, WittenTalk, Konstein:2000bi, Shaynkman:2001ip, Sezgin:2001zs, Vasiliev:2001zy, Mikhailov:2002bp,Sezgin:2002rt} 
focused on the correspondence between free CFTs with matrix-like fields and higher spin theories, stemming from the motivation of understanding 
the duality between ${\cal N}=4$ SYM theory and type IIB string theory in the small gauge coupling limit. In free CFTs with matrix-like fields
the single trace spectrum includes, similarly to the discussion of the previous sections, 
conserved HS currents $J_s \sim {\rm Tr}(\Phi \partial^s\Phi)$ which should be dual to massless HS gauge fields in AdS. However, in addition there is 
an exponentially growing number of single trace operators that are not conserved currents, and should be dual to massive fields in AdS. 
Vectorial CFTs provide much simpler realizations of the AdS/CFT duality, as elucidated in \cite{Klebanov:2002ja}: the spectrum of single trace 
operators in the singlet sector (\ref{single-trace}) just consists of the bilinears, 
and can be put in one-to-one correspondence to HS gravity theories of the Vasiliev type, as explained in the previous section.
 
While the duality between free vector model and HS gauge theory is already by itself 
a very interesting example of a potentially exact, 
non-supersymmetric version of AdS/CFT, a crucial observation made in \cite{Klebanov:2002ja} is that one can in fact 
easily generalize the duality to the case of large $N$ interacting vector models. Let us consider the standard $O(N)$ 
model with quartic interaction
\begin{equation}
S = \int d^dx\left(\frac{1}{2}(\partial_{\mu} \phi^i )^2 +\frac{m^2}{2}(\phi^i\phi^i)+\frac{\lambda}{4}(\phi^i\phi^i)^2\right)\,.
\label{quartic}
\end{equation} 
It is well-known that in $2<d<4$, this model has non-trivial interacting IR fixed points. Indeed, the interaction is relevant for $d<4$, 
and it triggers a RG flow from the free CFT in the UV to an interacting CFT in the IR (provided the bare mass term is suitably 
tuned to reach criticality). This RG flow is perturbative in the framework of Wilson-Fisher $\epsilon$ expansion. Working in $d=4-\epsilon$, 
one finds the familiar one-loop beta function
\begin{equation}
\beta_{\lambda} = -\epsilon \lambda +\frac{N+8}{8\pi^2}\lambda^2+\ldots 
\end{equation}
and there is a weakly coupled IR fixed point at $\lambda_* = \frac{8\pi^2}{N+8}\epsilon+O(\epsilon^2)$. One may then compute 
various physical quantities in the $\epsilon$ expansion, and provided sufficiently high order calculations are performed, one 
can obtain this way estimates in the physical dimension $d=3$ ($\epsilon=1$), where the IR CFT is strongly coupled.\footnote{This typically requires some resummation procedure, see e.g. \cite{Kleinert:2001ax} for a review.} The same interacting CFT may be also described, as it is well known 
(see e.g. \cite{Moshe:2003xn} for a review), by the UV fixed points 
of the $O(N)$ non-linear sigma model, which are weakly coupled in $d=2+\epsilon$ \cite{Brezin:1975sq, Bardeen:1976zh}. 

\begin{figure}
\begin{center}
\includegraphics[width=0.5\textwidth]{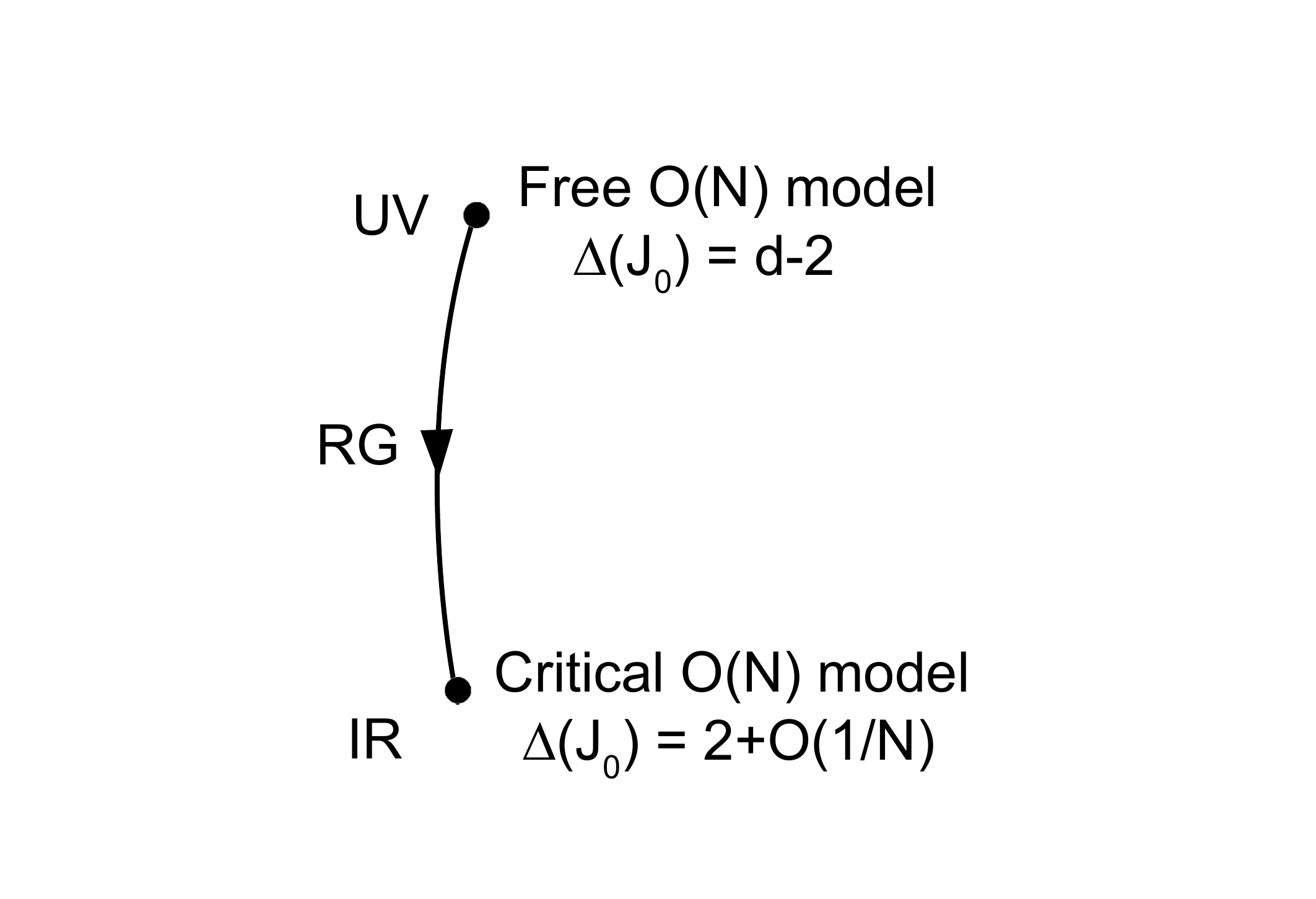}
\end{center}
\caption{RG flow from free to critical $O(N)$ model.}
\label{RG-ON}
\end{figure}

A complementary approach that can be developed at arbitrary dimension $d$, and is more natural for comparison with AdS, is the large $N$ expansion. 
A standard way to develop this expansion is based on introducing a Hubbard-Stratonovich auxiliary field $\sigma$ as
\begin{equation}
S = \int d^d x \left(\frac{1}{2}(\partial \phi^i)^2+\frac{1}{2}\sigma \phi^i\phi^i-\frac{\sigma^2}{4\lambda}\right)\,,
\label{hubbard}
\end{equation}
where we have set the mass term to zero since we are interested in the CFT.\footnote{We assume a regularization scheme such that mass is not generated 
if the bare mass is set to zero in the UV.} 
Integrating out $\sigma$ via its equation of motion $\sigma = \lambda \phi^i\phi^i$, one gets back the original lagrangian. However, integrating 
out the fundamental fields $\phi^i$ generates an effective non-local action for $\sigma$
\begin{equation}
\begin{aligned}
Z &= \int D\phi D\sigma\, e^{-\int d^d x\left(\frac{1}{2}(\partial \phi^i)^2+\frac{1}{2}\sigma \phi^i\phi^i-\frac{\sigma^2}{4\lambda}\right)}\\
&=\int D\sigma\, e^{\frac{1}{8}\int d^d x d^d y\, \sigma(x)\sigma(y)\, \langle \phi^i\phi^i(x)\phi^j\phi^j(y)\rangle_0+\int d^d x \frac{\sigma^2}{4\lambda}+{\cal O}(\sigma^3)}
\end{aligned}
\label{F-sigma}
\end{equation}
where we have assumed large $N$ and the subscript `$0$' denotes expectation values in the free theory. We have
\begin{equation}
\langle \phi^i\phi^i(x)\phi^j\phi^j(y)\rangle_0 = 2N [G(x-y)]^2\,,\qquad G(x-y)= \int \frac{d^dp}{(2\pi)^d} \frac{e^{ip(x-y)}}{p^2}\,.
\end{equation}
In momentum space, the square of the $\phi$ propagator reads
\begin{eqnarray}
&&[G(x-y)]^2 = \int \frac{d^d p}{(2\pi)^d} e^{ip(x-y)}\tilde G(p)\\
&&\tilde G(p)=\int \frac{d^dq}{(2\pi)^d} \frac{1}{q^2(p-q)^2} = -\frac{(p^2)^{d/2-2}}{2^d (4\pi)^{\frac{d-3}{2}}\Gamma(\frac{d-1}{2})\sin(\frac{\pi d}{2})} 
\equiv -\frac{2}{\tilde{C}_{\sigma}} (p^2)^{d/2-2}\nonumber 
\end{eqnarray}
and so from (\ref{F-sigma}) one finds the effective quadratic action for $\sigma$
\begin{equation}
S_2 = -\int d^dp \frac{1}{2}\sigma(p)\sigma(-p)\left[\frac{N}{\tilde{C}_{\sigma}}(p^2)^{d/2-2}+\frac{1}{2\lambda}\right]\,.
\label{sig-2}
\end{equation}
From this expression we see that for $d<4$ the induced kinetic term dominates in the IR limit compared to the bare quadratic term in (\ref{hubbard}), and hence 
the latter can be dropped at the IR fixed point. The conclusion is that, to develop the $1/N$ expansion of the critical IR theory, we may work with the action
\begin{equation}
S_{\rm crit} = \int d^dx \left(\frac{1}{2}(\partial \phi^i)^2+\frac{1}{2\sqrt{N}}\sigma \phi^i\phi^i\right)
\label{S-crit}
\end{equation}
where we have rescaled $\sigma$ for convenience, so that its two-point function scales as $N^0$, and $1/\sqrt{N}$ acts as a coupling constant. From (\ref{sig-2}), we thus 
find the two-point function of $\sigma$ in the IR to be
\begin{equation}
\langle \sigma(p)\sigma(-p)\rangle =  \tilde C_{\sigma}  (p^2)^{2-\frac{d}{2}}
\label{sigma-prop}
\end{equation}
or, in coordinate space:
\begin{equation}
\langle \sigma(x)\sigma(y) \rangle = \frac{2^{d+2}\Gamma\left(\frac{d-1}{2}\right)\sin(\frac{\pi d}{2})}{\pi^{\frac{3}{2}}\Gamma\left(\frac{d}{2}-2\right)}\,\frac{1}{|x-y|^4}
\equiv \frac{C_{\sigma}}{|x-y|^4}\,.
\label{sigma-prop-x}
\end{equation}
The large $N$ perturbation theory can then be developed using the propagator (\ref{sigma-prop})-(\ref{sigma-prop-x}) for $\sigma$, the canonical propagator for $\phi$ and the interaction term $\sigma\phi^i\phi^i$ in (\ref{S-crit}).
From (\ref{sigma-prop-x}) we see that $\sigma$, which plays the role of $J_0 = \phi^i \phi^i$ in this description, behaves as a primary scalar operator of dimension $\Delta=2$ (this value will of course be corrected at higher orders in the $1/N$ expansion). This means that under the RG flow, the dimension of the scalar operator $J_0 =\phi^i\phi^i$ has changed from $\Delta_{UV}=d-2$ to $\Delta_{IR} = 2+O(1/N)$, as shown in Figure \ref{RG-ON}. 

The quartic interaction in (\ref{quartic}) may in fact be viewed as a particular example of the double trace deformations studied in \cite{Gubser:2002vv}, where one considers a CFT perturbed by the square of a single trace operator
\begin{equation}
S_{\rm CFT} \rightarrow S_{\rm CFT}+\lambda \int d^d x O_{\Delta}^2 \,.
\label{d-trace}
\end{equation}
The interacting $O(N)$ model falls into this class, since it is a perturbation of the free CFT by the square of the $J_0$ operator. 
For $\Delta < d/2$ the interaction in (\ref{d-trace}) is relevant and there is a flow to a new CFT in the IR. One can then show \cite{Gubser:2002vv} that at large $N$ the dimension of $O_{\Delta}$ goes from $\Delta_{UV}$ at the UV fixed point to $\Delta_{IR} = d-\Delta_{UV}+O(1/N)$ at the IR fixed point. 

The observation that the interacting $O(N)$ model (\ref{quartic}) may be viewed as a double-trace deformation of the free CFT allows to immediately deduce its AdS dual interpretation. Indeed, the AdS dual of the general double-trace flows of the type (\ref{d-trace}) is well-understood \cite{Klebanov:1999tb, Witten:2001ua}. In AdS/CFT, a scalar 
operator $O_{\Delta}$ of dimension $\Delta$ is dual to a bulk scalar field $\varphi(z,\vec{x})$ with boundary behavior (in Poincare coordinates $ds^2 = (dz^2+d\vec{x}^2)/z^2$)
\begin{equation}
\varphi(z,\vec{x}) \stackrel{z\rightarrow 0}{\sim} z^{\Delta} A(\vec{x})
\label{b-behavior}
\end{equation}
and $\Delta$ is related to the mass of the scalar field by the familiar relation (we set the AdS scale to one):
\begin{equation}
\Delta(\Delta-d)=m^2~~\rightarrow ~~ \Delta_{\pm} =\frac{d}{2}\pm \sqrt{(\frac{d}{2})^2+m^2}\,.
\end{equation}
If we insist on unitarity, usually $\Delta_+$ is the only possible solution (recall the unitarity bound $\Delta \ge d/2-1$). However, for 
\begin{equation}
-(d/2)^2 < m^2 < -(d/2)^2+1
\label{m-ineq}
\end{equation} 
both $\Delta_+$ and $\Delta_-=d-\Delta_+$ are above the unitarity bound, and one may choose either one to specify the boundary condition (\ref{b-behavior}) of the bulk scalar field. Then, according to the dictionary developed in \cite{Klebanov:1999tb, Witten:2001ua}, one choice corresponds to the UV CFT, and the other to the IR CFT which sits at the endpoint of the RG flow triggered by the double-trace interaction (\ref{d-trace}). 

The inequality (\ref{m-ineq}) is precisely satisfied in the Vasiliev theory in AdS$_4$, where the scalar mass is given by $m^2=-2$. Therefore, both roots $\Delta=1$ and $\Delta=2$ are above unitarity and we can choose either one to quantize the bulk scalar field. The choice $\Delta=1$ (in general, $\Delta=d-2$) corresponds to the free 
CFT at the boundary, while $\Delta=2$ to the interacting one. In other words, the critical $O(N)$ model is dual to precisely {\it the same} Vasiliev theory, but with 
a different choice of boundary condition on the bulk scalar field. 

But what about the higher spin currents? They are of course still present in the spectrum, but since the CFT is now interacting, the $s>2$ currents should not be exactly conserved anymore. Indeed, one finds that the divergence of the HS operators is now non-zero, except for $s=2$, and at large $N$ takes the schematic form
\begin{equation}
\partial \cdot J_s = \frac{1}{\sqrt{N}} \sum_{s'=2}^{s-2} \sum_{k=0}^{s-s'-1} c_{s,s'k} \partial^{s-s'-k-1}
 J_{s'} \partial^k \sigma \equiv \frac{1}{\sqrt{N}}K_{s-1}
\label{divJbr}
\end{equation}
where the coefficients $c_{s,s'k}$ can be determined using the equation of motion (see \cite{Skvortsov:2015pea,Giombi:2016hkj} for the explicit result). Note 
that $K_{s-1}$ is a ``double-trace" operator of the model, dual to a two-particle state in the bulk. This equation implies 
that the higher spin symmetry is only {\it weakly broken} at large $N$. Indeed, using the fact that the 2-point function of a spin $s$ primary operator of dimension $\Delta_s=d-2+s+\gamma_s$ (recall that $\Delta_s=d-2+s$ is the dimension of a conserved current, so $\gamma_s$ denotes the anomalous dimension) is 
fixed by conformal invariance to be
\begin{equation}
\langle J_s(x_1,\epsilon_1)J_s(x_2,\epsilon_2)\rangle = C_s 
\frac{\left(\epsilon_1\cdot \epsilon_2-2 \frac{\epsilon_1 \cdot x_{12} \epsilon_2 \cdot x_{12}}{x_{12}^2}\right)^s}{(x_{12})^{2\Delta_s}}\,,
\end{equation} 
the non-conservation equation (\ref{divJbr}) implies 
$\frac{1}{N}\langle K_{s-1} K_{s-1}\rangle = \langle \partial \cdot J_s \partial \cdot J_s\rangle \propto \gamma_s$, 
and so the anomalous dimension $\gamma_s$ to order $1/N$ may be extracted as \cite{Anselmi:1998ms, Belitsky:2007jp, Skvortsov:2015pea, Giombi:2016hkj}
\begin{equation}
\gamma_s = \frac{\alpha_s}{N}+O(1/N^2)\,,\qquad \alpha_s \sim \frac{\langle K_{s-1} K_{s-1}\rangle}{\langle J_s J_s\rangle}\,.
\end{equation} 
For instance, in $d=3$, the explicit result takes the form \cite{Lang:1992zw, Giombi:2016hkj}
\begin{equation}
\gamma_s = \frac{16(s-2)}{3\pi^2(2s-1)}\frac{1}{N}\,.
\end{equation}

\begin{figure}
\begin{center}
\includegraphics[width=0.7\textwidth]{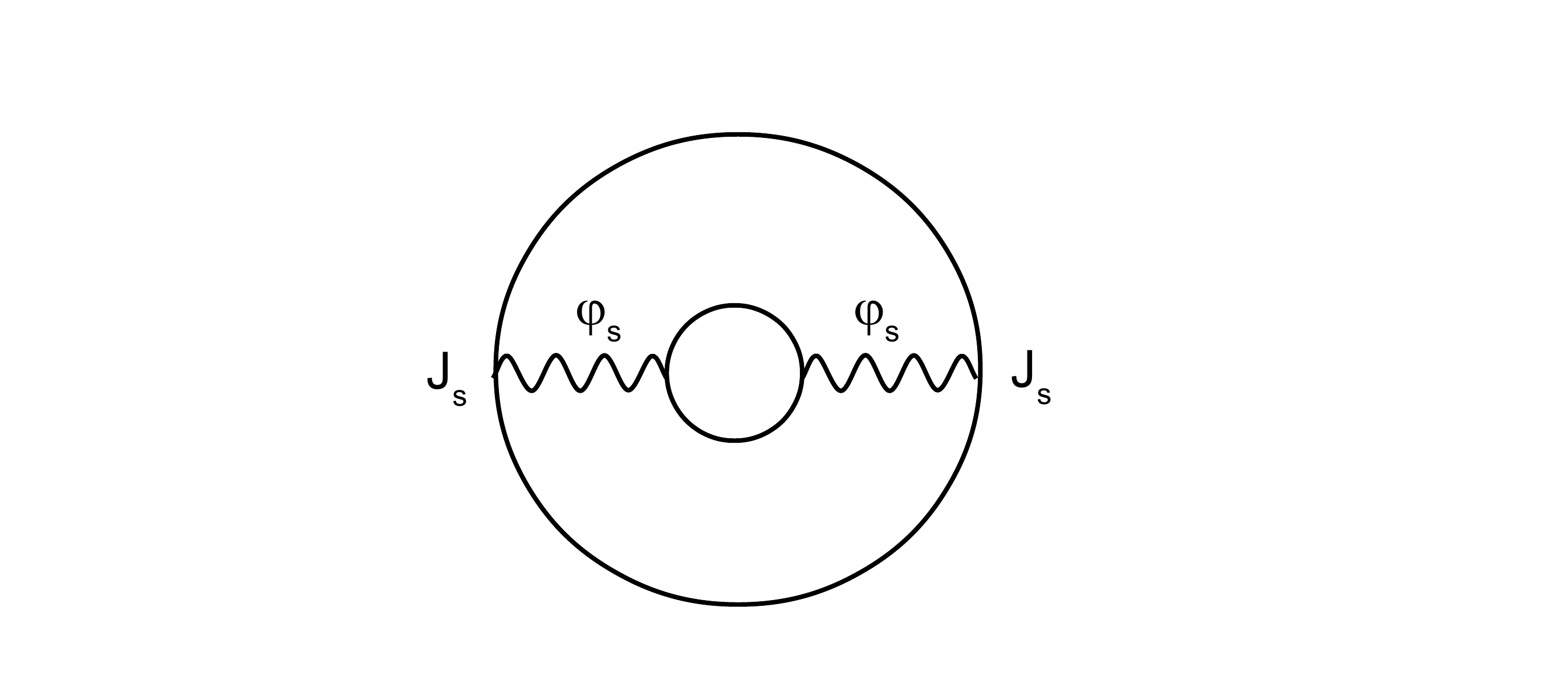}
\end{center}
\caption{Witten diagram contributing to the one-loop correction to the HS current two-point function. The loop involves all fields in the spectrum, in particular 
the scalar field with $\Delta=2$ boundary condition.}
\label{2pt-loop}
\end{figure}

To summarize, in the critical $O(N)$ theory, the dimension of the HS currents is only corrected starting at order $1/N$
\begin{equation}
\Delta_s = d-2+s+\frac{\alpha_s}{N}+\ldots 
\end{equation}
which means that the dual HS fields remain massless at the classical level, and receive masses through loop corrections \cite{Girardello:2002pp}.\footnote{The mass 
of a spin $s$ field is related to the dual dimension by $(\Delta+s-2)(\Delta+2-d-2)=m^2$, which implies $m^2\sim \alpha_s(2s+d-4)/N$.} 
The leading $1/N$ correction should arise from a one-loop diagram in the bulk as the one depicted in Figure \ref{2pt-loop}, where one of the loop lines involves the scalar field with $\Delta=2$ boundary condition (the bulk-to-bulk propagator for the scalar field does depend on the choice of $\Delta$). It was shown in \cite{Giombi:2011ya} that, assuming no anomalous dimensions are generated by loops with the $\Delta=1$ boundary condition, then the correct anomalous dimensions in agreement with the critical $O(N)$ model would indeed be reproduced by the bulk loop diagrams, to all orders in $1/N$. However, it remains to be shown that with the choice of boundary condition dual to free CFT the loop diagrams are indeed trivial (this should follow from the exact HS symmetry, but showing it explicitly by a bulk calculation is an open problem).    

\section{Fermionic CFT}
\label{ferm-HS-sec}

So far we discussed the case of scalar CFT, but there is of course another free CFT that we can write in any dimension, namely a free 
massless fermion. More generally, let us take $N$ free massless Dirac fermions
\begin{equation}
S= \int d^d x \bar\psi_i \gamma^{\mu}\partial_{\mu}\psi^i\,.
\label{free-fer}
\end{equation}
This theory has a $U(N)$ global symmetry under which $\psi^i, i=1,\ldots, N$ transforms in the fundamental representation. Depending on the 
dimension $d$, we may also impose a Majorana condition to obtain a CFT with $O(N)$ symmetry. As in the case of the free scalar CFT, 
the model (\ref{free-fer}) admits an infinit tower of exactly conserved currents of the form
\begin{equation}
\begin{aligned}
&J_{\mu_1 \cdots \mu_s} = \bar\psi_i \gamma_{\mu_1}\partial_{\mu_2}\cdots \partial_{\mu_s}\psi^i+\ldots \,,\quad s=1,2,3,\ldots \\
&\partial^{\mu}J_{\mu\mu_2\cdots \mu_s}=0\,,\qquad J^{\mu}_{\ \mu \mu_3\cdots\mu_s}=0 
\label{HS-fer}
\end{aligned}
\end{equation}
which are in the totally symmetric traceless representation of $SO(d)$. Of course, one also has conserved currents in non-trivial representations 
of $U(N)$, but above we just wrote the singlet ones since we will be interested in projecting onto the singlet sector, as we have done in the scalar 
case. Note that the conformal dimension of the operators (\ref{HS-fer}) is $\Delta_s = d-2+s$, as appropriate for a conserved current, since $\Delta_{\psi}=(d-1)/2$. 

The explicit form of the currents is most conveniently given using an auxiliary null polarization vector, as described above in the scalar case. Imposing 
conservation (or the condition that $J_s$ is a primary), one finds that the conserved currents are given by
\begin{equation}
\begin{aligned}
&J_s(x,\epsilon) = \epsilon^{\mu_1}\cdots \epsilon^{\mu_s} J_{\mu_1\cdots \mu_s} =
 \sum_{k=0}^{s-1} c_{sk} \bar\psi_i \epsilon \cdot \gamma (\epsilon\cdot \overleftarrow{\partial})^k 
(\epsilon \cdot \overrightarrow{\partial})^{s-1-k}\psi^i \\
& c_{sk} = \frac{(-1)^k}{k! (k+\frac{d-2}{2})! (s-k-1+\frac{d-2}{2})! (s-k-1)!}\,.
\end{aligned}
\end{equation}
The reader may check as an exercise that for $s=1$ and $s=2$ this reproduces respectively the familiar $U(1)$ current $J_1 \sim \bar\psi_i \gamma_{\mu}\psi^i$ and 
the stress tensor $T_{\mu\nu}=\frac{1}{2}\bar\psi_i \left(\gamma_{\mu}\overleftrightarrow{\partial}_{\nu}+
\gamma_{\nu}\overleftrightarrow{\partial}_{\mu}\right)\psi^i$. For $d=3$, similarly to (\ref{all-s-sc}), 
one may also derive a simple generating function encoding all spins \cite{Giombi:2011kc}
\begin{equation}
\begin{aligned}
&{\cal J}(x,\epsilon) = \bar\psi_i \epsilon \cdot \gamma 
f(\epsilon\cdot \overleftarrow{\partial},\epsilon\cdot \overrightarrow{\partial}) \,\psi^i\\ 
&f(u,v)=e^{u-v} \frac{\sin\left(2\sqrt{u v}\right)}{2\sqrt{u v}}\,.
\end{aligned}
\label{all-s-fer}
\end{equation}

Let us now consider the truncation of the model to its $U(N)$ singlet sector. If we restrict to the $d=3$ case, then it is not difficult to 
see that the HS currents (\ref{HS-fer}), together with the {\it parity odd}\footnote{In $d=3$, we may define ``parity" by 
the sign reversal on all coordinates, $\vec{x}\rightarrow -\vec{x}$.} scalar operator
\begin{equation}
J_0=\bar\psi_i \psi^i 
\end{equation}
of dimension $\Delta_0 = 2$ ($\Delta_0 = d-1$ in general $d$), exhaust the spectrum of single trace operators
\begin{equation}
\begin{aligned}
\mbox{single trace ($U(N)$ singlets):}~~~&J_0 +\sum_{s=1}^{\infty} J_s \\
(\Delta,S) = & (d-1,0)^-+\sum_{s=1}^{\infty}(d-2+s,s)\,.
\end{aligned}
\label{single-trace-fer}
\end{equation}
In $d=3$, it is also possible to impose a Majorana condition, so that the resulting theory has $O(N)$ invariance, and the $O(N)$ singlet sector 
involves only the even spin operators. Note that for general $d>3$, there would be more operators in the single trace spectrum, because one 
can construct fermion bilinears involving products of more than one $\gamma$ matrix (we will comment on this further in the next section).  

The single trace spectrum (\ref{single-trace-fer}) is 
very similar to the one in the free scalar theory, with the exception that the scalar operator is parity odd and has $\Delta=2$. The AdS$_4$ dual 
should then be a HS theory which includes a {\it pseudoscalar} with $m^2=-2$, together with the tower of HS gauge fields of all (even) spins. 
It is also easy to see that the interactions in this theory must be different from the one in the dual to the free scalar, because 
the correlation functions of HS currents are given by different tensor structures in scalar and fermionic CFTs, as in the case of 
$\langle TTT\rangle$, eq. (\ref{TTT}). 

So we learn that there should be two inequivalent HS theories in AdS$_4$, with almost identical spectrum (except for the parity 
of the bulk scalar), but with different interactions. Indeed, as we will see later, there are precisely two parity invariant HS theories in AdS$_4$, which 
have the required spectrum and interactions. They are usually referred to as ``type A" and ``type B" theories, the former including a 
parity even scalar and the latter a parity odd one. The conjecture that the type B theory should be dual to the fermionic vector model was 
first made in \cite{Sezgin:2003pt, Leigh:2003gk}. 

\begin{figure}
\begin{center}
\includegraphics[width=0.5\textwidth]{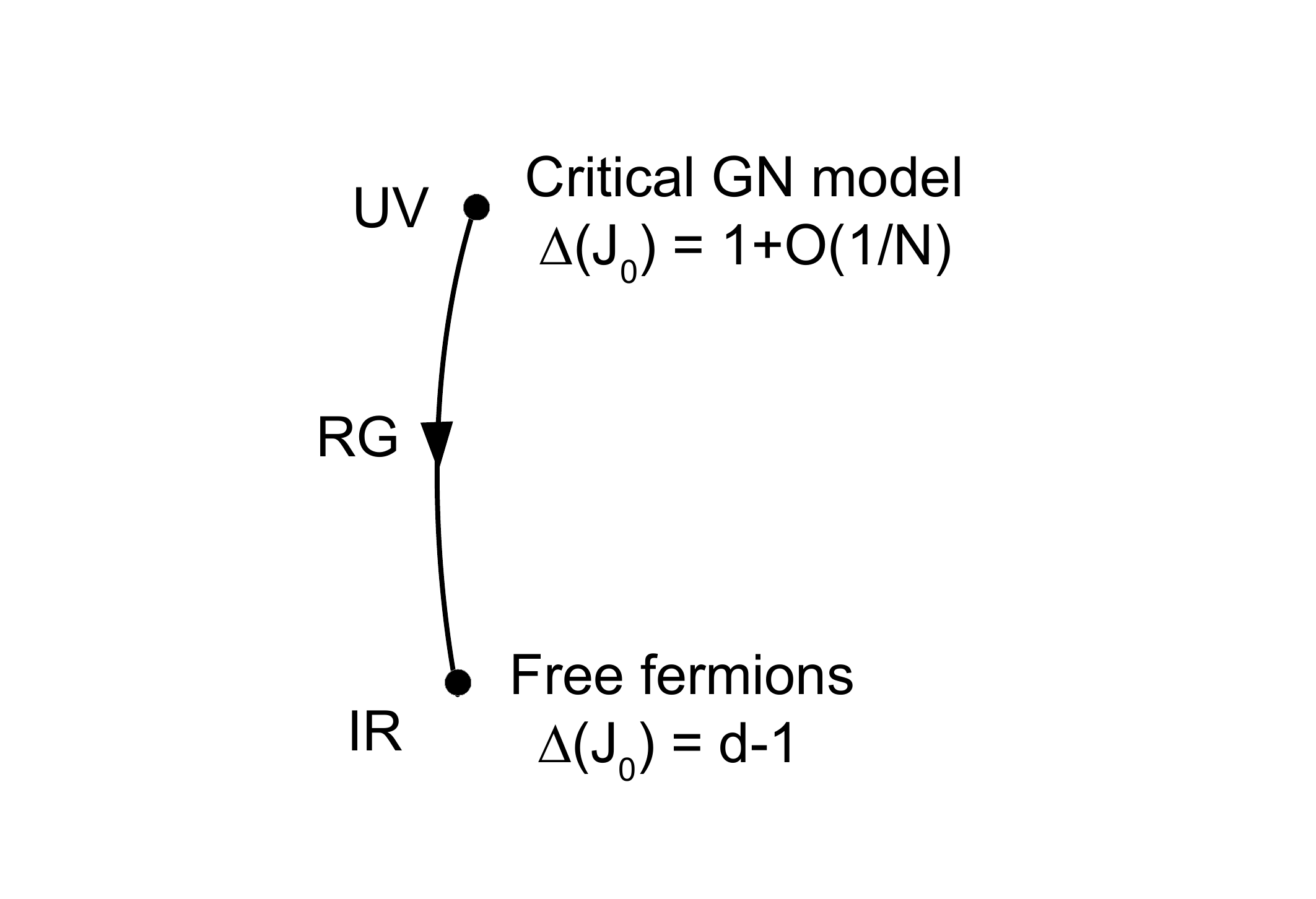}
\end{center}
\label{RG-GN}
\caption{RG flow from the UV fixed point of the Gross-Neveu model to free fermions.}
\end{figure}

Note that in the type B theory, duality with the free fermion CFT requires that the bulk scalar is assigned the $\Delta=2$ boundary condition. This 
corresponds to unbroken HS symmetry in this fermionic case. As in the scalar case, we expect that the alternate boundary condition $\Delta=1$ corresponds 
to an interacting CFT related to the free one by a double trace deformation. This is just the familiar Gross-Neveu model 
\begin{equation}
S= \int d^d x \left(\bar\psi_i \gamma^{\mu}\partial_{\mu}\psi^i+\frac{g}{2}(\bar\psi_i\psi^i)^2\right)\,.
\end{equation}
The interaction is irrelevant, but working in the large $N$ expansion one can show, by methods similar to the ones described above 
for the critical $O(N)$ model, that there is a non-trivial UV fixed point where the scalar operator (which may traded by a Hubbard-Stratonovich 
auxiliary field) has dimension $\Delta_{UV}=1+O(1/N)$. We may refer to this CFT as the critical Gross-Neveu or critical fermion theory. 
The free CFT, with $\Delta_{IR} = d-1$, sits now at the IR fixed point of the RG flow, as shown in Figure \ref{RG-GN}. 
While the large $N$ expansion may be developed formally for any $d$, it is clear that the UV fixed point 
is unitary only for $d<4$, and in particular in the physically interesting dimension $d=3$. Note that the UV fixed point may be also accessed perturbatively, at finite $N$, in the $d=2+\epsilon$ expansion, where it is well known that the Gross-Neveu model has a weakly coupled 
UV fixed point. Let us also mention that the same critical fermion CFT admits a ``UV complete" description as the IR fixed point 
of the ``Gross-Neveu-Yukawa" model \cite{Hasenfratz:1991it,ZinnJustin:1991yn,Moshe:2003xn}
\begin{equation}
S_{GNY}= \int d^d x \left(\bar\psi_i \gamma^{\mu}\partial_{\mu}\psi^i+\frac{1}{2}(\partial_{\mu}\sigma)^2 + 
g \sigma \bar\psi_i \psi^i +\frac{\lambda}{4}\sigma^4\right)\,,
\end{equation}
which has perturbative IR fixed points in $d=4-\epsilon$ that are expected to be equivalent to the UV fixed points of the Gross-Neveu model with 
quartic interaction (this can be checked explicitly at large $N$ by matching critical exponents computed in the two approaches). Note that the relation 
between the IR fixed points of the GNY model and the UV fixed points of the Gross-Neveu model is analogous to the relation between 
the Wilson-Fisher fixed points of the $\phi^4$ theory (\ref{quartic}) and the UV fixed points of the non-linear sigma model in $d=2+\epsilon$. 

The AdS dual of the large $N$ critical Gross-Neveu model may be deduced analogously to the scalar case. It is the {\it same} type B theory dual 
to the free fermion CFT, but with the $\Delta=1$ boundary condition assigned to the bulk pseudoscalar. In the interacting CFT, the HS currents 
are weakly broken at large $N$ and satisfy a non-conservation equation analogous to (\ref{divJbr}), which implies anomalous dimensions starting 
at order $1/N$. 
In the bulk, one expects then the HS fields to acquire masses via loop corrections, as in Figure \ref{2pt-loop}, when the scalar has $\Delta=1$ boundary condition (but not 
when it has $\Delta=2$ boundary condition).  

\section{Summary of parity invariant HS/CFT dualities}
\label{HS-summ}

Let us summarize in Table 1 the AdS$_4$/CFT$_3$ Higher Spin/Vector Model dualities \cite{Klebanov:2002ja,Leigh:2003gk,Sezgin:2003pt} discussed so far.
\begin{table}[ht]
\tbl{\label{tab:KPSS}Summary of AdS$_4$/CFT$_3$ Higher Spin/Vector Model dualities \cite{Klebanov:2002ja,Leigh:2003gk,Sezgin:2003pt}}{
\begin{tabular}{c|c|c}
  & Vasiliev HS$_4$ type A & Vasiliev HS$_4$ type B  \\
\hline
$\Delta=1$ scalar b.c. & Free Scalar &  Critical Fermion (Gross-Neveu) \\
\hline
$\Delta=2$ scalar b.c. & Critical Scalar (Wilson-Fisher) & Free Fermion 
\end{tabular}}
\label{T1}
\end{table}

In each of the cases shown in table, one may consider the $O(N)$ or $U(N)$ version of the vector models, which are respectively dual to the minimal (even spins only) 
and non-minimal (all integer spins) higher spin theories. An interesting feature of the non-minimal theories is that in addition to two possible boundary conditions one can impose on the bulk scalar, there is also a one-parameter family of conformally invariant boundary conditions one can impose on the bulk spin-1 gauge field 
\cite{Witten:2003ya, Chang:2012kt}. With the ordinary boundary condition ($\Delta_1=2$) on a spin-1 gauge field in AdS$_4$, the bulk gauge field is dual to a conserved spin 1 current on the boundary, as we have assumed above. On the other hand, one may impose the alternate boundary condition ($\Delta_1=1$),  which corresponds on the CFT side to {\it gauging} the global $U(1)$ flavor symmetry. 
More generally one can impose a (parity breaking) mixed boundary condition, which corresponds to setting a linear combination of the ``electric" field $F_{zi}$ ($z$ being the Poincar\'e radial coordinate) and the ``magnetic" field $\epsilon_{ijk} F_{jk}$ to vanish at the boundary. With the mixed boundary condition, the dual CFT is obtained from the original one by gauging the global $U(1)$ flavor symmetry and turning on Chern-Simons coupling at some level $k$. The case $k=\infty$ corresponds to the ordinary boundary condition, while the ``purely electric" boundary condition corresponds to $k=0$. In the latter case, while one gauges the boundary flavor current, the kinetic term for the boundary gauge field is entirely generated from integrating out the matter fields at one-loop, corresponding to the case of three-dimensional critical QED 
\cite{Appelquist:1981vg, 1988PhRvL..60.2575A}. So, to summarize, the three-dimensional critical QED's with $N$ bosonic or fermionic flavors, restricted to $U(N)$ singlet sector, are holographically dual to type A or type B non-minimal Vasiliev theory, with the alternate boundary condition imposed on the bulk spin-1 gauge field, and possibly including a $U(1)$ Chern-Simons term at level $k$. By further imposing the alternate boundary conditions on the bulk scalar, one may also obtain the dual to the $CP^{N-1}$ model, or the Gross-Neveu model coupled to a $U(1)$ gauge field. Let us also briefly mention that it is possible to impose alternate boundary conditions, $\Delta_s=2-s$, on the higher spin fields as well: this corresponds to gauging the HS symmetry at the boundary, and the resulting theory is a conformal higher spin theory \cite{Leigh:2003ez, Bekaert:2010ky, Vasiliev:2012vf, Giombi:2013yva}.

In the table we have focused on the AdS$_4$/CFT$_3$ case, which is the most well-understood so far, but it is natural to ask about higher dimensional generalizations 
of these dualities. In the free scalar case, as mentioned earlier, the spectrum of single trace operators is given by (\ref{single-trace}) in any $d$, and this matches the spectrum of the known Vasiliev theory in AdS$_{d+1}$ \cite{Vasiliev:2003ev}, which we may view as a generalization to all $d$ of the type A HS$_4$ theory. In the free 
fermion case, on the other hand, for $d>3$ the single trace spectrum is more complicated than (\ref{single-trace-fer}), as mixed symmetry 
representations of $SO(d)$ can appear \cite{Anselmi:1998bh, Anselmi:1999bb,Vasiliev:2004cm,dolan,Alkalaev:2012rg, Giombi:2014yra}. 
For instance, for the free fermion in $d=4$ restricted to the $U(N)$ singlet sector, in addition to (\ref{single-trace-fer}), there is an extra scalar and an extra tower of totally symmetric HS operators 
\begin{equation}
\tilde{J}_0 = \bar\psi_i \gamma_5 \psi^i\,,\qquad \tilde{J}_{\mu_1\ldots \mu_s} = \bar\psi_i \gamma_5 \gamma_{\mu_1}\partial_{\mu_2\cdots \mu_s}\psi^i+\ldots \,,\quad s=1,2,3,\ldots\,,
\end{equation}
and also a tower of operators in the mixed symmetry representation corresponding to a two-row Young diagram with $s$ boxes in the first row and 1 box in the second row
\begin{equation}
B_{\mu_1\ldots \mu_s,\mu} = \bar\psi_i \gamma_{\mu \mu_1}\partial_{\mu_2}\cdots \partial_{\mu_s}\psi^i+\ldots\,,\quad s\ge 1\,.
\label{mixed}
\end{equation}
The dual HS theory in AdS$_5$, which should be viewed as a generalization of the type B HS$_4$ theory, should then involve two scalars, two towers of totally symmetric 
HS fields, and one tower of mixed symmetry HS fields dual to (\ref{mixed}). While one can write free equations for these fields in AdS, the full theory describing their interactions has not yet been constructed. 

One may also ask if there is any interacting version of the duality in $d>3$ that one can obtain by changing boundary conditions of the bulk fields. In the scalar CFT case, 
where the bulk scalar field has $m^2=-2(d-2)$, the alternate boundary condition $\Delta=2$ is actually above unitarity for $d<6$. This suggests the possibility of 
a unitary interacting vector model in $d=5$, dual to Vasiliev ``type A" theory in AdS$_6$ with alternate $\Delta=2$ boundary condition on the bulk scalar \cite{Bekaert:2011cu,Bekaert:2012ux,Giombi:2014iua}. On the CFT side, since the quartic interaction is irrelevant for $d>4$, 
this interacting CFT should be viewed as a UV fixed point of (\ref{quartic}), whose existence can be seen formally in the large $N$ expansion \cite{Parisi:1975im,Fei:2014yja}. In \cite{Fei:2014yja}, 
it was shown that the following model with $N+1$ scalars and $O(N)$ invariant cubic interactions
\begin{equation}
S= \int d^d x\Big(  \frac{1}{2} 
\left(\partial_{\mu} \phi^{i}\right)^2  + \frac{1}{2} (\partial_{\mu} \sigma)^2  + \frac{g_1}{2} \sigma \phi^i \phi^i + \frac{g_2}{6} \sigma^3\Big)\,,
\end{equation}
posseses IR stable, perturbatively unitary fixed points in $d=6-\epsilon$ which provide a ``UV completion" 
of the large $N$ UV fixed points of the $O(N)$ model in $d>4$. This proposal has passed various non-trivial checks 
\cite{Fei:2014yja, Fei:2014xta, Gracey:2015tta}. These perturbative fixed points exist for $N>1038(1+O(\epsilon))$, and are expected to be unitary to all 
orders in $\epsilon$ and $1/N$ expansions. However, non-perturbative effects presumably render the vacuum metastable via instanton effects. Understanding the counterpart 
of this instability from the point of view of the dual HS theory in AdS$_6$ with $\Delta=2$ boundary condition is an interesting open problem.   

\section{Chern-Simons vector models}
\label{CS-duality}

One important feature of the HS/vector model dualities discussed so far is that they involve a projection to the $U(N)/O(N)$ singlet sector of the CFT. This is 
essential for the matching of bulk and boundary spectra to work. Without this projection, we would have many more primaries in the CFT (for instance, $\phi^i$ itself) that 
do not have a counterpart in the Vasiliev HS theory. As mentioned earlier, a natural way to impose the singlet constraint is to weakly gauge the $U(N)/O(N)$ symmetry, 
and then consider the zero gauge coupling limit. This decouples the gauge field, but we still retain the constraint that only gauge invariant operators are physical. 

In $d=3$, there is a nice way to gauge the symmetry without breaking conformal invariance: we can couple the vector model to a $U(N)/O(N)$ Chern-Simons gauge field 
\cite{Giombi:2011kc, Aharony:2011jz}. For 
instance, in the case of the fermionic $U(N)$ vector model, we consider the gauge theory
\begin{equation}
S=\frac{k}{4\pi} \int d^3x {\rm Tr}\left(A dA+\frac{2}{3}A^3\right)+\int d^3 x \bar\psi \gamma^{\mu}D_{\mu}\psi 
\label{CS-fer}
\end{equation}
where $\psi$ is in the fundamental representation of the $U(N)$ gauge group (similarly, one can consider the $O(N)$ version). The singlet sector of the free vector model can be recovered by sending $k\rightarrow \infty$, which decouples the gauge field. Since we are interested in the large $N$ limit, we can take the `t Hooft limit 
\begin{equation}
N,k \rightarrow \infty\,,\qquad \lambda \equiv \frac{N}{k} \quad {\rm fixed}\,.
\end{equation} 
The singlet sector of the large $N$ vector model is then obtained by sending the `t Hooft coupling $\lambda$ to zero.  Similarly, one may couple the scalar vector model 
to a $U(N)/O(N)$ Chern-Simons gauge field
\begin{equation}
S=\frac{k}{4\pi} \int d^3x {\rm Tr}\left(A dA+\frac{2}{3}A^3\right)+\int d^3 x D_{\mu}\phi^* D^{\mu}\phi\,. 
\label{CS-sc}
\end{equation}

This construction naturally suggests that it should be interesting to study the more general theories with $\lambda \neq 0$. As we will see below, it turns out that 
even when the gauge coupling $\lambda$ is turned on, these models possess approximate HS symmetry at large $N$ and should be dual to (parity breaking) HS theories in AdS$_4$, 
leading to a wide generalization of the conjectures of \cite{Klebanov:2002ja,Leigh:2003gk,Sezgin:2003pt}.  

As mentioned above, a nice feature of these models is that coupling the vector models to the Chern-Simons theory does not break conformal invariance. Essentially, the reason 
is that the Chern-Simons level $k$ is quantized, and it cannot run in perturbation theory (except for a possible integer shift at one loop). This is all we need to 
prove that the fermionic model (\ref{CS-fer}) defines a CFT for any $\lambda$, because there are no relevant interactions that can be generated, except for 
the fermion mass term that we can always tune to zero. In the scalar version of the model, one has to be more careful since there is a classically marginal coupling 
$\frac{\lambda_6}{N^2} (\phi^* \phi)^3$ that can be generated. However, one can show that in the large $N$ limit, with $\lambda$, $\lambda_6$ fixed, $\beta_{\lambda_6}=0$ 
\cite{Aharony:2011jz}. Hence, 
in the $N\rightarrow \infty$ limit, the CS-scalar model defines a CFT with two marginal parameters, $\lambda$ and $\lambda_6$. Away from infinite $N$, one finds a non-zero beta function $\beta_{\lambda_6}(\lambda, \lambda_6)$, and one can show that there are zeroes $\lambda_6^*(\lambda)$, at least for sufficiently large $N$ \cite{Aharony:2011jz}. 

\begin{figure}
\begin{center}
\includegraphics[width=0.7\textwidth]{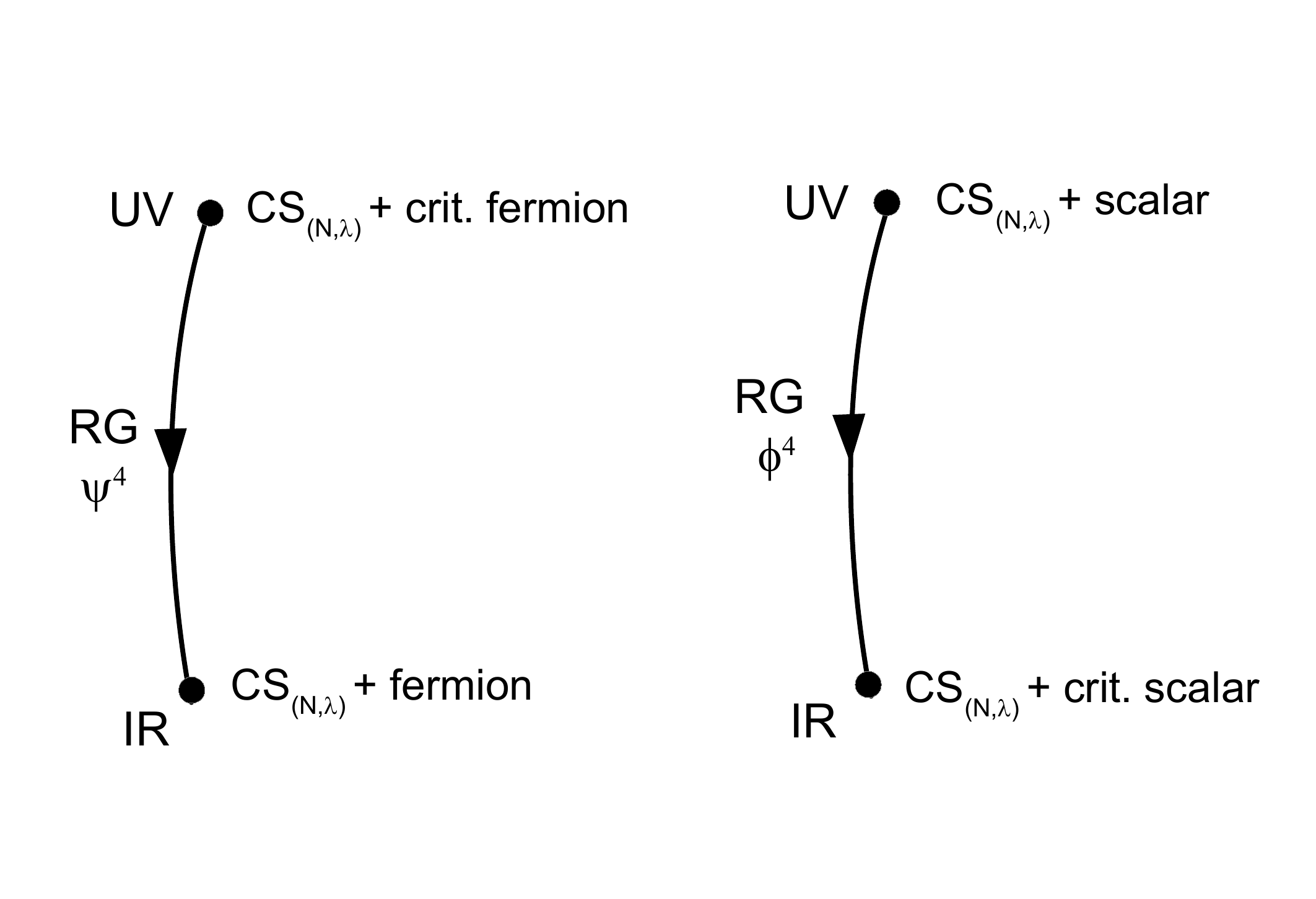}
\end{center}
\caption{RG flows between large $N$ Chern-Simons vector model CFTs. Each fixed point is labelled by $N$ and the Chern-Simons coupling $\lambda$. 
At infinite $N$, the CS-scalar and CS-critical-fermion CFTs admit an additional marginal deformation corresponding to sextic couplings.}
\label{RG-CSM}
\end{figure}

As in the ungauged case, we can deform the CFTs defined by (\ref{CS-fer}) and (\ref{CS-sc}) by adding double trace interactions $g_4 (\bar\psi\psi)^2$ and 
$\lambda_4 (\phi^* \phi)^2$. This allows to obtain one-parameter generalizations of the critical scalar (Wilson-Fisher) and critical fermion (Gross-Neveu) which include 
the Chern-Simons coupling. A schematic depiction of the RG flows is given in Figure \ref{RG-CSM}. Note that in the critical fermion case, at infinite $N$, we have 
an additional marginal deformation $g_6 (\bar\psi \psi)^3$ (since $\Delta_{\bar\psi \psi}=1+O(1/N)$ in the UV), analogous to the $(\phi^* \phi)^3$ term in the CS-scalar model. On the other hand, in the CS-critical-scalar, the $(\phi^* \phi)^3$ interaction becomes irrelevant (since $\Delta_{\phi^2}=2+O(1/N)$ in the IR).     

All these CFTs should have corresponding AdS$_4$ duals. Since we know that for $\lambda=0$ the vector models are dual to type A/type B Vasiliev theory in AdS$_4$, 
the duals must be deformations of these HS theories parameterized by $\lambda$ (in particular, the bulk theory should be parity breaking since the CS term breaks parity on the CFT side). A crucial observation is that, even when $\lambda$ is turned on, the HS symmetry is still only weakly broken at large $N$, for any $\lambda$. Let us sketch 
the argument to see this, focusing on the fermionic case for concreteness (the scalar case goes through in a similar way) \cite{Giombi:2011kc, Aharony:2011jz}. Recall that in the free theory ($\lambda=0$), the 
spectrum of single trace operators is given by
\begin{equation}
\begin{aligned}
&J_0 = \bar\psi \psi\,,\qquad J_s \sim  \bar\psi \gamma \partial^{s-1}\psi\,\\
&(\Delta,S) = (2,0)^{-}+\sum_{s=1}^{\infty}(s+1,s) \,.
\label{CS-fer-str}
\end{aligned}
\end{equation}
When we turn on the gauge coupling, we still have this set of primary operators, provided we make them gauge invariant by replacing derivatives with covariant ones. The 
crucial point is now that there are no additional single trace operators beside these. This is essentially because the CS theory is topological, and it does not provide 
any additional local operators. Operators that involve powers of the field strength inserted between two fermions, which naively look like single trace, in fact 
behave as multi-trace, due to the equation of motion
\begin{equation}
(F_{\mu\nu})^{i}_{\ j} = \frac{1}{k} \bar\psi_j \gamma^{\rho}\psi^i \epsilon_{\mu\nu\rho}\,,
\end{equation}     
which essentially ``breaks" the would-be single trace operator into a multi-trace one. Now, when interactions are turned on, we expect the HS currents to be not conserved 
and satisfy an equation of the form
\begin{equation}
\partial \cdot J_s = K_{s-1}
\end{equation} 
where the operator on the right-hand side has spin $s-1$ and, in the limit where the currents is conserved, it should be a conformal primary of dimension $\Delta=s+2$ in order to match the dimension of the operator on the left-hand side. But as we argued above, (\ref{CS-fer-str}) (suitably 
covariantized) are all the single-trace operators in the model, and among them there is no operator with the quantum numbers $(s+2,s-1)$! It follows that the operator appearing on the right-hand side of the non-conservation equation must be a multi-trace one, similarly to (\ref{divJbr}). Schematically, the structure of the non-conservation relation takes the following form 
\begin{equation}
\label{CS-divJ}
\partial \cdot J_s = \frac{f(\lambda)}{\sqrt{N}} \sum_{s_1+s_2<s} \partial^n J_{s_1} \partial^m J_{s_2} 
+  \frac{g(\lambda)}{N} \sum_{s_1+s_2+s_3 <s} \partial^n J_{s_1} \partial^m J_{s_2} \partial^p J_{s_3} 
\end{equation}
where the currents are assumed to be normalized so that $\langle J_s J_s\rangle \sim O(1)$ at large $N$. Note that the operator on the right-hand side is at most 
triple trace. This is because at infinite $N$ the current $J_s$ has twist $\Delta - s = 1$ and its divergence twist 3, and so can only be expressed 
in terms of a sum of products of two or three currents, but no more than three.

Eq. (\ref{CS-divJ}) imply that the HS currents are not broken at infinite $N$, for any $\lambda$, and the anomalous dimensions are only generated starting at order $1/N$
\begin{equation}
\Delta_s = s+1+\frac{\gamma_s(\lambda)}{N}+\ldots 
\end{equation}
It follows that the dual theory should still be a higher spin gravity theory involving classically massless HS fields, with the HS symmetry broken by quantum effects. The same conclusion holds for the CS coupled to scalar (\ref{CS-sc}). Thus, we are led to the following generalization of the Klebanov-Polyakov conjecture 
\cite{Giombi:2011kc, Aharony:2011jz}
\begin{equation}
\begin{aligned}
\mbox{Chern-Simons vector model} \quad \Leftrightarrow \quad \mbox{Parity breaking HS$_4$ gravity} 
\label{HS-vec-parity}
\end{aligned}
\end{equation}
where the HS theory on the right-hand side should be a parity breaking generalization of the type A/type B models. We will see that the Vasiliev construction of the HS 
equations of motion in AdS$_4$ indeed includes a family of parity breaking theories that are natural candidates to be dual to the CS vector model CFTs. Note that the arguments above do not directly show that the dimension of the scalar operator $J_0$ is not renormalized at planar level, 
but one can argue \cite{Giombi:2011kc, Aharony:2011jz} that this is the case as well, 
essentially because the scalar operator has to appear on the right-hand side of the non-conservation equation (\ref{CS-divJ}). Therefore, one has $\Delta_{J_0} = 2+O(1/N)$ in the CS-fermion theory, and $\Delta_{J_0} = 1+O(1/N)$ in the CS-scalar theory, for any $\lambda$. The dual bulk scalar has therefore $m^2=-2$ classically, like in the type A/type B models. Let us mention that, while the duality (\ref{HS-vec-parity}) makes sense and is consistent at the level 
of gauge invariant local operators and their correlation functions on the plane, which is our main focus here, 
on manifolds with non-trivial topology \cite{Banerjee:2012gh} it presumably requires extending the HS theory 
by some sort of topological sector to account for the non-trivial CS dynamics. 

Interestingly, the weakly broken HS symmetries can be used to derive non-trivial constraints on the correlation functions of the Chern-Simons vector models at large $N$ 
\cite{Maldacena:2012sf}. Using the structure of the non-conservation relation (\ref{CS-divJ}), Maldacena and Zhiboedov showed that 3-point functions of the HS currents 
in the planar limit are constrained to be, in the fermionic case
\begin{eqnarray}
&&\langle J_{s_1} J_{s_2} J_{s_3}\rangle= \frac{1}{\sqrt{\tilde N}} \left[
\frac{1}{1+\tilde \lambda^2} \langle J_{s_1} J_{s_2} J_{s_3}\rangle_{\rm free~fer}+
\frac{\tilde \lambda^2}{1+\tilde \lambda^2} \langle J_{s_1} J_{s_2} J_{s_3}\rangle_{\rm free~sc}\right.\cr 
&&\qquad \qquad \qquad \qquad ~~~
\left.+\frac{\tilde \lambda}{1+\tilde \lambda^2} \langle J_{s_1} J_{s_2} J_{s_3}\rangle_{\rm odd}
\right]
\label{jjj-CS}
\end{eqnarray}
where $\tilde N= N c(\lambda)$ is related to the normalization of the stress-tensor 2-point function, and $\tilde \lambda =\alpha_1 \lambda+\alpha_3 \lambda^3+\ldots$ is a function of 
coupling which is essentially related to $f(\lambda)$ in (\ref{CS-divJ}), up to normalization. The symmetry argument does not allow to fix 
$\tilde N$ and $\tilde \lambda$ in terms of $\lambda$, but one can fix them by an explicit calculation \cite{Aharony:2012nh, GurAri:2012is}, 
see eq. (\ref{tN-tlam}) below. In the CS-scalar case, one finds the same expression, with the role of fermion and scalar tensor structures exchanged 
(and in principle different functions $\tilde N(\lambda), \tilde \lambda(\lambda)$, which are not fixed by HS symmetry considerations).

\begin{figure}
\begin{center}
\includegraphics[width=0.9\textwidth]{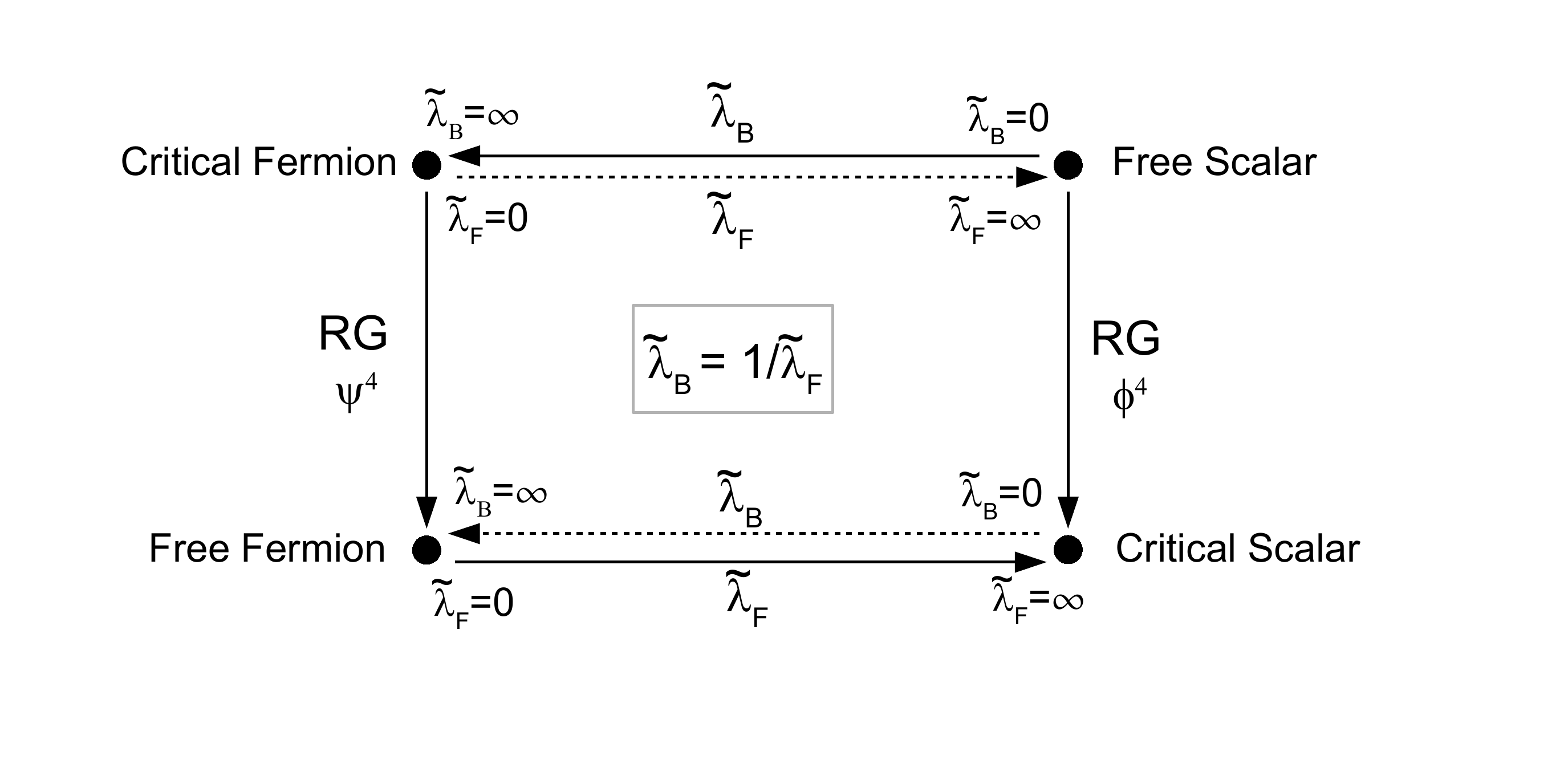}
\end{center}
\label{bose-fermi}
\caption{Schematic depiction of the ``3d bosonization" duality involving large $N$ scalar and fermion vector models coupled to 
$U(N)$ Chern-Simons theory. The vertical RG flow lines are present for all values of $\tilde\lambda_{B,F}$.}
\end{figure}

Note that from (\ref{jjj-CS}) we see that as we dial $\tilde \lambda$ from zero to infinity, we interpolate between free fermion correlators at $\tilde \lambda=0$, and {\it critical} 
scalar correlators at $\tilde \lambda = \infty$. The reason we get the critical and not free scalar at $\tilde \lambda = \infty$ is that the dimension of the scalar operator $J_0$ 
is $\Delta=2+O(1/N)$ for all $\lambda$. Similarly, if we start from the CS-scalar theory and dial $\tilde \lambda$ to infinity, we end up with the correlators of the 
critical fermion (the UV fixed point of the flow on the left hand side of Figure \ref{RG-CSM}). 
At generic values of $\tilde \lambda$, we map fermion to boson correlators,
and vice-versa, by $\tilde \lambda_F = \frac{1}{\tilde \lambda_B}$, where the subscripts `$F$' and `$B$' stand for fermion and boson 
respectively.  
This suggests a strong/weak ``3d bosonization" duality 
\cite{Giombi:2011kc, Aharony:2011jz, Maldacena:2012sf, Aharony:2012nh} between fermionic and scalar Chern-Simons vector models, as schematically shown in the diagram in Figure \ref{bose-fermi}.   

While the relation between $\tilde N$, $\tilde {\lambda}$ and the CS coupling is not fixed by the weakly broken HS symmetry arguments, remarkably 
all-orders calculations in the planar limit are possible in these models. An explicit calculation of 2-point and 3-point functions 
yields the results \cite{Aharony:2012nh, GurAri:2012is}
\begin{equation}
\tilde N = 2N \frac{\sin(\pi \lambda)}{\pi \lambda}\,,\qquad 
\tilde \lambda = \tan(\frac{\pi \lambda}{2})\,,
\label{tN-tlam}
\end{equation} 
both in the fermionic and scalar theories, in terms of the corresponding 't Hooft couplings.
The mapping $\tilde \lambda_F =1/\tilde {\lambda}_B$ then translates to 
$\lambda_F = 1-\lambda_B$ or more precisely, taking into account a sign flip of the CS level (see e.g. \cite{GurAri:2012is} for a discussion of this)
\begin{equation}
|\lambda_F| = 1-|\lambda_B|\,,\qquad \frac{N_F}{|\lambda_F|} = \frac{N_B}{|\lambda_B|} \,,\qquad {\rm sign}(\lambda_F) = -{\rm sign}(\lambda_B)\,. 
\end{equation}
In terms of $N$ and $k$, defined in the conventions where $\lambda = N/(N+k)$, we see that this relation acts as a generalization of the 
level-rank duality of pure CS theory \cite{Naculich:1990pa, Mlawer:1990uv, Camperi:1990dk}. To be a bit more precise,
 one should note that in the fermionic theory the CS level should be 
half-integer, due to the parity anomaly \cite{Niemi:1983rq, Redlich:1983dv, Redlich:1983kn}. Taking this fact into account, it was argued 
in \cite{Aharony:2012nh} that the precise form of the duality should be \footnote{Generalizations of this duality map involving the $SU(N)$ 
gauge group were also recently discussed in \cite{Aharony:2015mjs}, where the mapping of baryon and monopole operators was studied, see also 
\cite{Radicevic:2015yla}.}
\begin{equation}
U(N)_{k-1/2} ~~{\rm Fermion}\quad \Leftrightarrow \quad U(k)_{-N} ~~ {\rm Critical~Scalar}
\label{lev-rank}
\end{equation}
and similarly for the duality between CS-critical-fermion and CS-scalar (in this case, one may also derive \cite{GurAri:2012is} the mapping between 
the additional marginal couplings $g_6 (\bar\psi\psi)^3$ and $\lambda_6 (\phi^* \phi)^3$). In addition to 2-point and 3-point functions, further 
tests of the duality (\ref{lev-rank}) have been obtained by calculating the thermal free energy of the models on the plane. In the 't Hooft limit, 
the free energy takes the form
\begin{equation}
F = -\log Z_{S^1_{\beta}\times R^2} = N V_2 T^2 f_{\rm th}(\lambda)
\label{F-thermal}
\end{equation}
where $S^1_{\beta}$ is the thermal circle ($\beta=1/T$) and $V_2$ is the (infinite) volume of the plane; $f_{\rm th}(\lambda)$ is a non-trivial 
function of $\lambda$ that can be computed exactly in these models using an Euclidean version of light-cone gauge 
\cite{Giombi:2011kc, Jain:2012qi}. Provided 
the contribution of the holonomy of the gauge field is properly taken into account \cite{Aharony:2012ns}, the result for (\ref{F-thermal}) in the boson and fermion theories 
can be shown to be precisely consistent with (\ref{lev-rank}). The duality has been also generalized to theories involving bosons and fermions on the same 
side \cite{Jain:2013gza, Chang:2012kt}, which include as a special case supersymmetric theories, for which dualities closely related to (\ref{lev-rank}) are well-established \cite{Giveon:2008zn, Benini:2011mf}. In fact, it was argued in \cite{Gur-Ari:2015pca} that at large $N$ one may derive the non-supersymmetric dualities from the supersymmetric ones by RG flow, and show that all planar correlators map correctly under the duality relation. For additional work and tests of the duality, 
see for instance \cite{Jain:2013py, Moshe:2014bja, Inbasekar:2015tsa, Bedhotiya:2015uga}. Note that the level-rank duality of Chern-Simons theory, 
as well as the closely related supersymmetric dualities \cite{Giveon:2008zn, Benini:2011mf}, are valid at finite $N$ and $k$. It is therefore plausible 
that the non-supersymmetric duality (\ref{lev-rank}) also holds away from the large $N$ 't Hooft limit. For small $N$ and $k$, this duality may 
have interesting applications in condensed matter physics, see e.g. \cite{Karch:2016sxi, Murugan:2016zal, Seiberg:2016gmd} for recent closely related work.

\section{Fronsdal equations for free HS fields}
\label{Fronsdal}

We now leave the CFT side and review some basic aspects of the theory of massless HS fields. We start with 
the Frosdal equations describing the propagation of free HS gauge fields \cite{Fronsdal:1978rb, Fronsdal:1978vb}. We 
will focus on integer spins for simplicity, but the equations are also known for half-integer spins \cite{Fang:1978wz, Fang:1979hq}. 

\subsection{Flat spacetime}

Let us start with the case of flat $D$-dimensional spacetime, and recall the familiar examples of the 
free equation of motion for spin 1 (Maxwell theory) and spin 2 (linearized 
gravity). In the $s=1$ case, we have the vector field $A_{\mu}$ subject to the gauge symmetry 
\begin{equation}
\delta A_{\mu} =\partial_{\mu} \epsilon\,,
\end{equation}  
and the gauge invariant equation of motion is 
\begin{equation}
\partial_{\mu} F^{\mu\nu}=0\,,\qquad F_{\mu\nu}=\partial_{\mu}A_{\nu}-\partial_{\nu}A_{\mu}\,,
\end{equation}
or
\begin{equation}
\Box A_{\mu} -\partial_{\mu}\partial^{\nu}A_{\nu}=0\,.
\end{equation}
Of course, the equations of motion follow from the Maxwell lagrangian $S=\int d^d x \frac{1}{4}F_{\mu\nu} F^{\mu\nu}$. 

In the $s=2$ case, we start with the Einstein equations (with zero cosmological constant)
\begin{equation}
R_{\mu\nu}-\frac{1}{2}g_{\mu\nu}R=0\,,
\end{equation}
and writing $g_{\mu\nu} =\eta_{\mu\nu}+h_{\mu\nu}$, we can derive the linearized equations for the fluctuation $h_{\mu\nu}$:
\begin{equation}
\Box h_{\mu\nu}-\partial_{\mu}\partial^{\rho} h_{\rho\nu}-\partial_{\nu}\partial^{\rho}h_{\rho\mu}+\partial_{\mu}\partial_{\nu} h^{\rho}_{\ \rho}=0\,.
\end{equation}
It is not difficult to verify that this equation is invariant under the linearized gauge transformation 
\begin{equation}
\delta h_{\mu\nu} = \partial_{\mu}\epsilon_{\nu}+\partial_{\nu}\epsilon_{\mu}\,,
\end{equation}
which follows from the diffeomorphism invariance of Einstein equation. 

We want to generalize this to massless HS fields, which we take to be totally symmetric fields $\varphi_{\mu_1\mu_2\cdots \mu_s}$.
Historically, Fronsdal obtained the gauge invariant HS equations by taking the massless limit of the Singh-Hagen equations 
\cite{Singh:1974qz} for massive fields. We will 
not go into the details of how the equations were obtained, and just present the final result, which is relatively simple. First of all, Fronsdal's 
formulation requires that the totally symmetric HS field is constrained to be {\it double-traceless}
\begin{equation}
\varphi^{\mu \ \ \nu}_{\ \mu \ \ \nu \mu_5 \cdots \mu_s}=0\,.
\label{d-tr-cons}
\end{equation}
Note that this should be regarded as an ``off-shell" constraint on the field. The gauge invariance takes the form \footnote{We use round brackets to denote symmetrization of indices with ``strength one", i.e. 
$A_{(\mu} B_{\nu)} = \frac{1}{2}\left(A_{\mu}B_{\nu}+A_{\nu}B_{\mu}\right)$.}
\begin{equation}
\delta \varphi_{\mu_1\cdots \mu_s} = \partial_{(\mu_1}\epsilon_{\mu_2\cdots \mu_s)}\,,
\end{equation}
with the constraint that the spin-$(s-1)$ gauge parameter is traceless
\begin{equation}
\epsilon^{\mu}_{\ \mu \mu_3\cdots \mu_{s-1}} =0\,.
\end{equation}
Note that the double-traceless constraint (\ref{d-tr-cons}) is gauge invariant 
(it is important that the gauge parameter is traceless for this to be true). Generalizing the $s=1$ and $s=2$ cases given above, 
the gauge invariant free equation of motion for any spin is found to be
\begin{equation}
\begin{aligned}
{\cal F}_{\mu_1 \cdots \mu_s}\equiv  \Box \varphi_{\mu_1\cdots \mu_s} 
-s \partial_{(\mu_1} \partial^{\mu}\varphi_{\mu_2\cdots \mu_s)\mu}
+\frac{s(s-1)}{2} \partial_{(\mu_1}\partial_{\mu_2} \varphi_{\mu_3\cdots \mu_s)\mu}^{\ \ \ \ \ \ \ \ \ \ \mu} = 0\,.
\end{aligned}
\label{Fronsd}
\end{equation}
The reader may check as an exercise that the gauge variation of these equations is given by
\begin{equation}
\delta {\cal F}_{\mu_1\cdots \mu_s} \propto \partial_{(\mu_1} \partial_{\mu_2}\partial_{\mu_3} \epsilon_{\mu_4\cdots \mu_s)\mu}^{
\ \ \ \ \ \ \ \ \ \ \ \ \mu}
\end{equation}
and hence the equations are gauge invariant as desired when the gauge parameter is constrained to be traceless. 
One can also write down a Lagrangian that yields these equations of motion
\begin{equation}
S = \int d^D x \left(\varphi^{\mu_1\cdots \mu_s}{\cal F}_{\mu_1\cdots \mu_s}-\frac{1}{4}s(s-1)\varphi_{\mu}^{\ \ \mu \mu_3\cdots \mu_s}
{\cal F}^{\nu}_{\ \ \nu\mu_3\cdots \mu_s}\right)\,,
\label{Fronsd-L}
\end{equation}
which is gauge invariant provided the HS field is double-traceless. 

To extract the physical content of the Fronsdal equations, we need to gauge fix. It is convenient to choose the on-shell gauge where 
$\varphi_{\mu_1\cdots \mu_s}$ is transverse and traceless.\footnote{One way to reach this gauge is to start by fixing 
the de Donder gauge condition 
$\partial^{\mu}\varphi_{\mu\mu_2\cdots \mu_s} -\frac{s-1}{2}\partial_{(\mu_2}\varphi_{\mu_3\cdots \mu_s)\mu}^{\ \ \ \ \ \ \ \ \ \ \mu}=0$. 
Note that this leaves residual gauge transformations by a gauge parameter satisfying $\Box \epsilon_{\mu_1\cdots \mu_{s-1}}=0$. One 
can then use this residual gauge freedom to fix $\varphi^{\mu}_{\ \mu\mu_2\cdots \mu_s}=0$ on-shell, and 
hence $\partial^{\mu}\varphi_{\mu\mu_2\cdots \mu_s}=0$.}
Then, (\ref{Fronsd}) reduces to the Fierz-Pauli 
equations 
\begin{equation}
\begin{aligned}
&\Box \varphi_{\mu_1\cdots \mu_s} = 0\,,\\
&\partial^{\mu}\varphi_{\mu\mu_2\cdots \mu_s}=0\,, \\
&\varphi^{\mu}_{\ \ \mu\mu_3\cdots \mu_s}=0\,.
\end{aligned}
\label{FP}
\end{equation}
Note that the transverse-traceless condition does not completely fix the gauge. There are residual gauge transformations by the gauge parameters 
satisfying similar Fierz-Pauli equations
\begin{equation}
\begin{aligned}
&\Box \epsilon_{\mu_1\cdots \mu_{s-1}}  = 0\,,\\
&\partial^{\mu}\epsilon_{\mu\mu_2\cdots \mu_{s-1}}=0\,, \\
&\epsilon^{\mu}_{\ \ \mu\mu_3\cdots \mu_{s-1}}=0\,.
\label{FPg}
\end{aligned}
\end{equation}
The equations (\ref{FP}) describe the propagation of a number of polarizations $g_s$ equal to the dimension of the totally symmetric spin $s$ representation 
of the massive little group $SO(D-1)$
\begin{equation}
g_s = \frac{(2s+D-3)(s+D-4)!}{(D-3)!s!}\,.
\label{gs}
\end{equation}
Subtracting the number of degrees of freedom $g_{s-1}$ corresponding to the residual gauge transformations, the number 
of physical degrees of freedom is given by
\begin{equation}
\# {\rm d.o.f.}~ = g_s - g_{s-1} = \frac{(2s+D-4)(s+D-5)!}{(D-4)!s!}
\label{dof}
\end{equation}
which is indeed the correct number of physical polarization for a massless HS field (it corresponds to the dimension of the totally 
symmetric spin $s$ representation of $SO(D-2)$). Note that for $D=4$, (\ref{dof}) gives two degrees of freedom for any spin, corresponding to 
the two helicities $\pm s$, as expected. 

\subsection{(A)dS}

How do we generalize the Fronsdal equations (\ref{Fronsd}), or equivalently the action (\ref{Fronsd-L}), to curved spacetime? Naively, 
one may think we should just covariantize (\ref{Fronsd}) by replacing all derivatives by covariant derivatives, and assume that 
all tensor contractions are taken with the metric of the curved space. This amounts to minimally couple the HS fields to gravity. 
However, one finds that this procedure does not work for a general background. Essentially, due to the fact covariant derivatives do not commute, one finds that for $s\ge 3$ ($s\ge 5/2$ in the half-integer case) the gauge variation of (\ref{Fronsd}) 
involves the full Riemann tensor \cite{Aragone:1979hx}, which is unconstrained for general background. The lack of gauge invariance then implies that the mimally 
coupled HS massless fields do not propagate consistently in an arbitrary curved spacetime.

A solution to the problem \cite{Fronsdal:1978vb} can be obtained for very special backgrounds, namely the maximally symmetric ones. In this case, 
the Riemann tensor takes the simple form
\begin{equation}
R_{\mu\nu\rho\sigma} = -\frac{1}{\ell^2}\left(g_{\mu\rho}g_{\nu\sigma}-g_{\mu\sigma}g_{\nu\rho}\right)\,,
\end{equation}
where $\ell^2=\ell^2_{\rm AdS}$ for AdS, $\ell^2=-\ell^2_{\rm dS}$ for dS. In this case, the offending terms in the gauge variation 
greatly simplify, and it is possible to find a suitable compensating term to be added to the naively covariantized Fronsdal operator (\ref{Fronsd}). 
The result is that the gauge invariant equation of motion in (A)dS is given by
\begin{equation}
\begin{aligned}
&\nabla^2\varphi_{\mu_1\cdots \mu_s} 
-s \nabla_{(\mu_1} \nabla^{\mu}\varphi_{\mu_2\cdots \mu_s)\mu}
+\frac{s(s-1)}{2} \nabla_{(\mu_1}\nabla_{\mu_2} \varphi_{\mu_3\cdots \mu_s)\mu}^{\ \ \ \ \ \ \ \ \ \ \mu} \\
&
-\frac{1}{\ell^2}\left[\left((s-2)(s+D-3)-s\right)\varphi_{\mu_1\cdots\mu_s}
+\frac{s(s-1)}{4} g_{(\mu_1\mu_2}\varphi_{\mu_3\cdots \mu_s)\mu}^{\ \ \ \ \ \ \ \ \ \ \mu}\right]=0\,,
\end{aligned}
\label{Fronsd-AdS2}
\end{equation}
where the first line is just the covariantized version of (\ref{Fronsd}), and the second line is the compensating term needed for gauge invariance. 
Owing to this extra term, the equation 
of motion (\ref{Fronsd-AdS2}) is invariant under the gauge transformation
\begin{equation}
\delta \varphi_{\mu_1\cdots \mu_s} = \nabla_{(\mu_1}\epsilon_{\mu_2\cdots \mu_s)}\,,
\end{equation}
and describes the consistent propagation of massless fields of all integer spins in (A)dS. Note that, while the compensating term looks 
like a ``mass term", it is in fact just the required coupling to the (constant) curvature that ensures gauge invariance, as appropriate for a 
massless field. To check (\ref{Fronsd-AdS2}) in a familiar example, we can consider Maxwell equations in curved space. For general background, 
they are
\begin{equation}
\nabla_{\mu} F^{\mu\nu} = 0 = \nabla^2 A^{\nu}-\nabla_{\mu}\nabla^{\nu}A^{\mu} \,.
\end{equation} 
To compare with the form in (\ref{Fronsd-AdS2}), we have to commute the covariant derivatives 
\begin{equation}
0=\nabla^2 A^{\nu}-\nabla_{\mu}\nabla^{\nu}A^{\mu} =
\nabla^2 A^{\nu}-\nabla^{\nu}\nabla_{\mu} A^{\mu}-R^{\nu}_{\  \mu} A^{\mu} 
\label{Max-AdS}
\end{equation}
where we used $[\nabla_{\mu},\nabla^{\nu}]A^{\mu}=R^{\nu}_{\  \mu} A^{\mu}$. Recalling that in (A)dS we have 
$R_{\mu\nu}=-\frac{D-1}{\ell^2}g_{\mu\nu}$, (\ref{Max-AdS}) indeed reproduces (\ref{Fronsd-AdS2}). 

As in flat space, one can gauge-fix by imposing a transverse traceless gauge. 
The equation of motion then reduces to the curved space version of the Fierz-Pauli 
equations 
\begin{equation}
\begin{aligned}
&\left(\nabla^2-\frac{(s-2)(s+D-3)-s}{\ell^2}\right) \varphi_{\mu_1\cdots \mu_s} = 0\,,\\
&\nabla^{\mu}\varphi_{\mu\mu_2\cdots \mu_s}=0\,, \qquad 
\varphi^{\mu}_{\ \ \mu\mu_3\cdots \mu_s}=0\,,
\end{aligned}
\label{FP-AdS}
\end{equation}
and there are residual gauge transformations with the gauge parameter satisfying (see, for instance, \cite{Mikhailov:2002bp})
\begin{equation}
\begin{aligned}
&\left(\nabla^2-\frac{(s-1)(s+D-3)}{\ell^2}\right) \epsilon_{\mu_1\cdots \mu_{s-1}} = 0\,,\\
&\nabla^{\mu}\epsilon_{\mu\mu_2\cdots \mu_{s-1}}=0\,, \qquad 
\epsilon^{\mu}_{\ \ \mu\mu_3\cdots \mu_{s-1}}=0\,.
\end{aligned}
\label{FPg-AdS}
\end{equation}

The ``mass-like" term in (\ref{FP-AdS}) is precisely consistent with the duality with a conserved spin $s$ current at the boundary of AdS$_D$. Working in Poincare 
coordinates $ds^2 =\frac{\ell^2_{\rm AdS}}{z^2}\left(dz^2+\sum_{i=1}^d dx^i dx^i\right)$, with $D=d+1$
, one can show that a symmetric traceless transverse tensor 
obeying the wave equation $(\nabla^2-\kappa^2)\varphi_{\mu_1\cdots \mu_s}=0$ has the boundary behavior \cite{Mikhailov:2002bp, Giombi:2009wh}
\begin{equation}
\varphi_{i_1\cdots i_s}(z,\vec{x}) \sim z^{\Delta-s}\alpha_{i_1\cdots i_s}(\vec{x})
\end{equation}
with 
\begin{equation}
\Delta(\Delta-d)-s = \kappa^2\ell^2_{\rm AdS}
\end{equation}
Setting $\kappa^2 = \frac{(s-2)(s+D-3)-s}{\ell^2_{\rm AdS}}$ as given in (\ref{FP-AdS}), this yields $\Delta=d-2+s$, which is the dimension of a conserved current in the CFT$_d$ (the other root $\Delta =2-s$ corresponds to gauging the HS symmetry at the boundary \cite{Giombi:2013yva}). 

\section{Frame-like formulation of HS fields}
\label{fram-HS}

The Fronsdal approach to HS fields is often referred to as the ``metric-like" formulation, because the totally symmetric tensor $\varphi_{\mu_1\cdots \mu_s}$ generalizes 
the metric. A fully non-linear theory of interacting HS fields in this approach is not yet known at present 
(but see for instance \cite{Taronna:2012gb, Sleight:2016dba,Bekaert:2015tva} for some recent interesting progress in fixing 3-point and (partially) 4-point interactions). 

Vasiliev non-linear HS theory is instead based on the so-called ``frame-like" formulation, which essentially generalizes the vielbein approach to gravity and naturally makes use of the language of differential forms. In the case of gravity, it is well known that we can introduce the vielbein 
\begin{equation}
e^a_{\mu}
\end{equation}
where we use greek symbols to denote ``curved indices", and latin ones for ``flat indices". The vielbein is related to the metric by
\begin{equation}
g_{\mu\nu} = \eta_{ab} e^a_{\mu}e^b_{\nu}\,.
\end{equation}     
It is clear from this expression that we are allowed to make local Lorentz transformations on the vielbein, $\delta e^a = \epsilon^a_{\ b} e^b$, without changing the metric. Indeed, while the vielbein has $D^2$ components, the metric only has $D(D+1)/2$. The remaining $D(D-1)/2$ components are accounted for by the freedom of local Lorentz rotations, which act as gauge symmetries. The corresponding gauge field is the spin connection $\omega^{ab}_{\mu}$. 

It is natural to view $e_{\mu}^a$ and $\omega_{\mu}^{ab}$ as the components of one-forms
\begin{equation}
e^a = e^a_{\mu} dx^{\mu}\,, \qquad \omega^{ab} = \omega^{ab}_{\mu} dx^{\mu}\,.
\end{equation}
In terms of these differential forms, we can write the so-called Cartan structure equations 
\begin{equation}
\begin{aligned}
&de^a + \omega^a_{\ b} \wedge e^b = T^a \\
&d\omega^{ab} +\omega^a_{\ c} \wedge \omega^{cb} = R^{ab}\,.
\label{Cartan}
\end{aligned}
\end{equation}
Here $T^a$ is the torsion two-form, and $R^{ab}=\frac{1}{2}R_{\mu\nu}^{ab} dx^{\mu}\wedge dx^{\nu}$ is the Riemann tensor two-form. When the constraint $T^a=0$ is 
imposed, one can solve for the spin connection in terms of the vielbein, and the curvature two-form $R^{ab}$ then coincides with the usual Riemann tensor
\begin{equation}
R_{\mu\nu\rho\sigma} = R_{\mu\nu}^{ab} e_{\rho a} e_{\sigma b} \,.
\end{equation}

It is natural to combine $e^a$ and $\omega^{ab}$ into a single one-form and ask whether we can view it as a gauge field of some Lie algebra. We have 
$D^2+D(D-1)/2=D(D+1)/2$ one-forms, so we need a Lie algebra with this dimension. We know that $\omega^{ab}$ must be the gauge field of local 
Lorentz transformations, so the algebra must contain the Lorentz generators $M_{ab}$ satisfying the usual commutation relations. We also know 
that $e^a$ transforms as a vector under Lorentz rotations, and so the corresponding generator, call it $P_a$, must also transform as a vector, which fixes 
its commutator with $M_{ab}$. To get a closed algebra, the only possibility is then $[P_a, P_b] \propto M_{ab}$. By suitably rescaling $P_a$, we can 
put the algebra in the form
\begin{equation}
\begin{aligned}
&[M_{ab}, M_{cd}] = i\left(\eta_{bc}M_{ad}\pm {\rm 3~terms}\right)\\
&[M_{ab},P_c]=i\left(\eta_{bc}P_a-\eta_{ac}P_b\right)\\
&[P_a, P_b]=\frac{i}{\ell^2}M_{ab}\,.
\end{aligned}
\end{equation}
For $\ell\rightarrow \infty$, this is the Poincare algebra $ISO(D-1,1)$; for $\ell^2=-\ell^2_{\rm dS}$ it is the de Sitter algera SO(D,1); 
and for $\ell^2=+\ell^2_{\rm AdS}$, it is the AdS algebra $SO(D-1,2)$ (isomorphic to the conformal algebra in $d=D-1$). The generators 
$P_a$ correspond to local translations. We can now build the one-form
\begin{equation}
W = -i \left(e^a P_a +\frac{1}{2}\omega^{ab}M_{ab}\right)
\label{W-form}
\end{equation}
and view it as a gauge field of one of the three algebras above. This construction is familiar in the $D=3$ case, where one can show \cite{Witten:1988hc} 
that Einstein gravity with zero, positive or negative cosmological constant is (classically) equivalent to Chern-Simons gauge theory with 
gauge group respectively $ISO(2,1)$, $SO(3,1)$ or $SO(2,2)$. In the higher dimensional case, things do not work as nicely, but it is still useful 
to consider the object (\ref{W-form}) to understand the structure of curvatures, gauge transformations etc., and in particular because it leads 
to a natural generalization to the HS case. 

The curvature of (\ref{W-form}) is 
\begin{equation}
\begin{aligned}
dW+W\wedge W = &-i\left( (de^a+\omega^a_{\ b} \wedge e^b)P_a\right.\\
&\left.+\frac{1}{2}\left(d\omega^{ab}+\omega^a_{\ c} \wedge \omega^{cb}+\frac{1}{\ell^2} e^a\wedge e^b\right)M_{ab}\right)\\
= & -i \left( T^a P_a+\frac{1}{2} \left(R^{ab}+\frac{1}{\ell^2} e^a\wedge e^b\right)M_{ab}\right)\,.
\end{aligned}
\end{equation}
Note that the condition that $W$ is a flat connection gives
\begin{equation}
\begin{aligned}
dW+W\wedge W =0 \quad \Leftrightarrow \quad & T^a=0 \\
& R^{ab}=-\frac{1}{\ell^2}e^a \wedge e^b\,.
\label{Flat-W}
\end{aligned}
\end{equation}
The first condition is the zero torsion constraint, and the second equation is equivalent to
\begin{equation}
R_{\mu\nu\rho\sigma} = -\frac{1}{\ell^2} \left(g_{\mu\rho}g_{\nu\sigma}-g_{\mu\sigma} g_{\nu \rho}\right)\,.
\end{equation}
Hence, a flat connection corresponds to a maximally symmetric background. 

The gauge transformations on $W$ are dictated by the usual gauge theory rules, and read
\begin{equation}
\begin{aligned}
&\delta W = d\epsilon + [W,\epsilon] \\
&\epsilon=-i\left(\epsilon^a P_a +\frac{1}{2}\epsilon^{ab}M_{ab}\right)\,,
\end{aligned}
\end{equation}
where $\epsilon^a$ and $\epsilon^{ab}$ are the gauge parameters for local translations and local Lorentz respectively. After solving
for $\omega=\omega(e)$ and translating to the metric formalism, these reduce to the diffeomorphisms $\delta g_{\mu\nu} = \nabla_{(\mu}\epsilon_{\nu)}$. 
Let us note that in the AdS case it is also natural to use manifestly covariant $SO(D-1,2)$ (or $SO(D,1)$ in the dS case) 
notation by combining $P_a,M_{ab}$ into the $D(D+1)/2$ generators $T_{AB}, A,B=0,\ldots , D$, and $W=-i \omega^{AB} T_{AB}$. 

We now want to generalize the ``frame-like" approach to the HS case \cite{Lopatin:1987hz}. It is first useful to recall the linearized relation between vielbein and metric 
fluctuations. Writing 
\begin{equation}
\begin{aligned}
&g_{\mu\nu}=g^{(0)}_{\mu\nu}+\varphi_{\mu\nu} \\
&e_{\mu}^a = e_{(0)\,\mu}^a+\hat e^a_{\mu}\,,
\end{aligned}
\end{equation}
the relation $\eta_{ab}e^a_{\mu}e^b_{\nu}=g_{\mu\nu}$ yields, to linear order in the fluctuation
\begin{equation}
\begin{aligned}
&\varphi_{\mu\nu} = \hat e_{\mu, \nu}+\hat e_{\nu,\mu}\,,\\
&\hat e_{\mu, \nu} \equiv \hat e_{\mu}^a e_{(0)\, \nu}^b \eta_{ab}\,.
\end{aligned}
\end{equation}
Therefore we see that the totally symmetric Fronsdal field $\varphi_{\mu\nu}$ is equal to the symmetric part of the vielbein fluctuation (of course, 
this is as expected: the antisymmetric part of the vielbein fluctuation can be gauged away by the local Lorentz symmetry). The generalization to 
HS is natural. The analog of the vielbein is given by the one-form
\begin{equation}
e^{a_1\cdots a_{s-1}} = e^{a_1\cdots a_{s-1}}_{\mu}dx^{\mu}
\label{HS-frame}
\end{equation}
which is totally symmetric and traceless in the fiber indices 
\begin{equation}
\eta_{ab} e^{aba_3 \cdots a_{s-1}} = 0\,.
\label{viel-tr}
\end{equation} 
The Fronsdal field is then identified with the ``symmetric part" of the HS vielbein
\begin{equation}
\varphi_{\mu_1\cdots \mu_s} = e_{(\mu_1,\mu_2\cdots \mu_s)}
\label{vphi}
\end{equation}
where it is understood that the fiber indices are lowered with the background vielbein $e_{(0)\, \mu}^a$. Note that, as a consequence of (\ref{viel-tr}), 
the totally symmetric field (\ref{vphi}) is automatically double traceless, as required in Frosdal approach. Clearly the frame-like HS field 
(\ref{HS-frame}) has 
more components than the Fronsdal field. In terms of representations of $SO(D)$, the tensor 
$e^{a_1\cdots a_{s-1}}_{\mu}$ decomposes according to the tensor product\footnote{We use the notation $[n_1,n_2,\ldots]$ to denote a Young tableaux 
with $n_1$ boxes in the first row, $n_2$ in the second, etc., and n}
\begin{equation}
[1,0,\ldots] \otimes [s-1,0,\ldots] = [s,0,\ldots]+[s-2,0,\ldots]+[s-1,1,0,\ldots]\,.
\end{equation}
The first two representations on the right-hand side make up a totally symmetric, double-traceless field. The third, hook-type, representation corresponds 
to gauge redundancies analog to local Lorentz in the $s=2$ case, and so one needs a corresponding gauge field $\omega_{\mu}^{a_1\cdots a_{s-1},b}$ 
which generalizes the spin connection. A novel feature compared to the $s=2$ case is that the analog of the torsion constraint does not 
determine $\omega_{\mu}^{a_1\cdots a_{s-1},b}$ uniquely, it has its own gauge redundancy which requires a new HS spin connection 
$\omega_{\mu}^{a_1\cdots a_{s-1},bc}$, and so on. The final result is that the frame-like formulation of HS fields requires the vielbein-like 
field (\ref{HS-frame}) plus a tower of spin connection-like fields \cite{Lopatin:1987hz}
\begin{equation}
\begin{aligned}
&e^{a_1\ldots a_s}_{\mu}\\
&\omega_{\mu}^{a_1\ldots a_s, b_1\ldots b_t}\,,\quad t=1,2,\ldots, s-1
\label{HS-fr-all}
\end{aligned}
\end{equation}
and there is a corresponding tower of torsion-like constraints that allow to solve for the HS ``spin connections" in terms of the HS vielbein
\begin{equation}
\begin{aligned}
&R^{a_1\cdots a_{s-1}}=0\\
&R^{a_1\cdots a_{s-1},b}=0\\
&\vdots \\
&R^{a_1\cdots a_{s-1},b_1\cdots b_{s-1}} = e_{(0)c}\wedge e_{(0)d} C^{a_1\cdots a_{s-1} c, b_1 \cdots b_{s-1} d}
\end{aligned}
\label{curv-HS}
\end{equation}
where $R^{a_1\cdots a_{s-1},b_1\cdots b_t}, t=0,\ldots , s-1$ are curvature two-forms that generalize (\ref{Cartan}), and in the last line the 0-form 
$C^{a_1\cdots a_{s-1} c, b_1 \cdots b_{s-1} d}$ is the HS generalization of the Weyl tensor: it transforms in the representation $[s,s,0,\ldots]$ and is 
built out of $s$-derivatives of the Fronsdal field. With some work (see for instance \cite{Didenko:2014dwa}, to which we refer the reader 
for a more detailed review of the above construction), one can show that these equations are equivalent to the Fronsdal equations (\ref{Fronsd-AdS2}) in 
terms of the symmetric field (\ref{vphi}).                                                                                                                                                                  

Recall that in the $s=2$ case one can combine $e^a,\omega^{ab}$ into a single gauge field $\omega^{AB}$, $A,B=0,\ldots, D$ in the adjoint 
of $SO(D-1,2)$ (from now on we focus on the AdS case). It is a group theory exercise to show that one may analogously combine the tower of 
fields (\ref{HS-fr-all}) into a single irreducible representation of $SO(D-1,2)$
\begin{equation}
\left \{ e^{a_1\cdots a_{s-1}},\omega^{a_1\cdots a_{s-1},b_1\cdots b_t}|_{t=1,\ldots s-1}  \right \}\rightarrow 
\omega^{A_1\cdots A_{s-1},B_1 \cdots B_{s-1}}\,,
\end{equation}
which is in the representation $[s-1,s-1,0,\ldots]$ of the AdS algebra, or equivalently of the conformal algebra. Each gauge field 
is then associated to a generator $T_{A_1\cdots A_{s-1},B_1\cdots B_{s-1}}$ in the same representation. Recall from Section \ref{HS-free-scalar} 
that this is precisely the representation appearing in the discussion of the HS algebra on the CFT side, where we reviewed the construction 
of the HS algebra generators in terms of conformal Killing tensors. The  $T_{A_1\cdots A_{s-1},B_1\cdots B_{s-1}}$ appearing here are the 
generators of precisely the same HS algebra, which becomes a gauge algebra from AdS point of view. It is then natural to combine all HS fields 
into a single one-form
\begin{equation}
W = -i \sum_s \omega^{A_1\cdots A_{s-1},B_1 \cdots B_{s-1}}\, T_{A_1\cdots A_{s-1},B_1\cdots B_{s-1}}\,.
\end{equation}
Expanding around the AdS background to linear order in the fluctuations $\omega_1$:
\begin{equation}
W = -i \left(\omega_0^{AB}T_{AB}+\sum_s \omega^{A_1\cdots A_{s-1},B_1 \cdots B_{s-1}}_1 \, T_{A_1\cdots A_{s-1},B_1\cdots B_{s-1}} \right)\,,
\end{equation}
and the linearized curvatures appearing in (\ref{curv-HS}) may be obtained according to the usual gauge theory rules 
\begin{equation}
R_1 = \left(dW+W\wedge W\right)_{\rm linearized} = d\omega_1+[\omega_0, \omega_1]\,,
\end{equation}
where the commutators $[T_{(2)}, T_{(s)}]$ needed in this expression are simply fixed by the $SO(D-1,2)$ algebra; the 
linearized gauge transformations are similarly given by
\begin{equation}
\delta \omega_1 = d\epsilon+[\omega_0,\epsilon]
\end{equation}
where $\omega_0 = \omega_0^{AB}T_{AB}$, $\omega_1 = \sum_s 
\omega_1^{A_1\cdots A_{s-1},B_1 \cdots B_{s-1}} \, T_{A_1\cdots A_{s-1},B_1\cdots B_{s-1}}$, and the gauge parameter is 
$\epsilon = \sum_s \epsilon^{A_1\cdots A_{s-1},B_1 \cdots B_{s-1}} T_{A_1\cdots A_{s-1},B_1 \cdots B_{s-1}}$. Note 
that at linearized level the gauge transformations do not mix fields of different spins. 

\section{Vasiliev equations in AdS$_4$}
\label{Vas-Equations}

The frame-like approach described above at linearized level is a natural starting point for the non-linear generalization 
(for comprehensive reviews on Vasiliev HS gauge theory see e.g. \cite{Vasiliev:1995dn, Vasiliev:1999ba, Bekaert:2005vh, Iazeolla:2008bp, Didenko:2014dwa}). One combines 
all HS gauge fields in the one-form $W=-i \omega_{(s)}T_{(s)}$, where $T_{(s)}$ are the generators of the HS algebra, and the non-linear 
gauge symmetry is given by
\begin{equation}
\delta W = d\epsilon + [W, \epsilon]\,.
\end{equation}
The goal is to write down fully non-linear equations invariant under this gauge symmetry and reproducing at linearized level 
the free HS equations described in the previous sections. The first step in Vasiliev approach is to find a convenient realization 
of the HS algebra generators $T_{(s)}$. This is done by introducing an auxiliary internal space endowed with a non-commutative star product. 
While the construction can be carried out in any $D$, it greatly simplifies in $D=4$, where one can use the spinorial formalism. 
We briefly review this in the next section, and then move on to write down the Vasiliev equations. 

\subsection{Tensor-spinor dictionary}

It is well known that in $D=4$ a vector index can be replaced by a bispinor index. This is essentially the isomorphism
$SO(4)=SU(2)\times SU(2)$. The explicit map can be carried out by using the matrices
\begin{equation}
(\sigma^a)_{\alpha\dot\alpha} = \left(\mathbf{1}, \vec{\sigma}\right)_{\alpha\dot\alpha}\,,\qquad 
a=0,\ldots, 3\,,\quad \alpha, \dot\alpha = 1,2\,,
\end{equation}
where $\vec{\sigma}$ are the usual Pauli matrices. Using this, we can associate to a vector $v^a$ a bispinor $v_{\alpha \dot\alpha}$ as
\begin{equation}
v^a \rightarrow v_{\alpha \dot\alpha} = v^a (\sigma_a)_{\alpha\dot\alpha}\,.
\label{vec-spinor}
\end{equation}
In group theory language, this is the correspondence between the fundamental representation $[1,0]$ of $SO(4)$ and the representation 
$\left(\frac{1}{2},\frac{1}{2}\right)$ 
of $SU(2)\times SU(2)$. \footnote{We use the notation $(s_1,s_2)$ for the representation of $SU(2)\times SU(2)$ of dimension $(2s_1+1)(2s_2+1)$, 
corresponding to the multispinor $v^{\alpha_1\cdots \alpha_{2 s_1}\dot\alpha_1\cdots \dot\alpha_{2 s_2}}$ which is totally symmetric in the 
dotted and undotted indices.} Spinor indices may be raised and lowered using the $\epsilon$-tensor
\begin{equation}
\epsilon_{\alpha\beta} = \begin{pmatrix}0&1\\-1&0\end{pmatrix}\,,\qquad 
\epsilon^{\alpha\beta} = \begin{pmatrix}0&1\\-1&0\end{pmatrix}
\end{equation}
with the conventions
\begin{equation}
v^{\alpha} = \epsilon^{\alpha\beta}v_{\beta}\,,\qquad 
v_{\alpha} = v^{\beta} \epsilon_{\beta\alpha}
\end{equation}
and similarly for dotted indices. We may also define the matrix
\begin{equation}
\left(\bar\sigma^a\right)^{\dot\alpha\alpha}= \epsilon^{\alpha\beta}\epsilon^{\dot\alpha\dot\beta}\left(\sigma_a\right)_{\beta\dot\beta}\,.
\end{equation}
We will often drop the `bar', since it is understood that the two matrices are related by raising and lowering indices with $\epsilon$'s.  
Useful relations include 
\begin{equation}
\begin{aligned}
&\left(\sigma^a\right)_{\alpha\dot\alpha} \left(\sigma_a\right)^{\dot\beta \beta} = -2 \delta^{\beta}_{\alpha} \delta^{\dot\beta}_{\dot \alpha} \\
&\left(\sigma^a\right)_{\alpha}^{\ \dot\alpha} \left(\sigma^b\right)_{\dot\alpha}^{\  \beta} 
=\eta^{ab}\delta_{\alpha}^{\beta}+\left(\sigma^{ab}\right)_{\alpha}^{\ \beta}\\
&\left(\sigma^a\right)_{\dot\alpha}^{\ \alpha} \left(\sigma^b\right)_{\alpha}^{\  \dot\beta} 
=\eta^{ab}\delta_{\dot \alpha}^{\dot \beta}+\left(\bar \sigma^{ab}\right)_{\dot\alpha}^{\ \dot\beta}
\end{aligned}
\end{equation}
where
\begin{equation}
\left(\sigma^{ab}\right)_{\alpha}^{\ \beta} = \frac{1}{2}\left(
(\sigma^a)_{\alpha}^{\ \dot\alpha} (\sigma^b)_{\dot \alpha}^{\ \beta} 
- (\sigma^b)_{\alpha}^{\ \dot\alpha} (\sigma^a)_{\dot \alpha}^{\ \beta}
\right)
\end{equation}
and similarly for $\bar\sigma^{ab}$. 

Analogously to (\ref{vec-spinor}), one may show that an antisymmetric tensor $F_{ab}$ in the $[1,1]$ of $SO(4)$ corresponds to 
the representation $(1,0)\oplus (0,1)$ of $SU(2)\times SU(2)$
\begin{equation}
F_{ab} \quad \rightarrow \quad  F_{\alpha\beta}\,,\quad \bar F_{\dot\alpha\dot\beta}
\end{equation}
which we can see explicitly from 
\begin{equation}
\begin{aligned}
&F^{ab} (\sigma_a)_{\alpha\dot\alpha} (\sigma_b)_{\beta\dot\beta} = \epsilon_{\dot\alpha\dot\beta} 
F_{\alpha\beta}+\epsilon_{\alpha\beta} F_{\dot\alpha\dot\beta}\\
&F_{\alpha\beta} =-\frac{1}{2} F_{ab}(\sigma^{ab})_{\alpha\beta}\,,\qquad 
\bar F_{\dot\alpha\dot\beta} = -\frac{1}{2} F_{ab}(\bar\sigma^{ab})_{\dot\alpha\dot\beta}\,.
\end{aligned}
\end{equation}
Note that $F_{\alpha\beta}$ and $\bar F_{\dot\alpha\dot\beta}$ are related to the selfdual and anti-selfdual parts of $F_{ab}$. In particular, 
the $s=2$ vielbein and spin connection read, in spinorial language
\begin{equation}
e^a, \omega^{ab} \quad \rightarrow \quad e^{\alpha\dot\alpha}\,,\quad \omega^{\alpha\beta}\,,\quad \bar\omega^{\dot\alpha \dot\beta}\,.
\end{equation}

The general map between $SO(4)$ and $SU(2)\times SU(2)$ representations read
\begin{equation}
[k,l] \quad \rightarrow \quad \left(\frac{k+l}{2},\frac{k-l}{2}\right)\oplus \left(\frac{k-l}{2},\frac{k+l}{2}\right)
\label{4-to-2-map}
\end{equation}
or, explicitly in terms of tensors
\begin{equation}
v^{a_1\cdots a_k, b_1\cdots b_l}\quad \rightarrow \quad v^{\alpha_1\cdots \alpha_{k+l},\dot\alpha_1\cdots \dot \alpha_{k-l}}
\,,\quad \bar v^{\alpha_1\cdots \alpha_{k-l},\dot\alpha_1\cdots \dot \alpha_{k+l}}\,.
\end{equation}
Therefore, we find that the tower of fields (\ref{HS-fr-all}) needed in the frame-like approach to HS fields corresponds to 
the collection of multispinors
\begin{equation}
\omega^{\alpha_1 \cdots \alpha_{s-1+n},\dot\alpha_1\cdots \dot\alpha_{s-1-n}}\,,\qquad n = -(s-1),\ldots, (s-1)\,.
\label{HS-spinorial}
\end{equation}
In particular, the spin $s$ vielbein field $e^{a_1\cdots a_{s-1}}$ corresponds to the multispinor 
$e^{\alpha_1 \cdots \alpha_{s-1},\dot \alpha_1 \cdots \dot\alpha_{s-1}}$, i.e. $n=0$ above, and all other fields are spin connection-like. 

Let us finally note that, according to (\ref{4-to-2-map}), the HS Weyl tensor $C_{a_1\cdots a_s,b_1\cdots b_s}$ in the $[s,s]$ of $SO(4)$ corresponds to the totally symmetric 
multispinors $C_{\alpha_1\cdots \alpha_{2s}}$ and $\bar C_{\dot\alpha_1\cdots \dot\alpha_{2s}}$ in the $(s,0)\oplus (0,s)$. For instance, 
in the $s=2$ case, the selfdual and antiselfdual parts of the Weyl tensor $C_{abcd}$ correspond to the multispinors $C_{\alpha\beta\gamma\delta}$ 
and $\bar C_{\dot\alpha\dot\beta\dot\gamma\dot\delta}$. 

\subsection{Twistor variables and star-product realization of HS algebra}

As anticipated above, in Vasiliev construction the HS algebra generators are realized in terms of an auxiliary space with a non-commutative 
star product. We introduce the twistor-like variables 
\begin{equation}
y_{\alpha}\,,\quad \bar y_{\dot\alpha}
\end{equation}
in the spinor representations $(1/2,0)$ and $(0,1/2)$. These variables are {\it bosonic}, 
in particular $y_{\alpha}y^{\alpha} = \epsilon^{\alpha\beta}y_{\alpha}y_{\beta} = 0$. 
We then introduce the star-product rule
\begin{equation}
f(y,\bar y) * g(y,\bar y) = f(y,\bar y) 
e^{\epsilon^{\alpha\beta}\frac{\overleftarrow{\partial}}{\partial y^{\alpha}}\frac{\overrightarrow{\partial}}{\partial y^{\beta}}
+\epsilon^{\dot\alpha\dot\beta}\frac{\overleftarrow{\partial}}{\partial \bar y^{\dot \alpha}}
\frac{\overrightarrow{\partial}}{\partial \bar y^{\dot \beta}}}
g(y,\bar y)\,.
\end{equation}
In particular, for the monomials of $y,\bar y$, we have
\begin{equation}
\begin{aligned}
&y_{\alpha}*y_{\beta} = y_{\alpha} y_{\beta}+\epsilon_{\alpha\beta}\\
&\bar y_{\dot \alpha}*\bar y_{\dot \beta} = \bar y_{\dot \alpha} \bar y_{\dot \beta}+\epsilon_{\dot \alpha\dot \beta}\\
& y_{\alpha}*\bar y_{\dot \alpha} = y_{\alpha} \bar y_{\dot \alpha} \,.
\label{star-mono}
\end{aligned}
\end{equation}
Note that we use here the conventions of \cite{Didenko:2009td, Giombi:2009wh, Giombi:2012ms}, where no factors of $i$ appear in front of $\epsilon$'s. 
The following integral representation of the star product is often useful
\begin{equation}
f*g = \int d^2 u d^2 v d^2\bar u d^2 \bar v\, e^{uv+\bar u \bar v} f(y+u,\bar y+\bar u)g(y+v,\bar y+\bar v)
\end{equation}
where $uv \equiv u^{\alpha}v_{\alpha}$, $\bar u \bar v=\bar u^{\dot\alpha}\bar v_{\dot\alpha}$ and a choice of integration contour and measure 
normalization is assumed so that $1*f=f*1=f$. For instance we assume
\begin{equation}
\int d^2 u e^{uv} = \delta^2(v)
\end{equation}
and, in particular, the elementary star products (\ref{star-mono}) are reproduced. 
Useful identities that follow from the definition of the star product include 
\begin{equation}
\begin{aligned}
&[y_{\alpha}, f]_* = 2 \frac{\partial f}{\partial y^{\alpha}} \,,\qquad 
[y^{\alpha}, f]_{*} = 2 \epsilon^{\alpha\beta}\frac{\partial f}{\partial y^{\beta}} \\
&[y_{\alpha}y_{\beta}, f]_* = 2 y_{\alpha}\frac{\partial f}{\partial y^{\beta}}+ 2 y_{\beta}\frac{\partial f}{\partial y^{\alpha}}\\
&[y_{\alpha} \bar y_{\dot\beta}, f]_* = 2 y_{\alpha}\frac{\partial f}{\partial \bar y^{\dot \beta}}+ 
2 \bar y_{\dot \beta}\frac{\partial f}{\partial y^{\alpha}}\\
&\{y_{\alpha} \bar y_{\dot \alpha} , f\}_{*} = 2 y_{\alpha} \bar y_{\dot \alpha} f
+2 \frac{\partial^2 f}{\partial y^{\alpha} \partial \bar y^{\dot \alpha}}
\label{star-ident}
\end{aligned}
\end{equation}
where $[A,B]_* = A*B-B*A$ and $\{A,B\}_* = A*B+B*A$, and $f=f(y,\bar y)$ is an arbitrary function of the twistor variables.  

Now one can see that the bilinears in $y,\bar y$ 
\begin{equation}
P_{\alpha\dot\alpha} = y_{\alpha}\bar y_{\dot\alpha} \,,\quad 
M_{\alpha\beta} = y_{\alpha} y_{\beta} \,, \quad 
\bar M_{\dot\alpha \dot \beta} = \bar y_{\dot\alpha} \bar y_{\dot \beta}
\end{equation}
satisfy an algebra under star commutators that (up to normalization) is precisely the $SO(3,2)$ algebra $[M,M]_* \sim M$, 
$[M,P]_* \sim P$, $[P,P]_* \sim M$, with $P_{\alpha\dot\alpha}$ corresponding to the local translations $P_a$ and $M_{\alpha\beta}$, 
$\bar M_{\dot \alpha\dot \beta}$ the selfdual and anti-selfdual parts of the Lorentz generators $M_{ab}$. 
Note that we have implicitly set $\ell^2_{\rm AdS}=1$ in the definition of the star product.\footnote{We can reintroduce the AdS scale by 
taking $y, \bar y$ to have dimension $1/2$, with $y_{\alpha} *y_{\beta} = y_{\alpha}y_{\beta}+\frac{1}{\ell}\epsilon_{\alpha\beta}$ 
and similarly for $\bar y$. Then $P \sim y \bar y$ and $M \sim \ell y y$. } 

Hence, we can realize the spin 2 gauge field 
as a function of homogeneous degree two in $y, \bar y$
\begin{equation}
W_{s=2} = e_{\alpha\dot\alpha}(x) y^{\alpha}\bar y^{\dot \alpha} 
+\omega_{\alpha\beta}(x)y^{\alpha}y^{\beta}
+\bar\omega_{\dot\alpha\dot\beta}(x)\bar y^{\dot\alpha} \bar y^{\dot \beta} \,.
\end{equation}
The generalization to higher spins is straightforward. We simply take a one-form $W(x|y,\bar y)=dx^{\mu}W_{\mu}(x|y,\bar y)$ which is a general function of the auxiliary 
twistor variables
\begin{equation}
W(x|y,\bar y) = \sum_{n,m=0}^{\infty} \frac{1}{n!m!} W_{\alpha_1 \cdots \alpha_n, \dot \alpha_1 \cdots \dot \alpha_m}^{(n,m)}(x)
y^{\alpha_1}\cdots y^{\alpha_n} \bar y^{\dot \alpha_1}\cdots \bar y^{\dot \alpha_m}\,. 
\label{W-y}
\end{equation}
The Taylor expansion of the ``master field" above encodes the physical HS gauge fields of all spins. 
From (\ref{HS-spinorial}), we see that the spin $s$ gauge fields are given by 
\begin{equation}
{\rm spin-}s:\quad W^{(s-1+n,s-1-n)}_{\alpha_1 \cdots \alpha_{s-1+n}, \dot \alpha_1 \cdots \dot \alpha_{s-1-n}}\,,\quad -(s-1)\le n \le s-1
\label{s-W}
\end{equation}
i.e., by the terms in (\ref{W-y}) of total degree $2(s-1)$ in $y, \bar y$. Note that the $(y,\bar y)$-independent part of $W$ is a $s=1$ field 
(keep in mind that each component in the expansion (\ref{W-y}) is a one-form). The Fronsdal field is related to the HS vielbein which is 
the $W^{(s-1,s-1)}$ component
\begin{equation}
W_{\mu}^{\alpha_1 \cdots \alpha_{s-1}, \dot \alpha_1 \cdots \dot \alpha_{s-1}}\quad \rightarrow \quad 
e_{\mu}^{a_1\cdots a_{s-1}}\quad \rightarrow \quad \varphi_{\mu_1\mu_2\cdots \mu_s} = e_{(\mu_1,\mu_2\cdots \mu_s)}\,.
\end{equation}
Note that the expansion (\ref{W-y}) includes, as written, fields of both integer and half-integer spins (for example, 
$W_{\mu, \alpha}^{(1,0)}$ and $W_{\mu , \dot\alpha}^{(0,1)}$ make up a $s=3/2$ field). One may truncate to the integer spins only by 
imposing the constraint that $W$ is even in the auxiliary variables, $W(x|y,\bar y)=W(x|-y,-\bar y)$. 

The set of monomials in $y,\bar y$ of total degree $2(s-1)$, corresponding to the gauge fields (\ref{s-W}), are then interpreted as the 
generators of the HS algebra
\begin{equation}
T_{s}(y,\bar y) = \{ y^{\alpha_1}\cdots y^{\alpha_n} \bar y^{\dot \alpha_1}\cdots \bar y^{\dot \alpha_m}\}|_{n+m=2(s-1)}
\label{Ts}
\end{equation}
which may be packaged into the $SO(3,2)$ representation $T_{A_1\cdots A_{s-1},B_1\cdots B_{s-1}}$ as explained in the previous section. The 
HS algebra is then the algebra of the generators (\ref{Ts}) under star-commutators, with the star product defined above. It has the structure, 
schematically
\begin{equation}
[T_{s_1}, T_{s_2}]_* = \sum_{s=|s_1-s_2|+2}^{s_1+s_2-2} T_s \,.
\label{TT-com}
\end{equation}
For instance $[T_2,T_2]_* \sim T_2$; $[T_3,T_3]_* \sim T_4+T_2$;  $[T_4, T_4]_* \sim T_6+T_4+T_2$ etc. The $s\le 2$ generators form the only 
finite subalgebra $SO(3,2)\oplus U(1)$. As already reviewed in Section \ref{HS-free-scalar}, commutators of generators with $s>2$ produce 
generators of higher spin, and hence the whole tower must be present. The bosonic HS algebra generated by all the integer spins $T_s$, $s\ge 1$ 
is denoted $hu(1,0|4)$ \cite{Vasiliev:1999ba}. Note that the structure of the commutators (\ref{TT-com}) indicates that a consistent 
subalgebra may be obtained by projecting out the odd spins. The corresponding HS algebra is denoted $ho(1,0|4)$ or sometimes $hs(4)$. 

While the one-form $W(x|y,\bar y)$ encodes all gauge fields with $s\ge 1$, we know at least from a CFT perspective that the HS theory should 
include a scalar field. This suggests that we need an additional master zero-form 
\begin{equation}
B(x|y,\bar y) = \sum_{n,m=0}^{\infty} \frac{1}{n!m!} B_{\alpha_1\cdots \alpha_n \dot\alpha_1\cdots \dot\alpha_m}^{(n,m)}
y^{\alpha_1}\cdots y^{\alpha_n}\bar y^{\dot \alpha_1}\cdots \bar y^{\dot \alpha_m}\,.
\end{equation}
Indeed, this is correct, and it turns out by analyzing the Vasiliev equations reviewed in the next section that $B^{(0,0)}$ is the bulk scalar field, 
and the tower of fields $B^{(n,n)}$, $n>0$ is auxiliary and expressed in terms of $B^{(0,0)}$ by the equations of motion. The fields $B^{(2s,0)}$ 
and $B^{(0,2s)}$ are related to the selfdual and anti-selfdual parts of the HS Weyl tensors, and $B^{(2s+n,n)}$, $B^{(n,2s+n)}$ with $n>0$ are related 
to the bottom components $n=0$ by the equations of motion. 

\subsection{Non-linear equations}

The master fields $W(x|y,\bar y)$ and $B(x|y,\bar y)$ introduced in the previous section are sufficient to encode all physical fields and to write 
down the free equations of motion. However, to construct the fully non-linear theory, a technical trick which turns out to be useful is to {\it double} 
the auxiliary twistor space by introducing a new set of variables $z, \bar z$
\begin{equation}
(y_{\alpha},\bar y_{\dot\alpha})\quad \rightarrow \quad 
(y_{\alpha},\bar y_{\dot\alpha}, z_{\alpha},\bar z_{\dot\alpha})\,.
\end{equation}
The new variables are purely auxiliary, and the physical fields will be contained in the $(z,\bar z)$-independent part of the master fields introduced 
below. On the doubled twistor space, we then introduce the extended star product
\ie
&f(Y,Z)* g(Y,Z) = f(Y,Z) \exp\left[ \epsilon^{\A\B} \left(\overleftarrow\partial_{y^\A} + \overleftarrow\partial_{z^\A}\right) \left(\overrightarrow\partial_{y^\B} - \overrightarrow\partial_{z^\B}\right) \right.
\\
&\left.~~~~~~+ \epsilon^{\da\db} \left(\overleftarrow\partial_{\bar y^\da} + \overleftarrow\partial_{\bar z^\da}\right) \left(\overrightarrow\partial_{\bar y^\db} - \overrightarrow\partial_{\bar z^\db}\right) \right] g(Y,Z).
\fe
where we have introduced the shorthand notation $(Y,Z) = (y_{\alpha},\bar y_{\dot\alpha}, z_{\alpha},\bar z_{\dot\alpha})$. In 
addition to (\ref{star-mono}), the definition above implies the elementary star products 
\begin{equation}
\begin{aligned}
&z^{\alpha}*z^{\beta}=z^{\alpha}z^{\beta}-\epsilon^{\alpha\beta} \\
&y^{\alpha}*z^{\beta}=y^{\alpha}z^{\beta}-\epsilon^{\alpha\beta} \\
&z^{\alpha}*y^{\beta} = z^{\alpha}y^{\beta}+\epsilon^{\alpha\beta}
\label{star-mono-yz}
\end{aligned}
\end{equation}
and similarly for dotted spinors. The star commutators are then given by
\ie{}
[y^\A , y^\B]_* = 2 \epsilon^{\A\B},~~~ [z^\A , z^\B]_* = - 2\epsilon^{\A\B},~~~ [y^\A , z^\B]_* = 0\,.
\fe
It is important to note that while $y$ and $z$ star-commute, their star product (\ref{star-mono-yz}) is non-trivial.  More 
generally, one finds the identities
\begin{equation}
\begin{aligned}
&z^{\alpha}* f(y,\bar y, z,\bar z) = z^{\alpha} f+\epsilon^{\alpha\beta}
\left(\frac{\partial f}{\partial y^{\beta}}-\frac{\partial f}{\partial z^{\beta}}\right)\,,\\
&[z^{\alpha}, f(y,\bar y, z,\bar z)]_* = -2 \epsilon^{\alpha\beta} \frac{\partial f}{\partial z^{\beta}}\,.
\end{aligned}
\end{equation}
One has the following integral representation of the extended star product 
\ie
&f(Y, Z) * g(Y,Z) \\
&= \int d^4U d^4V \, e^{ uv+\bar u\bar v }
f(y+u,\bar y + \bar u,z+u,\bar z+\bar u) g(y+v,\bar y+\bar v,z-v,\bar z-\bar v)
\fe
where as explained in the previous section one assumes an integration contour and measure normalization such that $f*1=1*f=f$, 
and we have used the shorthand $d^4U=d^2u d^2\bar u$ and similarly for $v,\bar v$.   

An important property of the extended star product is that it admits the ``Kleinian operators"
\ie
\kappa =e^{z^\alpha y_\alpha} ,~~~~ \bar\kappa = e^{\bar z^\da \bar y_\da}
\label{Kleinians}
\fe
which satisfy, as the reader may check as an exercise:
\begin{equation}
\kappa * \kappa = 1\,,\qquad \bar \kappa *\bar \kappa =1 \,.
\end{equation}
Given an arbitrary function $f(y,z)$, one may also derive the following results
\begin{equation}
\begin{aligned}
&f(y,\bar y, z,\bar z) * \kappa =f(-z,\bar y, -y,\bar z)\kappa\,, \\
&\kappa *f(y,\bar y,z, \bar z)=\kappa  f(z,\bar y , y,\bar z).
\end{aligned}
\end{equation}
and similarly for the star products with $\bar\kappa$. In particular, it follows that 
\begin{equation}
\begin{aligned}
&\kappa * f(y,\bar y,z,\bar z)*\kappa = f(-y,\bar y, -z,\bar z)\\
&\bar \kappa * f(y,\bar y,z,\bar z)*\bar\kappa = f(y,-\bar y,z,-\bar z)\,.
\end{aligned}
\end{equation}

We now introduce the full set of master fields required to write down Vasiliev equations. They are:
\begin{enumerate}
\item A master one-form $W(x|y,\bar y,z,\bar z)=dx^{\mu}W_{\mu}(x|y,\bar y,z,\bar z)$. This is the same one-form introduced in the previous section, 
except it now depends on the enlarged twistor space variables. As explained above, $W$ is to be viewed as a gauge field of the HS algebra. The gauge 
transformation is
\begin{equation}
\delta W = d_x \epsilon+[W,\epsilon]_*
\end{equation} 
where the gauge parameter $\epsilon=\epsilon(y,\bar y, z,\bar z)$ is a function of all twistor variables. 
\item A zero-form $B(x|y,\bar y,z,\bar z)$, which is postulated to transform in the ``twisted adjoint" representation, with the gauge transformation
\begin{equation}
\begin{aligned}
&\delta B = -\epsilon * B+B*\pi(\epsilon)\\
&\pi(\epsilon)\equiv \epsilon(x|-y,\bar y,-z,\bar z) = \kappa * \epsilon * \kappa \,.
\label{del-B}
\end{aligned}
\end{equation}
Note that we can map the twisted adjoint to the ordinary adjoint by star multiplication with the Kleinian. Indeed, 
from (\ref{del-B}) it follows that $\Phi = B*\kappa$ transforms in the adjoint  
\begin{equation}
\delta \Phi  =-\epsilon * \Phi+\Phi * \epsilon \,.
\end{equation}
\item An additional gauge field which is a one-form in the $(z,\bar z)$-direction
\begin{equation}
S(x|y,\bar y, z,\bar z)=dz^{\alpha} S_{\alpha}(x|y,\bar y, z,\bar z)+
d\bar z^{\dot \alpha} \bar S_{\dot \alpha}(x|y,\bar y, z,\bar z)
\end{equation}
with the gauge transformation 
\begin{equation}
\delta S = d_Z \epsilon+[S,\epsilon]_{*}\,,
\label{del-S}
\end{equation}
where $d_Z=dz^{\alpha}\frac{\partial}{\partial z^{\alpha}}+d\bar z^{\dot\alpha}\frac{\partial}{\partial \bar z^{\dot \alpha}}$. 
\end{enumerate}
The physical degrees of freedom are contained in $W$ and $B$ projected to the $z=\bar z=0$ subspace, and $S$ is purely auxiliary.\footnote{The original Vasiliev's system of \cite{Vasiliev:1992av} actually contains a doubled set of master fields compared to the ones listed above. We will not discuss explicitly this model here.} 
The following constraints are imposed on the master fields
\begin{equation}
\begin{aligned}
&W(Y,Z)=W(-Y,-Z)\,,\qquad B(Y,Z)=B(-Y,-Z)\\
&S_{\alpha}(Y,Z) = -S_{\alpha}(-Y,-Z)\,,\quad \bar S_{\dot\alpha}(Y,Z)=-\bar S_{\dot\alpha}(-Y,-Z) 
\label{const-bose}
\end{aligned}
\end{equation}
and similarly on the gauge parameter $\epsilon(Y,Z)=\epsilon(-Y,-Z)$. 
These constraints are important for consistency of the non-linear equations, and they imply that the physical spectrum includes only integer spins. 

We are now ready to write down the fully non-linear equations of motion. To write them in the most compact form, it is convenient to define the one-form 
in $(x,Z)$ space
\begin{equation}
{\cal A}(x|Y,Z) = dx^{\mu}W_{\mu}(x|Y,Z)+dz^{\alpha} S_{\alpha}(x|Y,Z)+
d\bar z^{\dot \alpha} \bar S_{\dot \alpha}(x|Y,Z)
\label{master-A}
\end{equation}
which transforms under the gauge symmetry as 
\begin{equation}
\delta {\cal A} = d\epsilon+[A,\epsilon]_{*}\,,\qquad d\equiv  d_x+d_Z\,.
\label{del-A}
\end{equation}
The Vasiliev non-linear equations are then given by
\begin{equation}
\begin{aligned}
&d{\cal A}+{\cal A}*{\cal A} = f_*(B*\kappa)dz^2+\bar f_*(B*\bar\kappa)d\bar z^2 \\
&d B+{\cal A}*B-B*\pi({\cal A})=0\,.
\label{VasEq}
\end{aligned}
\end{equation}
Here $\pi$ is defined in general as the operation that flips the signs of $(y,z,dz)$ while preserving the signs of $(\bar y,\bar z,d\bar z)$. If no $z$-differential 
is involved, this is equivalent to conjugation by $\kappa$, as in (\ref{del-B}). Similarly one may define the conjugate operation $\bar\pi$ that flips the sign of $(\bar y,\bar z,d\bar z)$. Note, however, that due to the constraints (\ref{const-bose}), $\pi(\cal A)=\bar\pi(\cal A)$ and $\pi(B)=\bar\pi(B)$. In 
(\ref{VasEq}), $f(X)$ is an analytic function of $X$, and $\bar f$ its complex conjugate. $f_*(X)$ is the corresponding star-function obtained by replacing all products 
of $X$ in the Taylor series of $f(X)$ by star-products.  The function $f(X)$ reflects some freedom in the interactions allowed by the HS gauge symmetry, as will be discussed below. 

One can see that the Vasiliev equations (\ref{VasEq}) are manifestly gauge invariant under (\ref{del-A}), (\ref{del-B}). Note that the right-hand side of the first equation in (\ref{VasEq}) transforms in the adjoint, as appropriate. The consistency of these equations depends crucially on the fact that we have only two $z^\A$'s and two $\bar z^\da$'s (so that there is no holomorphic 3-form in $z$), and the truncation condition (\ref{const-bose}). For instance, in showing that the second equation is consistent with Bianchi identities and the first equation, one has to use $\pi(W)=\bar\pi(W)$, $\pi(S_{\alpha})=-\bar\pi(S_{\alpha})$, etc. 

Let us also note that one can impose a consistent truncation of (\ref{VasEq}) by imposing, in addition to (\ref{const-bose}), the constraints 
\ie
& W(x,iy,i\bar y,-iz,-i\bar z) = -W(x,y,\bar y,z,\bar z)\,, \\
& S (x,iy,i\bar y,-iz,-i\bar z,-idz,-id\bar z) = -S (x,y,\bar y,z,\bar z,dz,d\bar z)\,,\\ 
& B(x,iy,-i\bar y,-iz,i\bar z) = B(x,y,\bar y,z,\bar z)\,.
\label{mintrunc}
\fe
The resulting theory is the ``minimal bosonic" HS theory, and its spectrum involves only even spins. For instance, note that the constraint on $W$ in (\ref{mintrunc}) retains 
only the terms in (\ref{W-y}) with total degree in $(y,\bar y)$ equal to $4k-2$  ($k=1,2,\ldots$), corresponding to spin $s=2k$.\footnote{We refer the reader to \cite{Giombi:2012ms} for a proof that (\ref{mintrunc}) defines a truncation consistent with (\ref{VasEq}).}

Let us end this section by writing out the Vasiliev equations (\ref{W-y}) more explicitly in terms of the master fields $W,B,S$. They are
\begin{equation}
\begin{aligned}
&d_x W+W*W=0\\
&d_Z W+d_x S+W*S+S*W=0\\
&d_Z S+S*S=f_*(B*\kappa)dz^2+\bar f_*(B*\bar\kappa)d\bar z^2\\
&d_x B+W*B-B*\pi(W)=0\\
&d_Z B+S*B-B*\pi(S)=0\,.
\label{VasEq2}
\end{aligned}
\end{equation}
Note that it is possible to remove the $Z$-derivatives from the system (\ref{VasEq2}) or (\ref{VasEq}) by redefining $S =\hat S+\frac{1}{2}z_{\alpha}dz^{\alpha}
+\frac{1}{2}\bar z_{\dot\alpha} d\bar z^{\dot\alpha}$. Indeed, due to $[z_{\alpha},f]_* = -2\frac{\partial f}{\partial z^{\alpha}}$ (and similarly for $\bar z_{\dot\alpha}$), 
the terms involing $d_Z$ in (\ref{VasEq2}) and the gauge transformation (\ref{del-S}) can be absorbed into the star commutators with $\frac{1}{2}z_{\alpha}dz^{\alpha}
+\frac{1}{2}\bar z_{\dot\alpha} d\bar z^{\dot\alpha}$, and the resulting equations in terms of $W, \hat S, B$ do not involve the $Z$-differential. In what follows, we 
will always use the form (\ref{VasEq2}), or (\ref{VasEq}), of Vasiliev equations. 

\subsection{Type A, Type B and parity breaking theories}

It turns out that not all choices of the function $f(X)$ in (\ref{VasEq}) define physically distinct theories. It is possible to show that,
 by using the freedom of making field redefinitions consistent with the equations of motion, gauge transformations and reality conditions, one can put $f(X)$ in the form \cite{Vasiliev:1992av, Sezgin:2002ru}, see also \cite{Chang:2012kt, Giombi:2012ms}
\begin{equation}
f(X) = X \exp(i\theta(X))
\label{f-Th}
\end{equation}
where 
\begin{equation}
\theta(X)=\sum_{n=0}^\infty \theta_{2n} X^{2n}
\label{ThetaX}
\end{equation}
is a real even analytic function.  The function (\ref{f-Th}), or the phase $\theta(X)$, characterizes the interactions in the most general 4d Vasiliev theory. One 
can restrict $f(X)$ by imposing a parity symmetry. This acts on $(Y,Z)$ by $y_\A\leftrightarrow \bar y_\da$, $z_\A\leftrightarrow \bar z_\da$, and so exchanges the two terms $f(B*K)dz^2$ and $\bar f(B*\bar K)d\bar z^2$ in the equation of motion. We may assign the scalar $B$ to be either parity even or parity odd. If we choose $B$ to be parity 
even, then invariance under parity requires $f(X)=\bar f(X)$ and hence, from (\ref{f-Th}), $\Theta(X)=0$. On the other hand, if $B$ is chosen to be parity odd, then 
parity invariance requires $f(X) = \bar f(-X)$ and hence $\Theta(X)=\pi/2$. To summarize, we have two parity invariant theories 
\begin{equation}
\begin{aligned}
&B~{\rm parity~even}\,,\qquad f_A(X)=X\\
&B~{\rm parity~odd}\,,\qquad f_B(X)=iX\,.
\end{aligned}
\end{equation}
These define respectively the so-called ``type A" and ``type B" models, conjecturally dual to scalar and fermionic vector models as summarized in Table \ref{T1}. 

If $\Theta(X)$ is not identically equal to $0$ or $\pi/2$, we have the more general parity breaking HS theories. Note that at the level of the classical equations discussed 
so far, these theories are labelled by the infinite set of parameters $\theta_0,\theta_2,\theta_4,\ldots$. The 
conjectured duality (\ref{HS-vec-parity}) then would require a mapping between these phases and the 't Hooft coupling of the Chern-Simons 
vector models. We will come back to this in Section \ref{3pt-tests} below.

\subsection{AdS$_4$ background and linearized equations}

The Vasiliev equations (\ref{VasEq}) are formulated in a background independent way. To see how the AdS spacetime arises, we now show 
that the theory has a vacuum solution corresponding to AdS$_4$, and then derive the linearized equations that describe the propagation of 
the free HS gauge fields on the AdS background. 

The maximally symmetric, vacuum, solution is obtained by setting $B=0$, $S=0$ and taking $W$ to be quadratic in $y$, $\bar y$
\begin{equation}
\begin{aligned}
W_0(x|Y) &= e_0(x|Y) + \omega_0(x|Y)\,,\qquad B=0\,,\quad S=0 \\
&= (e_0)_{\A\db} y^\A\bar y^\db + (\omega_0)_{\A\B}y^\A y^\B + (\omega_0)_{\da\db} \bar y^\da \bar y^\db\,.
\label{AdS4-master}
\end{aligned}
\end{equation}
Here $e_0$ and $\omega_0$ are the vielbein and spin connection 1-forms of the background solution. In our 
conventions, the relation to the usual $e^a$, $\omega^{ab}$ in vector notations is 
\begin{equation}\label{trans}
(e_0)_{\alpha {\dot \beta}}= \frac{1}{4} e^a \sigma^a_{\alpha {\dot \beta}}\,,\quad 
(\omega_0)_{\alpha \beta}=\frac{1}{16} \omega^{ab} \sigma^{ab}_{\alpha \beta}\,,\quad 
(\omega_0)_{\da \db}=-\frac{1}{16} \omega^{ab} {\bar \sigma}^{ab}_{\da \db}.
\end{equation}
With $B=0$, $S=0$, the only nontrivial  component of Vasiliev equations (\ref{VasEq2}) is
\begin{equation}
d_x W_0+W_0*W_0=0\,.
\end{equation} 
Collecting the independent terms in $(y,\bar y)$, one finds the equations
\begin{equation}
\begin{split}
y^{\alpha} \bar{y}^{\dot{\alpha}} &:~~~~ d_x e_{\alpha\dot{\beta}} + 4 \omega_{\alpha}^{~\beta}\wedge e_{\beta\dot{\beta}} 
          - 4 e_{\alpha\dot{\gamma}} \wedge \omega^{\dot{\gamma}}_{~\dot{\beta}} = 0, \\
y^{\alpha} y^{\beta} &:~~~~ d_x \omega_{\alpha}^{~\beta} - 4 \omega_{\alpha}^{~\gamma}\wedge w_{\gamma}^{~\beta} 
                  - e_{\alpha\dot{\alpha}}\wedge e_{\beta\dot{\beta}}\epsilon^{\dot{\alpha}\dot{\beta}} = 0, \\ 
y^{\dot{\alpha}}y^{\dot{\beta}} &:~~~~ d_x \omega^{\dot{\alpha}}_{~\dot{\beta}} 
                 + 4 \omega^{\dot{\alpha}}_{~\dot{\gamma}}\wedge \omega^{\dot{\gamma}}_{~\dot{\beta}} 
               - e_{\alpha\dot{\alpha}}\wedge e_{\beta\dot{\beta}}\epsilon^{\alpha\beta} = 0. \\ 
\end{split}
\label{3eqn}
\end{equation} 
As explained around eq. (\ref{Flat-W}) above, these flatness equations translate into the vanishing of torsion and on the condition 
that the Riemann tensor is that of the maximally symmetric space AdS$_4$. 

In Poincar\'e coordinates, with the metric given in Euclidean signature by
\ie
ds^2 = {d\vec x^2+dz^2\over z^2},
\fe
we can write $\omega_0$ and $e_0$ explicitly as
\ie
& e_0(x|Y) = -{1\over 4} {dx^\mu\over z} y\sigma_{\mu} \bar y\,,\\
&\omega_0(x|Y) = -{1\over  8} {dx^i\over z}\left( y\sigma^{iz}y + \bar y\sigma^{iz}\bar y \right)\,,
\label{e-omega}
\fe
where $x^{\mu} = (z,x^i)$, $i=1,2,3$ (we use Euclidean signature, and identify $x^4$ with $z$). Note that, while the $\sigma$ matrices naturally carry flat indices, in this expression we identify flat and curved indices on $\sigma$ 
by using the diagonal vielbein, i.e. for instance 
\begin{equation}
(e_0)_{\alpha\dot\alpha} = \frac{1}{4}e_0^a (\sigma_a)_{\alpha\dot\alpha} = \frac{1}{4} dx^{\mu} e_{0,\mu}^a  (\sigma_a)_{\alpha\dot\alpha}
=\frac{dx^{\mu}}{4z}\delta^a_{\mu}(\sigma_a)_{\alpha\dot\alpha} \,.
\end{equation}
Note also that in our conventions  we have 
$y\sigma^\mu \bar y = y^\A (\sigma^\mu)_\A{}^\db \bar y_\db = -(\sigma^\mu)_{\A{}\db}y^\A \bar y^\db=\bar y^\db (\sigma^\mu)^\A{}_{\db} y_\A = \bar y \sigma^\mu y$, etc.

To extract the physical content of the theory, we can develop a perturbative expansion for fluctuations around the AdS vacuum. One 
writes
\ie
W = W_0(x|Y) + \widehat W(x|Y,Z),~~~ S = S(x|Y,Z), ~~~ B = B(x|Y,Z),
\fe
and solves Vasiliev equations order by order in the fluctuations $\widehat W$, $S$, $B$. At the linearized level, the equations are
\ie
\label{lineareqn}
& D_0\widehat W_1 = 0,
\\
& d_Z \widehat W_1 + D_0 S_1 = 0,
\\
& d_Z S_1 = e^{i\theta_0} B_1*\kappa dz^2 + e^{-i\theta_0} B_1*\bar\kappa d\bar z^2,
\\
& \widetilde D_0 B_1 = 0,
\\
& d_Z B_1 = 0,
\fe
where we have introduced the subscript `1' to denote the linearized fields. Here $D_0$ and $\widetilde D_0$ 
are the covariant and twisted covariant differential with respect to $W_0$, namely
\ie
D_0 = d_x + [W_0, \cdot\,]_*,~~~~ \widetilde D_0 = d_x + [\omega_0, \cdot\,]_* + \{e_0, \cdot\,\}_*.
\fe
Note that the anticommutator in $\widetilde D_0$ arises because of the $\pi$-operation in the twisted adjoint covariant derivative, see 
the second line of (\ref{VasEq}). Using 
the identities (\ref{star-ident}), we may also write 
\begin{equation}
\begin{aligned}
&D_0 = \nabla^L +2 e_0^{\alpha\dot\alpha}\left(
y_{\alpha} \frac{\partial}{\partial \bar y^{\dot\alpha}}
+\bar y_{\dot \alpha} \frac{\partial}{\partial y^{\alpha}}
\right)\,,\\
&\widetilde D_0 = \nabla^L + 2 e_0^{\alpha\dot\alpha}\left(
y_{\alpha}\bar y_{\dot\alpha} 
+\frac{\partial}{\partial y^{\alpha}}\frac{\partial}{\partial \bar y^{\dot\alpha}}\right)\,,\\
&\nabla^L=d_x+[\omega_0,\cdot]_* \\
&~~~~= d_x+2 \omega_0^{\alpha\beta}\left(
y_{\alpha} \frac{\partial}{\partial y^{\beta}}
+ y_{\beta} \frac{\partial}{\partial y^{\alpha}}\right)
+2 \bar\omega_0^{\dot\alpha\dot\beta}\left(
\bar y_{\dot \alpha} \frac{\partial}{\partial \bar y^{\dot \beta}}
+ \bar y_{\dot \beta} \frac{\partial}{\partial \bar y^{\dot \alpha}}\right)\,.
\label{Ds}
\end{aligned}
\end{equation}

Recall that $Z$-twistor variable is entirely auxiliary, and the physical degrees of freedom are contained in the master fields $\widehat W$ and $B$ 
restricted to $Z = (z_\A,\bar z_\da)=0$. To solve the linearized equations, we write 
\begin{equation}
\begin{aligned}
&W_1(x|Y,Z)=\Omega(x|y,\bar y)+W'(x|y,\bar y,z, \bar z)\,,\qquad W'|_{Z=0}=0\,,\\
&B_1(x|Y,Z) = C(x,y,\bar y)\,,
\end{aligned}
\end{equation}
where by writing the second line we have already solved the last equation in (\ref{lineareqn}). 
The master field $S$ is purely auxiliary, as one may impose the gauge condition \cite{Vasiliev:1995dn, Vasiliev:1999ba}
\begin{equation}
S_{\alpha}|_{Z=0}=S_{\dot\alpha}|_{Z=0}=0\,,
\end{equation}
which can always be achieved with a $Z$-dependent gauge transformation $\epsilon(x|Y,Z)$. In this gauge, one may solve the third equation in 
(\ref{lineareqn}) to find
\ie
\label{sbrel}
S_1 &= - e^{i\theta_0} z_\A dz^{\A} \int_0^1 dt \,t (C*\kappa)|_{z_\A\to t z_\A} +c.c \\
&= - e^{i\theta_0} z_\A dz^{\A} \int_0^1 dt \,t\, C(x|-t z,\bar y) e^{t zy}  +c.c.\,.
\fe
Plugging this into the second equation in (\ref{lineareqn}) allows to determine the $Z$-dependent part of $W$, denoted $W'(x|y,\bar y,z, \bar z)$ above;  
see \cite{Giombi:2012ms} for the detailed calculation that we skip here for brevity. Finally, after plugging 
the result for $W'$ into the first equation in (\ref{lineareqn}), one can compute 
\ie
D_0 \Omega = -D_0 W' = - (D_0 W')|_{Z=0}\,,
\fe
and the end result is that the linearized equations for the ``physical master fields" $\Omega(x|Y)$ and $C(x|Y)$ are given by 
\begin{equation}
\begin{aligned}
&D_0 \Omega 
= 2 e^{i\theta_0} (e_0)_\A{}^\db\wedge (e_0)^{\A \dot\gamma} \partial_{\bar y^\db} \partial_{\bar y^{\dot\gamma}}  C(x| 0,\bar y) \\
&~~~~~~~+2 e^{-i\theta_0} (e_0)^\B{}_\da \wedge (e_0)^{\gamma\dot\alpha} \partial_{y^\B} \partial_{y^{\gamma}}  C(x| y,0)\,,\\
&\widetilde D_0 C = 0\,.
\label{lin-final}
\end{aligned}
\end{equation}
One can show that these equations describe the free propagation of a tower of HS fields, one for each integer spin,\footnote{Or one for each even spin in the minimally truncated theory.} plus a scalar with mass squared $m^2=-2$ in AdS units \cite{Vasiliev:1992av}. 
We will only sketch the derivation of this in the remaining of this section, and refer the reader to the literature \cite{Vasiliev:1995dn, Vasiliev:1999ba, Bekaert:2005vh, Iazeolla:2008bp, Didenko:2014dwa, Giombi:2012ms} for more details. First, note that the $\nabla^L$ operator in (\ref{Ds}) does 
not change the degree in $y,\bar y$, it is just a Lorentz covariant derivative acting on each component of the master field. On the other hand, 
the operator $e_0 (y\partial_{\bar y}+\bar y \partial_{y})$ in $D_0$ changes the degree of $y$ and $\bar y$ by one unit, while keeping the total 
degree in $y,\bar y$ fixed. Therefore, we see that the first equation in (\ref{lin-final}) splits into decoupled equations for the 
set $\Omega(x|Y)|_{y^{s-1+k}\bar y^{s-1-k}}\equiv \Omega^{(s-1+k,s-1-k)}$, with $-(s-1)\leq k\leq (s-1)$; the equations relate ``neighboring" components until the bottom 
components $\Omega^{(2s-2,0)}$ and $\Omega^{(0,2s-2)}$ are reached, which are respectively connected by (\ref{lin-final}) to $C^{(2s,0)}$ and $C^{(0,2s)}$. 
As for the second equation, we see that the operator $e_0(y\bar y+\partial_{y}\partial_{\bar y})$ appearing in $\widetilde D_0 C = 0$ changes the total 
degree in $y,\bar y$ by two, while keeping the {\it difference} of the degrees in $y$ and $\bar y$ fixed. Therefore the second equation splits in decoupled 
sets for the fields $C^{(n,m)}$, $|n-m|=2s$. Since the components $C^{(2s,0)}$, $C^{(0,2s)}$ are related to the HS gauge fields by the first equation, 
we conclude that (\ref{lin-final}) decompose into subsets of equations for each spin $s$, involving the components
\begin{equation}
{\rm spin~} s: \begin{cases}
\Omega^{(s-1+k,s-1-k)}, \qquad k=-(s-1),\ldots, (s-1)\,,\\
C^{(2s+n,n)},~~ C^{(n,2s+n)}\,\qquad n\ge 0\,.
\end{cases}
\end{equation}
Of course, the equation for the scalar is entirely contained in the second line of (\ref{lin-final}). Writing out that equation in terms of the 
scalar components $C^{(n,n)}=\frac{1}{(n!)^2}C_{\alpha_1\ldots \alpha_n\dot\alpha_1\cdots \dot\alpha_n}$, $n>0$, we get 
\begin{equation}
\nabla_{\mu}^L C^{(n,n)}+2 e_{0,\mu}^{\alpha\dot\alpha}\left(
y_{\alpha}\bar y_{\dot \alpha}C^{(n-1,n-1)}+\frac{\partial}{\partial y^{\alpha}\partial \bar y^{\dot\alpha}}C^{(n+1,n+1)}\right)=0\,.
\label{sc-tower}
\end{equation}
This tower of first order differential equations represents the so-called unfolded form of the usual Klein-Gordon equation. To see this, we can 
write the first two equations in the tower
\ie
\label{ceqnb}
& \nabla_\mu C^{(0,0)} + 2e_{0,\mu}^{\A\db} C^{(1,1)}_{\alpha\dot\alpha}  = 0\,,\\
& \nabla_\mu C^{(1,1)}_{\alpha\dot\alpha} +2 (e_{0,\mu})_{\A\dot\alpha} C^{(0,0)} 
+ 2e_{0,\mu}^{\beta \db}  C^{(2,2)}_{\alpha\beta\dot\alpha\dot\beta}  = 0\,.
\fe
Taking a covariant derivative of the first line, and solving for $\nabla_{\mu}B^{(1,1)}_{\alpha\dot\alpha}$ from the second line, one finds 
\begin{equation}
\begin{aligned}
&\nabla^2 C^{(0,0)}+2 e_{0,\mu}^{\alpha\dot\alpha} \nabla^{\mu}C_{\alpha\dot\alpha}^{(1,1)}\\
&= \nabla^2 C^{(0,0)}-2 e_{0,\mu}^{\alpha\dot\alpha}\left( 2 (e_0^{\mu})_{\alpha\dot\alpha}C^{(0,0)}
+2 (e_0^{\mu})^{\beta\do\beta}C^{(2,2)}_{\alpha\beta\dot\alpha\dot\beta}\right)=0\,.
\end{aligned}
\end{equation}
Now using that in our conventions 
$e^{\A\db}_\mu e^\mu_{\C\dd} = -{1\over 8} \delta^\A_\C\delta^\db_\dd$, and $e^{\A\db}_\mu e^\nu_{\A\db} = -{1\over 8}\delta^\nu_\mu$, and 
that $\epsilon^{\alpha\beta}\epsilon^{\dot\alpha\dot\beta}C^{(2,2)}_{\alpha\beta\dot\alpha\dot\beta}=0$, we finally find the Klein-Gordon equation 
\begin{equation}
\left(\nabla^2+2\right)C^{(0,0)}=0\,.
\end{equation}
Hence, as anticipated in our CFT discussion, we see that the scalar field in Vasiliev AdS$_4$ theory has $m^2=-2/\ell^2_{\rm AdS}$. The higher 
equations in the tower (\ref{sc-tower}) do not yield new information, they just express all higher components $C^{(n,n)}$ as derivatives of the bottom 
component $C^{(0,0)}$. 

Similarly, we can analyze the $s=1$ equations. In this case, there is a single equation coming from the first line of (\ref{lin-final})
\begin{equation}
\nabla_{[\mu}\Omega_{\nu]}^{(0,0)} = 2 e^{i\theta_0} (e_{0,[\mu})_{\alpha}^{\dot\beta} (e_{0,\nu]})^{\alpha\dot\gamma} C^{(0,2)}_{\dot\beta\dot\gamma}
+2 e^{-i\theta_0} (e_{0,[\mu})^{\beta}_{\dot\alpha} (e_{0,\nu]})^{\gamma\dot\alpha} C^{(2,0)}_{ \beta \gamma}\,.
\end{equation} 
This equation essentially just says that $C^{(2,0)}$ and $C^{(0,2)}$ are related to the field strength $F_{\mu\nu}$ 
of the gauge field $\Omega_{\mu}^{(0,0)}$, up to the phase factors 
\begin{equation}
C^{(2,0)}_{\alpha\beta} = e^{i\theta_0} F_{\alpha\beta}\,,\qquad 
C^{(0,2)}_{\dot\alpha\dot\beta}=e^{-i\theta_0} \bar F_{\dot \alpha\dot \beta}\,.
\end{equation}
The second equation in (\ref{lin-final}) then yields the Maxwell equations 
\begin{equation}
\nabla_{\alpha\dot\alpha}F^{\alpha}_{\ \beta}=0\,,\quad \nabla_{\alpha\dot\alpha} \bar F^{\dot\alpha}_{\ \dot\beta}=0\,,
\end{equation}
which are equivalent to $\nabla_{\mu}F^{\mu\nu}=0$, $\nabla_{[\mu}F_{\nu\rho]}=0$. 

For $s>1$, one gets from the first line of (\ref{lin-final}) the tower of equations 
\begin{equation}
\begin{aligned}
\nabla^L \Omega^{(s-1-k,s-1+k)} 
&+ 2e_0^{\A\db}\wedge \left[ y_\A\partial_{\bar y^\db} \Omega^{(s-2-k,s+k)} + \partial_{y^\A} \bar y_\db \Omega^{(s-k,s-2+k)} \right] = 0\,,\\
&\qquad\qquad \qquad \qquad \quad -(s-1)<k<s-1
\end{aligned}
\label{tempom}
\end{equation}
and for $k=\pm(s-1)$ 
\begin{equation}
\begin{aligned}
&\nabla^L \Omega^{(2s-2,0)} 
+ 2e_0^{\A\db}\wedge  y_\A\partial_{\bar y^\db} \Omega^{(2s-3,1)} 
= 2e^{-i\theta_0} (e_0)^\B{}_\da \wedge (e_0)^{\gamma\dot\alpha} \partial_{y^\B} \partial_{y^{\gamma}}  C^{(2s,0)}\\
&\nabla^L \Omega^{(0,2s-2)} 
+2e_0^{\A\db}\wedge  \partial_{y^\A} \bar y_\db \Omega^{(1,2s-3)} 
=2 e^{i\theta_0} (e_0)^\B{}_\da \wedge (e_0)^{\gamma\dot\alpha} \partial_{y^\B} \partial_{y^{\gamma}}  C^{(0,2s)}\,.
\label{Weyl-HS}
\end{aligned}
\end{equation}
Eq. (\ref{tempom}) and (\ref{Weyl-HS}) are the correct HS free equations in the frame-like formalism, essentially equivalent to the tower of curvature constraints (\ref{curv-HS}). 
The Fronsdal equations (\ref{Fronsd-AdS2}) for the totally symmetric field 
\begin{equation}
\varphi_{\mu_1\cdots \mu_s} = \Omega_{(\mu_1}^{\alpha_1\cdots \alpha_{s-1}\dot\alpha_1\cdots \dot\alpha_{s-1}} 
(e_{0,\mu_2})_{\alpha_1\dot\alpha_1}\cdots (e_{0,\mu_s)})_{\alpha_{s-1}\dot\alpha_{s-1}}
\end{equation}
are in fact already contained in (\ref{tempom}) for $k=0,\pm 1$. Note that they are independent of the phase factor $\theta_0$. 
Finally, the equations (\ref{Weyl-HS}) imply that $C^{(2s,0)}$ and $C^{(0,2s)}$ are 
related to the selfdual and anti-selfdual parts of the HS Weyl tensor $C^{(\pm)}_s$, up to the phase factor 
\begin{equation}
C^{(2s,0)}_{\alpha_1\cdots \alpha_{2s}}= e^{i\theta_0} C^{(+)}_{\alpha_1\cdots \alpha_{2s}} \,,\quad 
C^{(0,2s)}_{\dot\alpha_1\cdots \dot\alpha_{2s}}= e^{-i\theta_0} C^{(-)}_{\dot\alpha_1\cdots \dot\alpha_{2s}} \,.
\label{HSWeyl}
\end{equation}
The second equation in (\ref{lin-final}) in this case does not yield new information, it just corresponds to the Bianchi identities needed 
for consistency of the system. 

\subsection{Perturbation theory and correlation functions}
\label{3pt-tests}

The perturbative expansion around AdS described in the previous section can be carried out 
order by order in perturbation theory, and one should in principle be able to extract this way the cubic, quartic, etc. 
interaction vertices of the HS fields. In practice, this procedure is very complicated and 
has not been carried out beyond quadratic 
level (cubic vertices) \cite{Sezgin:2002ru,Giombi:2009wh, Boulanger:2015ova, Vasiliev:2016xui}.

In the context of AdS/CFT, the main object of interest are the correlation functions of the boundary HS 
currents, computed via bulk Witten diagrams as in Figure \ref{3ptW}. A direct calculation of the 3-point 
functions using Vasiliev equations was carried out in \cite{Giombi:2009wh,Giombi:2010vg} 
(see also \cite{Didenko:2012tv, Didenko:2013bj}). In this section, we briefly review the main idea and results of the 
calculation, and refer the reader to the review \cite{Giombi:2012ms} for a detailed summary 
of the more technical aspects of the computations. 

Let us first review some useful notation to describe the HS currents in CFT$_3$. 
In (\ref{Js-no-ind}), we have introduced the auxiliary null polarization vector $\epsilon^{\mu}$ which 
allows to greatly simplify tensor manipulations. While this works in any dimension, in $d=3$ a further 
simplification arises because we can trade a null polarization vector by a ``polarization spinor" $\lambda_{\alpha}$
by 
\begin{equation}
\epsilon_{\alpha\beta} = \vec{\epsilon}\cdot \vec{\tau}_{\alpha\beta}=\lambda_{\alpha}\lambda_{\beta}
\end{equation}
where we denote by $\vec{\tau}_{\alpha\beta}$ the usual Pauli matrices\footnote{We use a different symbol 
to avoid confusion with the $\sigma$ matrices used in the 4d bulk, which naturally carry a dotted and undotted 
index.}, and the last equality follows from the fact that $\vec{\epsilon}$ is null and so $\epsilon_{\alpha\beta}$ 
has vanishing determinant. In other words, we can encode the HS currents as 
\begin{equation}
J_s(x,\lambda) =J_{\alpha_1\cdots \alpha_{2s}}\lambda^{\alpha_1}\cdots \lambda^{\alpha_{2s}}
\label{Js-lam}
\end{equation}
where the totally symmetric multispinor $J_{\alpha_1\cdots \alpha_{2s}}$ is equivalent to the totally 
symmetric rank-$s$ tensor of $SO(3)$. 

The correlation functions $\langle J_{s_1}J_{s_2}J_{s_3}\rangle$ 
can be extracted directly from the Vasiliev equations in AdS. 
One picks two of the currents, say $J_{s_1}, J_{s_2}$ to play the role of boundary sources, and then 
solves for the field $\varphi_{s_3}(\vec x,z)$, dual to $J_{s_3}$, sourced by those boundary 
operators, working to second order in perturbation theory. The 3-point function is then obtained 
by taking the boundary limit $z\rightarrow 0$ of the second order field $\varphi_{s_3}(\vec x,z)$, after 
stripping off the factor $z^{\Delta_3}$ that is related to the scaling dimension of the boundary operator. This 
is schematically depicted in Figure 8. 
\begin{figure}
\begin{center}
\includegraphics[width=0.6\textwidth]{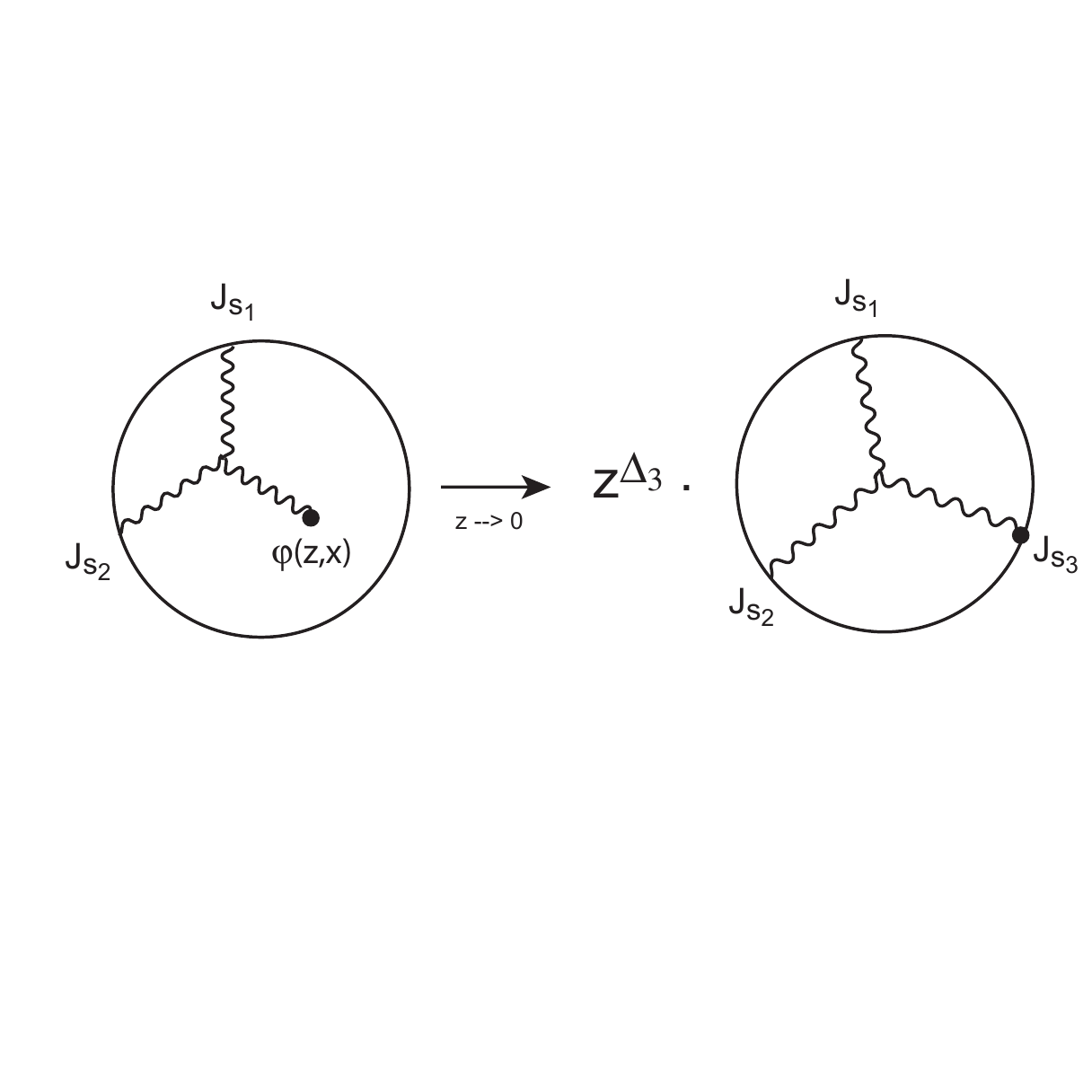}
\label{3pt-EOM}
\vskip -2cm
\caption{Extracting the 3-point functions from the boundary limit of the second order field.}
\end{center}
\end{figure}

The spin-$s$ degrees of freedom are contained both in the gauge field $\Omega(x|Y)=W(x|Y,Z)|_{Z=0}$ 
and the zero-form $C(x|Y)=B(x|Y,Z)|_{Z=0}$, as described in the previous section. 
In practice, it is more convenient to focus on the zero-form, and extract 
the HS fields from the (self-dual part of) the spin-$s$ Weyl curvature, $C^{(2s,0)}$. 
This is because in the boundary limit $z\to 0$, $C^{(2s,0)}$ is proportional to the metric-like rank-$s$ symmetric traceless tensor field up to a power of $z$. 
In fact, the linearized equations allow to derive a relation between the spin-$s$ Weyl curvature 
and gauge field of the form $C^{(2s,0)}(x|y) = c_s z^{1-s} e^{z^2 y \sigma^{\mu}\partial_{\mu} \partial_{\bar y}} z^{-1}\Phi_s(x|y\sigma^{\mu}\bar y)|_{\bar y =0}$, 
where $c_s$ is a normalization constant and $\Phi_s(x|y\sigma^{\mu}\bar y)$, which is related to the vielbein-like field $\Omega^{(s-1,s-1)}_{\mu}$, encodes all the totally symmetric Fronsdal fields, see eq.~(3.59) and (3.60) of \cite{Giombi:2009wh}. 
In the $z\rightarrow 0$ limit, the leading behavior of the spin-$s$ gauge field is $\Phi_s \sim z^{s+1} \varphi_{i_i\cdots i_s}(\vec{x}) (y\sigma^{i_1}\bar y)\cdots (y\sigma^{i_s}\bar y)$, where $\varphi_{i_i\cdots i_s}(\vec{x})$ is the boundary profile dual to the current $J_{i_1\cdots i_s}$. In the relation between $C^{(2s,0)}$ and $\Phi_s$ above, at leading order in small $z$ one can replace $  \sigma^{\mu}\partial_{\mu} \sim \sigma^z \partial_z$, so that upon setting $\bar y=0$ one obtains $C^{(2s,0)} \sim z^{s+1} \varphi_{i_i\cdots i_s}(\vec{x})(y\sigma^{i_1z}y)\cdots(y\sigma^{i_s z}y)$ in the $z\rightarrow 0$ limit. 
In other words, $C^{(2s,0)}_{\alpha_1\cdots \alpha_{2s}}$ 
may be thought as dual to the spin-$s$ current $J_{\alpha_1\cdots \alpha_{2s}}$ in bispinor notation 
(\ref{Js-lam}), with the twistor variables $y_{\alpha}$ naturally identified with the boundary polarization spinors (note that $\sigma^{iz}_{\alpha\beta}$ 
plays the role of the boundary Pauli matrices $\tau^i_{\alpha\beta}$).\footnote{Even though the linearized relation between $C^{(2s,0)}$ and $\Phi_s$ 
will receive corrections at higher orders in perturbation theory, we expect that it should still 
hold at leading order at small $z$ away from boundary sources. See Appendix A of \cite{Giombi:2009wh} for a related discussion.} 
To summarize, after solving for the second order zero-form $B_2(x|Y,Z)$, we extract $C_2(x|Y)$ by setting $Z=0$, 
and then take the boundary limit of the selfdual part to read off the 3-point function 
\begin{equation}
\begin{aligned}
&B_2(z,\vec{x}|y_{\alpha},\bar y_{\dot\alpha}=0, z_{\alpha}=0,\bar z_{\dot\alpha}=0)
\stackrel{z\rightarrow 0}{\sim} 
z^{s+1} C_{\alpha_1\cdots \alpha_{2s}} (\vec{x})y^{\alpha_1}\cdots y^{\alpha_s}\\
&C_{\alpha_1\cdots \alpha_{2s}} =\langle J_{\alpha_1\cdots \alpha_{2s}}\rangle_{J_{s_1}, J_{s_2}} \quad 
\rightarrow \quad \langle J_{s_1}J_{s_2}J_{s_3}\rangle_{\rm CFT}
\end{aligned}
\end{equation} 

The explicit calculation sketched above has been carried out in two approaches: 
a ``physical spacetime approach" \cite{Giombi:2009wh}, where one 
works out the perturbation theory directly in the physical AdS spacetime; and a ``gauge function approach" \cite{Giombi:2010vg} 
where one utilizes the fact that Vasiliev equations (\ref{VasEq2}) allow for a formal gauge transformation (``$W=0$ gauge") 
that removes the spacetime dependence of the master fields, so that one can solve for them entirely in the $(Y,Z)$ space 
which is technically simpler. In the latter method, assuming a certain contour prescription to carry out the twistor space 
integrals, one obtains a rather compact representation of the 3-point functions for all spins
\begin{equation}
\begin{aligned}
&\langle J(\vec{x}_1,\lambda_1) J(\vec{x}_2,\lambda_2)J(\vec{0},\lambda_3)\rangle \\
&=\int d^4U d^4V B_1(U;\vec{x}_i,\lambda_i)B_1(V;\vec{x}_i,\lambda_i){\cal K}(U,V;\lambda_3)
\label{JJJ-tw}
\end{aligned}
\end{equation}
where $U=(u_{\alpha},\bar u_{\dot\alpha})$, and similarly for $V$. In this expression, ${\cal K}(U,V;\lambda_3)$ 
is a certain ``kernel" whose explicit expression can be found in \cite{Giombi:2010vg, Giombi:2012ms}; and $B_1$ is a linearized field 
which is related to the bulk-to-boundary propagators in the presence of the two sources at positions $\vec{x}_{1,2}$ and 
polarization $\lambda_{1,2}$. Allowing for the general parity breaking phase $\theta_0$, its expression takes the form 
(see eq. (\ref{HSWeyl}) above)
\begin{equation}
B_1(U;\vec{x}_i,\lambda_i) =e^{i\theta_0} C_1^{(+)}(U;\vec{x}_i,\lambda_i)+e^{-i\theta_0} C_1^{(-)}(U;\vec{x}_i,\lambda_i)
\label{B1-tw}
\end{equation}
where we refer to \cite{Giombi:2012ms} for the explicit expressions of the bulk-to-boundary propagators $C_1^{(\pm)}$. Plugging (\ref{B1-tw}) into (\ref{JJJ-tw}), the result is seen 
to split into three terms, as expected from weakly broken HS symmetry considerations \cite{Maldacena:2012sf}, see (\ref{jjj-CS})
\begin{equation}
\langle JJJ\rangle = \cos^2\theta_0 \langle JJJ\rangle_{A} + \sin^2\theta_0 \langle JJJ\rangle_{B}
+\sin \theta_0\cos \theta_0 \langle JJJ\rangle_{\rm odd}\,,
\label{jjj-ABodd}
\end{equation}
where $\langle JJJ\rangle_A$ and $\langle JJJ\rangle_B$ are parity preserving, and were shown in \cite{Giombi:2010vg, Giombi:2012ms} to precisely 
match the generating functions \cite{Giombi:2011rz} of 3-point functions of all spins in free scalar and free boson respectively, 
thus providing a non-trivial test of the HS/CFT duality in the parity preserving type A/type B cases. On the other hand, 
the term  $\langle JJJ\rangle_{\rm odd}$, which would be parity odd, 
apparently vanishes \cite{unpub} after carrying out the integration in (\ref{JJJ-tw}) with the contour prescription of \cite{Giombi:2012ms}, 
while it should definitely be non-zero from CFT point of view. Presumably 
this puzzle has to do with some subtlety in the contour prescription or with some gauge ambiguity that was not properly fixed in the calculation, 
see \cite{Giombi:2012ms} for a discussion. Note that the splitting into three structures (\ref{jjj-ABodd}) was also obtained recently 
in \cite{Vasiliev:2016xui}.  

In the ``physical spacetime approach", the calculation of \cite{Giombi:2009wh} was generalized to the parity breaking case in \cite{Giombi:2012ms}, 
where the correlator $\langle J_{s_1} J_{s_2}J_0\rangle$ was derived, assuming $\Delta=1$ boundary condition on the scalar (it should be possible 
to extract the result for $\Delta=2$ by similar methods). The final result (see Section 5.2 of \cite{Giombi:2012ms}) 
takes the form 
\begin{equation}
\langle J_{s_1} J_{s_2} J_0\rangle = \cos\theta_0 \langle J_{s_1}J_{s_2}J_{0}\rangle_{\rm free~sc}
+\sin\theta_0 \langle J_{s_1}J_{s_2}J_0\rangle_{\rm odd} \,,
\label{JJO}
\end{equation}
where the parity odd structure is correctly non-vanishing and was shown to match the available known CFT results \cite{Giombi:2011rz,Maldacena:2012sf}. 
Note that in this case there is no free fermion structure, since the scalar operator has $\Delta=1$. This result is precisely 
consistent with the weakly broken HS symmetry analysis \cite{Maldacena:2012sf}.\footnote{In writing the result in (\ref{JJO}), the scalar $J_0$ is assumed to be normalized so that the 2-point function is $\theta_0$-independent. One should take this into account when comparing 
to \cite{Maldacena:2012sf, Aharony:2012nh}. Also note that in (\ref{JJO}) and (\ref{jjj-ABodd}) we have 
omitted the overall bulk coupling constant $g_{\rm bulk} \sim G_N^{-1/2}$. Comparison with the results of \cite{Maldacena:2012sf, Aharony:2012nh}, 
see eq. (\ref{tN-tlam}) above, fixes $G_N^{-1} \sim \tilde N = 2N \frac{\sin(\pi\lambda)}{\pi \lambda}$.}
Combining the conjectured duality 
(\ref{HS-vec-parity}) with the explicit results of \cite{Aharony:2012nh}  
indicates that the phase parameter should be mapped to the 't Hooft coupling in the dual Chern-Simons vector model by
\ie
\theta_0 = {\pi\over 2}\lambda
\fe
for CS-scalar vector model, and
\ie
\theta_0 = {\pi\over 2}(1-\lambda)
\fe
for CS-fermion vector model. Note that, if the higher phases $\theta_{2n}, n>0$ in (\ref{ThetaX}) are absent \cite{Vasiliev:2015mka}, 
the conjectured duality (\ref{HS-vec-parity}) with $\theta_0(\lambda)$ given above would manifestly imply the 3d bosonization duality (\ref{lev-rank}) at 
large $N$, not just at the level of 3-point functions, but to all orders.  

\subsection{Generalizations: Chan-Paton factors and supersymmetry}

In the previous sections we have presented the simplest bosonic HS theory in AdS$_4$. For completeness, we briefly mention here 
two generalizations of the Vasiliev equations: Chan-Paton factors and supersymmetry 
\cite{Konstein:1989ij,Vasiliev:1990bu, Vasiliev:1990vu, Vasiliev:1999ba,Chang:2012kt}. 

Vasiliev system (\ref{VasEq}) admits an obvious extension to non-abelian 
higher spin fields, through the introduction of ``Chan-Paton" factors. We simply replace the master fields by 
$M\times M$ matrix valued fields, $W^{a}_{\ b}, B^{a}_{\ b}, S^{a}_{\ b}$, and replace the star-algebra in the gauge 
transformations and equations of motion by its tensor product with the algebra of $M\times M$ complex matrices. The spectrum 
then includes matrix valued HS fields of all spins, transforming in the adjoint of $U(M)$.\footnote{One may similarly consider 
orthogonal and symplectic cases, see e.g. \cite{Vasiliev:1999ba}.} On the CFT side, this corresponds to taking 
a CFT with $N\cdot M$ complex fields, say the free scalar theory (the fermionic case may be discussed in a similar way)
\begin{equation}
S = \int d^3 x \left(\partial_{\mu}\phi^*_{ia})(\partial^{\mu} \phi^{ia}\right)\,,\qquad i=1,\ldots, N\,,\quad a=1,\ldots,M
\end{equation} 
and imposing a $U(N)$ singlet constraint. Then there are $M^2$ conserved currents for each spin 
\begin{equation}
(J_{s})^a_{\ b} = \phi^*_{ib}\partial^s \phi^{ia}\,,\qquad a,b=1,\ldots,M\,.
\label{Jab}
\end{equation}
Note that while the bulk theory involves $M^2$ ``colored" gravitons, when interactions are turned on at the boundary (e.g. by 
double-trace deformations, corresponding to changing the boundary conditions in the bulk), only a single graviton, corresponding to the 
singlet in (\ref{Jab}), will remain massless \cite{Chang:2012kt}. Perturbative calculations like the ones done in 
\cite{Giombi:2009wh, Giombi:2010vg} can be extended in a straightforward way to 
the non-abelian case, as long as $M$ is kept fixed in the large $N$ limit, so that the bulk perturbative expansion still makes sense. 
Note that the bulk 't Hooft coupling is $\sim M/N$, since $G_N^{-1}\sim N$ and the bulk fields are $M\times M$ matrices.

Vasiliev system also admits a supersymmetric extension 
\cite{Vasiliev:1986qx,Konstein:1989ij,Engquist:2002vr,Chang:2012kt} that can be defined by simply introducing
Grassmannian auxiliary variables $\psi_i$, $i=1,\cdots,n$, that obey the
Clifford algebra $\{\psi_i, \psi_j\} = 2\delta_{ij}$, and commute with all 
the twistor variables $(Y,Z)$. The master fields $W, S, B$, as 
well as the gauge transformation parameter $\epsilon$,  are then
functions of the $\psi_i$'s as well as of $(x^\mu, y_\A, \bar y_\da, z_\A, \bar z_\da)$. The constraint that generalizes (\ref{const-bose}) 
now requires that the components of the master fields that are even functions of $\psi_i$ are also even functions of 
the spinor variables $Y, Z$, whereas odd functions of $\psi_i$ are also odd functions of 
$Y, Z$. The latter give rise to fermionic fields in AdS$_4$ of half-integer spins, consistently 
with the spin-statistics theorem. Note that the bilinears in $y,\bar y, \psi_i$ 
generate the superalgebra $OSp(4|n)$. While $n$ is arbitrary at the level of the equations of motion, 
it was shown in \cite{Chang:2012kt} that in the general parity violating theories (which are dual to interacting 
theories at the boundary) the maximum preserved supersymmetry is in fact ${\cal N}=6$. 

\section{One-loop tests of Higher Spin/CFT duality}
\label{OneLoop}

In Section \ref{3pt-tests} we have reviewed the existing tests of the higher spin/vector model dualities at the level of 3-point correlation 
functions. In this section, we review some more recent tests based on comparing bulk and boundary partition functions 
\cite{Giombi:2013fka, Giombi:2014iua, Jevicki:2014mfa, Beccaria:2014xda, Beccaria:2014zma, Beccaria:2014qea,Hirano:2015yha, Bae:2016rgm}. For brevity, we 
mainly focus on the original calculation in AdS$_4$/CFT$_3$ done in \cite{Giombi:2013fka}, and mention towards 
the end of the section a few generalizations. 

In a CFT$_3$, an interesting physical observable is given by the sphere partition function, or free energy $F=-\log Z$ of the theory 
on a round $S^3$. In a conformal theory, this is a finite, well-defined number that plays the role of a $c$-function in 3d: it satisfies 
$F_{\rm UV} > F_{\rm IR}$ under RG flow \cite{Myers:2010tj, Casini:2011kv, Jafferis:2011zi, Klebanov:2011gs, Casini:2012ei}, and is related to the universal 
term in the entanglement entropy across a circle. In the free CFT, the sphere free energy is easily computed by a one-loop path integral. Focusing on 
the scalar model, one has 
\begin{eqnarray}
&&F=-\log Z \qquad \qquad Z = \int D\phi e^{-S}\cr 
&& S = \int d^3x \sqrt{g}\left(\partial_{\mu}\phi^i\partial^{\mu}\phi^i+\frac{{\cal R}}{8}\phi^i\phi^i\right)
\end{eqnarray}
where the coupling to the scalar curvature  ${\cal R}=6$  of the 3-sphere is required by conformal invariance (we set the radius to 1). Then, one has 
\begin{equation}
F = \frac{N}{2}\log \det \left(-\nabla^2+\frac{3}{4}\right)\,.
\end{equation}
where $N$ is the number of real scalars. This determinant can be computed for instance by zeta-function techniques \cite{Klebanov:2011gs},  
yielding
\begin{eqnarray}
F = \frac{N}{2}\sum_{n=0}^{\infty} (n+1)^2\log[(n+1/2)(n+3/2)] = N\left(\frac{\log 2}{8}-\frac{3\zeta_3}{16\pi^2}\right).
\label{F-sc}
\end{eqnarray}
Importantly, the $N$ dependence of $F$ is trivial, since we have a free CFT. 

How do we reproduce the CFT result from the bulk? In particular, can we see that all the $1/N$ corrections are trivial? This is 
certainly not obvious from the bulk point of view, since the HS gravity theory is a highly non-linear theory. According to the AdS/CFT dictionary, 
we should have
\begin{equation}
Z_{\rm bulk} = Z_{\rm CFT}\,,
\end{equation}
where $Z_{\rm bulk}$ is the partition function of the bulk theory on the vacuum Euclidean AdS$_4$ (hyperbolic space), with metric
\begin{equation}
ds^2=d\rho^2+\sinh^2\rho\, d\Omega_3\,,
\label{AdS4}
\end{equation}
where $d\Omega_3$ is the metric of a unit round 3-sphere. 
On general grounds, the bulk partition function takes the form
\begin{equation}
Z_{\rm bulk} = e^{-\frac{1}{G_N} F^{(0)}-F^{(1)}-G_N F^{(2)}+\ldots} = e^{-F_{\rm bulk}}\,,
\label{Zbulk}
\end{equation}
where $G_N$ denotes the bulk coupling constant. As explained in Section \ref{CFT-AdS}, 
a simple argument based on matching the large $N$ scaling of correlation functions, one sees that $G_N^{-1} \propto N$ in the large $N$ limit. 
Then we see that in principle the bulk free energy may involve an infinite set of $1/N$ corrections coming from loops. 
The leading term in (\ref{Zbulk}) corresponds to the value of the suitably regularized classical bulk action computed on the vacuum solution with metric
(\ref{AdS4}), and with all other fields set to zero. Since we do not currently know an action for Vasiliev theory which reduces to the
standard quadratic actions at the free level,\footnote{Actions for Vasiliev theory were proposed in
\cite{Boulanger:2011dd,Doroud:2011xs,Boulanger:2012bj,Boulanger:2015kfa,Bonezzi:2016ttk}. 
They do not appear to reduce to ordinary quadratic lagrangians when expanded around the AdS$_4$ vacuum, 
and it is not clear at the moment how to extract the tree level free energy $F^{(0)}$ from those actions.} 
a direct calculation of $F^{(0)}$ appears to be out of reach at present. An alternative approach to compute $F^{(0)}$ would 
be to exploit the relation to the circle entanglement entropy, and try to find a suitable generalization of the Ryu-Takayanagi 
prescription \cite{Ryu:2006bv} to the HS gravity case. This would involve finding a suitable HS invariant 2-form to be integrated 
over a minimal surface ending on the circular contour at the boundary (for recent progress on constructing HS invariant 
observables in HS theory, see \cite{Sezgin:2011hq, Vasiliev:2015mka}).

The subleading terms in (\ref{Zbulk}), which arise from the quantum corrections, are obtained by computing the vacuum diagrams of the bulk fields 
in the AdS$_4$ background. 
In particular the calculation of the one loop term $F^{(1)}$ depends only on the spectrum of the theory, 
and may be computed by evaluating the appropriate functional determinants in AdS$_4$. For the type A theory in AdS$_{d+1}$, after 
gauge fixing \cite{Gaberdiel:2010ar, Gaberdiel:2010xv, Gupta:2012he} one may express the one-loop partition function as 
\begin{equation}
\begin{aligned}
Z_{\rm 1-loop} =&\frac{1}{\left[{\rm \det}\left(-\nabla^2-2(d-2)\right)\right]^{\frac{1}{2}}}\times \\
&\times \prod_{s=1}^{\infty}\frac{\left[{\rm det}^{STT}_{s-1}\left(-\nabla^2+(s-1)(s+d-2)\right)\right]^{\frac{1}{2}}}{\left[{\rm det}^{STT}_{s}\left(-\nabla^2+(s-2)(s+d-2)-s\right)\right]^{\frac{1}{2}}} \,.
\end{aligned}
\end{equation}
The label STT indicates that the determinants are evaluated on the space of symmetric traceless transverse tensors, and the structure 
of the kinetic terms are the ones coming from the gauge fixing of Fronsdal equations,  see eq. (\ref{FP-AdS}) and (\ref{FPg-AdS}) above. In 
the infinite product above, we have assumed that we are working in the non-minimal theory with all spins; in the minimal case, the product runs over 
the even spins. 

These one loop determinants can be computed with the aid of the spectral zeta function \cite{Hawking:1976ja} 
which was derived for all integer spins and all dimensions by Camporesi and Higuchi \cite{Camporesi:1993mz,Camporesi:1994ga}. 
Given the differential operator $(-\nabla^2+\kappa^2)$ (with $\kappa$ a constant) acting on the space of STT spin $s$ fields, 
one can define an associated spectral zeta function as the Mellin transform of the corresponding heat kernel. 
Explicitly, in general boundary dimension $d$ one finds \cite{Camporesi:1993mz,Camporesi:1994ga}
\begin{eqnarray}
&&\zeta_{(\Delta,s)}(z)=\left({ \int {\rm vol}_{AdS_{d+1} } \over \int {\rm vol}_{S^{d}}} \right) {2^{d-1} \over \pi} g_s \int_0^\infty d u \, { \mu_s(u) \over \left[ u^2 + \left( \Delta - {d \over 2} \right)^2 \right]^z } \cr
&&\left(\Delta-\frac{d}{2}\right)^2=\kappa^2+s+\frac{d^2}{4}\,.
\label{zeta-d}
\end{eqnarray}
Here ${\rm vol}_{AdS_{d+1}}$ is the (regularized) volume of Euclidean AdS, $g_s$ is the number of degrees of freedom (\ref{gs}) 
of a STT spin $s$ field in $d+1$ dimensions, and $\mu_s(u)$ is the so-called spectral density. In the present case of $d=3$, we have
\begin{eqnarray}
&&{\rm vol}_{AdS_{4}} = \frac{4}{3}\pi^2\,,\qquad {\rm vol}_{S^3}=2\pi^2 \cr
&&\mu_s(u) =\frac{\pi u}{16}\left[u^2+\left(s+\frac{1}{2}\right)^2\right]\tanh\pi u\,,\qquad g_s=2s+1\,.
\end{eqnarray}
Given the spectral zeta function (\ref{zeta-d}), the contribution to the one loop free energy $F^{(1)}=-\log Z_{\rm 1-loop}$  
of the spin $s$ field with kinetic operator $(-\nabla^2+\kappa^2)$ is obtained as \cite{Hawking:1976ja}
\begin{equation}
F^{(1)}_{(\Delta,s)} = -\frac{1}{2}\zeta'_{(\Delta,s)}(0)-\frac{1}{2}\zeta_{(\Delta,s)}(0) \log\left(\ell^2_{\rm AdS}\Lambda^2_{\rm UV}\right)\,,
\label{F-one-loop}
\end{equation}
where $\Lambda_{\rm UV}$ is a UV renormalization scale.  
The term proportional to $\zeta_{(\Delta,s)}(0)$ corresponds to a logarithmic divergence which arises in even dimensional spacetimes, 
in particular AdS$_4$. 
Computing the value of the spectral zeta function at $z=0$ from the expression (\ref{zeta-d}), one finds that 
the full logarithmic term of the one loop free energy is given by
\begin{eqnarray}
F^{(1)}\Big{|}_{\rm log-div} &=& -\frac{1}{2} \left(\zeta_{(1,0)}(0)+\sum_{s=1}^{\infty}\left(\zeta_{(s+1,s)}(0)-\zeta_{(s+2,s-1)}(0)\right)\right)\log\left(\ell^2_{\rm AdS}\Lambda^2_{\rm UV}\right)\cr
&=& \left(\frac{1}{360}+\sum_{s=1}^{\infty} \left(\frac{1}{180}-\frac{s^2}{24}+\frac{5s^4}{24}\right)\right)\log\left(\ell^2_{\rm AdS}\Lambda^2_{\rm UV}\right)\,.
\label{log-div}
\end{eqnarray}
Thus we see that each spin contributes a non-zero coefficient to the bulk one-loop divergence (for instance, for $s=2$, 
(\ref{log-div}) gives $\frac{571}{180}\log\left(\ell^2_{\rm AdS}\Lambda^2_{\rm UV}\right)$, which is the correct coefficient 
of the logarithmic divergence in pure Einstein gravity in AdS$_4$ \cite{Christensen:1979iy}). However, performing the sum 
by zeta function regularization,\footnote{One may obtain the same result by summing over all spins first and then 
analytically continuing the spectral parameter $z$ to zero \cite{Giombi:2014iua}.} and using $\zeta(0)=-\frac{1}{2}$ and $\zeta(-2n)=0$ for $n>0$, 
we see that the coefficient of the logarithmic divergence in the theory involving all spins actually vanishes, due to cancellation between 
the scalar and the infinite tower of HS fields. This is a non-trivial consistency 
check, since if the duality with the CFT is exact, one would expect the bulk theory to be ``UV complete", and not have any UV divergences. Note 
that the same result also holds in the minimal theory, and independently of the boundary conditions on the scalar field. 

Having shown that the logarithmic piece vanishes, one can then turn to the computation of the finite contribution to $F^{(1)}$, which 
is given by $\zeta'(0)$. This calculation is considerably more involved, and we refer the reader to \cite{Giombi:2013fka, Giombi:2014iua} 
for the technical details. 
The final result is that, in the non-minimal theory with all integer spins, and with $\Delta=1$ boundary condition on the bulk scalar, 
the one-loop bulk free energy precisely vanishes
\begin{equation}
F^{(1)} = 0\,.
\end{equation}
This is consistent with the duality with $N$ free complex scalars in the $U(N)$ singlet sector, 
and the simplest identification of the bulk coupling $G_N^{-1}\sim N$, with no order one shifts of $N$. 
The results of \cite{Diaz:2007an,Klebanov:2011gs,Giombi:2013yva} then imply that 
when the scalar is quantized with $\Delta=2$ boundary condition, the final result is $F^{(1)} = - \frac {\zeta(3)} {8 \pi^2}$, 
consistent with the field theory containing the double-trace interaction $\sim (\bar \phi^i \phi^i)^2$.

One can perform the analogous calculation in the case of the ``minimal" type A Vasiliev's theory which contains 
one scalar and one HS field of each even spin. As reviewed in Section \ref{CFT-AdS}, for $\Delta=1$ boundary condition 
this is conjecturally dual to $N$ real free scalars in the $O(N)$ singlet sector. Summing up the one-loop determinants of 
all fields in the spectrum, one finds the non-vanishing result \cite{Giombi:2013fka} 
\begin{equation}
F^{(1)}_{{\rm min}} = \frac{\log 2}{8}-\frac{3\zeta(3)}{16\pi^2}\,.
\label{min-loop}
\end{equation}
Remarkably, this is precisely equal to the value (\ref{F-sc}) of the 3-sphere free energy $F$ for a real conformally coupled scalar field! 
This result is therefore consistent with the duality with $N$ free real scalars in the $O(N)$ singlet sector, 
provided the identification between the bulk coupling $G_N$ and $N$ involves a shift $N\rightarrow N-1$, i.e. $G_N \sim \frac{1}{N-1}$, so that the classical piece
\begin{equation}
\frac{1}{G_N}F^{(0)}_{\rm min} = (N-1)\left(\frac{\log 2}{8}-\frac{3\zeta(3)}{16\pi^2}\right)\,,
\end{equation}
when combined with the one-loop piece (\ref{min-loop}), would give the expected CFT result proportional to $N$. Note that this intepretation, 
if correct, implies that all higher loop corrections in the bulk beyond one-loop should vanish. This simple integer shift is presumably 
related to a quantization condition of the coupling constant in the HS theory \cite{Maldacena:2011jn}, and is 
somewhat reminiscent of the integer one-loop shift of the level that appears in the CS gauge theory. For a discussion 
of this shift from a different point of view, see also \cite{Jevicki:2014mfa}. 

The calculation outlined above can be generalized to the type A theory in AdS$_{d+1}$ for all $d$ \cite{Giombi:2014iua}, as well as to more general 
HS theories \cite{Beccaria:2014xda, Beccaria:2014zma, Beccaria:2014qea, Zhong, Pang:2016ofv, Gunaydin:2016amv, Giombi:2016pvg} 
whose spectrum is dual to vector models with scalars, fermions and/or (in even $d$) $(d/2-1)$-forms, or combinations thereof.\footnote{Such theories 
involve in general half-integer spins as well as mixed symmetry fields. The non-linear equations are not known, but the spectrum can be 
uniquely identified from CFT considerations.} In odd $d$, the calculation is similar to the one above: 
$F$ is a finite number on the CFT side, and in the bulk we may have a UV 
divergence that is expected to cancel. In even $d$, the CFT free energy is logarithmically divergent and the coefficient of the divergence is 
the Weyl $a$-anomaly. In the bulk, this is reflected in the IR divergence of the regularized AdS volume, see \cite{Giombi:2014iua}. 
In all cases in even $d$, one finds agreement with the 
expected dualities, up to possible simple shifts of $N$ analogous to the one described above. For odd $d$, one also finds 
agreement with this pattern, except for an open puzzle in the case of the type B/fermionic vector model 
duality, where one gets a finite one-loop free energy which is not explained by a simple integer shift of $N$, 
see \cite{Giombi:2013fka, Pang:2016ofv, Gunaydin:2016amv, Giombi:2016pvg}. For all HS theories in AdS$_{d+1}$ 
with $d$ odd, the one-loop UV logarithmic divergences appear to vanish after summing over the relevant spectrum, 
which is an interesting result in its own right, even beyond the relevance of these models in the AdS/CFT context. 

\section{Summary and some open questions}

As reviewed in these notes, consistent fully non-linear theories of massless higher spin fields can be 
explicitly constructed if one assumes a non-zero cosmological constant. They involve infinite towers of massless 
fields of all spins, and a corresponding infinite dimensional gauge symmetry. While the existence of such 
theories is remarkable and fascinating in its own right, it has a natural interpretation in the context 
of the AdS/CFT correspondence: Vasiliev HS gauge theories have precisely the right spectrum 
to be holographically dual to large $N$ vector models. In the simplest versions of the duality, the vector model 
is a free CFT (restricted to the singlet sector), 
but the HS/CFT duality can be also extended to a variety of interacting theories, 
including the critical $O(N)$ model, Gross-Neveu model, 3d QED coupled to massless flavors, 
$CP^{N}$ model, and vector models coupled to Chern-Simons gauge fields. In the interacting vector models, the higher spin 
symmetry is weakly broken at large $N$, generating small anomalous dimensions for the HS current; 
in the corresponding AdS duals, the HS symmetry is broken by the boundary 
conditions and the gauge fields acquire masses via loop corrections. It is a crucial feature of these dualities that they 
involve vector models, rather than matrix-type theories. What is special about vector models is that their single trace 
spectrum is highly restricted, consisting only of operators which are bilinears in the fundamental fields. This is why 
a vector model can be dual to a pure HS gauge theory of Vasiliev type. While the emergence of conserved HS currents 
is a generic feature of weakly coupled field theories, the single trace spectrum of a free CFT with matrix-like fields 
would include, in addition to the conserved HS currents, infinite towers of operators 
that are not currents and are dual to massive fields in AdS. For example, in the case of free ${\cal N}=4$ SYM theory, this implies 
that the tensionless limit of the dual type IIB string theory on ${\rm AdS}_5 \times {\rm S}^5$ should be a higher spin gauge theory, 
coupled to infinite towers of massive fields. When the CFT interactions are turned on, 
conserved currents combine with operators in the appropriate representations to yield non-conserved currents; in the bulk, this should 
correspond to a HS version of the Higgs mechanism \cite{Beisert:2003te,Beisert:2004di}. 
In the Yang-Mills type theories, the anomalous dimension are generated at planar level, 
and the dual bulk HS fields should acquire masses already at classical level. In the case of vector models, on the other hand, 
the Higgsing happens at quantum level \cite{Girardello:2002pp}, since there are no single trace/single particle states in the spectrum 
with the appropriate quantum numbers. Studying the constraints on how to consistently couple the Vasiliev theories to matter fields 
is a very interesting open question, and would be a necessary step in order to understand the AdS dual of weakly coupled theories 
with adjoint fields. 

A distinguishing feature of the higher spin/vector model duality is that both sides 
of the correspondence are in principle under computational control in the same regime, 
since $1/N$ is the expansion parameter on both sides. 
This allows not only for direct quantitative tests of the correspondence, 
but it also suggests that for this class of models one might be able to prove 
the AdS/CFT correspondence, at least to all orders in perturbation theory. 
Several suggestions on how to derive the higher spin/vector model duality from first 
principles have appeared in  
\cite{Douglas:2010rc,Koch:2010cy,Jevicki:2011ss,Jevicki:2011aa,Leigh:2014tza, Leigh:2014qca, Mintun:2014gua}
(see also \cite{Das:2003vw} for relevant earlier work).

One outstanding issue that should be addressed in order to make progress in the HS/CFT duality 
is to understand how to properly quantize the Vasiliev theory. While this is currently not fully understood, 
partly due to the lack of a conventional action principle for Vasiliev equations, the recent one-loop results
reviewed in Section \ref{OneLoop} give preliminary evidence that quantum calculations in higher spin theories make 
sense, at least in a perturbative approach. It would be interesting to extend the existing 
one-loop calculations to the case of correlation functions. For example, the one-loop correction to the 
``boundary-to-boundary" two-point functions of the HS fields should encode the anomalous dimensions of the 
CFT currents: it should vanish in the case in which the dual is a free theory, but should be non-trivial 
when the duals are interacting models. This is especially interesting in the case of the parity violating HS theories, 
where one should reproduce the anomalous dimensions of the Chern-Simons 
vector models. Such calculations of bulk loop corrections to correlation functions may soon become 
feasible due to the progress \cite{Sleight:2016dba,Bekaert:2015tva} 
in perturbatively reconstructing the explicit cubic and quartic terms in the bulk Lagrangian. 

It is tempting to speculate that HS theories, due to the infinite dimensional symmetry, 
may provide examples of UV finite models of quantum gravity. The recent one-loop calculations of bulk partition functions 
reviewed above provide some partial evidence for this, as they show that UV divergences vanish at one-loop, in any 
spacetime dimensions and without supersymmetry.  If this persists to higher loops, HS theories may be examples 
of consistent theories of quantum gravity, 
and it is natural to ask whether such theories exist independently of string theory or if they may be derived from it. 
A concrete embedding of Vasiliev theory in type IIA string theory was proposed in \cite{Chang:2012kt} for a supersymmetric 
version of Vasiliev theory in AdS$_4$. It would be interesting to understand what happens in higher dimensions, and without supersymmetry. 

Much of the progress in understanding higher spin holography so far has involved studying small perturbations over the AdS vacuum, 
such as in the matching of correlation functions or in the calculation of one-loop partition functions. 
It is of clear interest to also study exact solutions of the Vasiliev equations, 
and understand their role in the holographic duality. In the context of the AdS$_3$ higher spin duality of 
\cite{Gaberdiel:2010pz, Gaberdiel:2012uj}, considerable progress has been made in constructing and studying black hole solutions. 
Much less is known in higher dimensions. In \cite{Shenker:2011zf} (see also \cite{Giombi:2014yra}) it was argued
from a calculation of the free energy of the 
singlet sector vector model on $S^1 \times S^2$, that the dual HS theory should not possess AdS-Schwarzschild black hole solutions in global AdS$_4$, 
essentially because the singlet constraint prevents order $N$ free energy at temperature $T\sim 1$ (in units of the radius of the sphere). 
On the other hand, an exact solution of the 4d HS theory 
was found in \cite{Didenko:2009td} (and generalized in \cite{Iazeolla:2011cb}) 
whose graviton sector appears to behave like a spherically symmetric charged black hole. 
It would be very interesting to understand the meaning of this solution in the context of the HS/CFT duality. 
Various deformations of the vector models, such as for instance adding a mass term in the boundary theory, 
should also correspond to some exact solutions of Vasiliev theory, and it would be interesting to find such solutions 
and study their implications for the holographic duality. 

\section*{Acknowledgements}

I would like to thank the organizers of TASI 2015 for the kind invitation and the 
opportunity to give these lectures. I also thank
the organizers of the 2013 GGI School on Higher Spins, String and Dualities; PiTP 2014; the 2014 
Mathematica Summer School on Theoretical Physics; and the 2015 Dynasty Summer School, 
where versions of these lectures were presented. 
I am deeply grateful to O. Aharony, L. Fei, G. Gur-Ari, V. Kirilin, 
I. Klebanov, J. Maldacena, S. Minwalla, S. Prakash, S. Pufu, B. Safdi, 
G. Tarnopolsky, S. Trivedi, A. Tseytlin, R. Yacoby, S. Wadia
and in particular X. Yin, for the fruitful collaborations on topics related to these lectures. I also 
thank I. Klebanov and A. Tseytlin for useful comments on earlier versions of these notes. 
This work is supported in part by the US NSF under Grant No.~PHY-1318681. 

\bibliographystyle{ws-rv-van}
\bibliography{HS-chapter}
\printindex
\end{document}